\documentclass[longauth]{aa} 

\usepackage{amsmath,amsfonts,amssymb}
\usepackage[dvipsnames]{xcolor}
\usepackage{color}
\usepackage{url}
\usepackage[varg]{txfonts}
\usepackage{graphicx,multirow}
\usepackage{enumitem} 
\usepackage{array}
\usepackage{dblfloatfix}
\usepackage{multicol}
\usepackage[normalem]{ulem} 
\usepackage{float}  
\usepackage{caption} 
\usepackage{rotating}
\usepackage{tikz}
\usetikzlibrary{shapes.geometric, arrows}

\usepackage{natbib,twoopt}
\usepackage[breaklinks=true]{hyperref} 
\bibpunct{(}{)}{;}{a}{}{,} 
\makeatletter
\newcommandtwoopt{\citeads}[3][][]{\href{http://adsabs.harvard.edu/abs/#3}%
{\def\hyper@linkstart##1##2{}%
\let\hyper@linkend\@empty\citealp[#1][#2]{#3}}}
\newcommandtwoopt{\citepads}[3][][]{\href{http://adsabs.harvard.edu/abs/#3}%
{\def\hyper@linkstart##1##2{}%
\let\hyper@linkend\@empty\citep[#1][#2]{#3}}}
\newcommandtwoopt{\citetads}[3][][]{\href{http://adsabs.harvard.edu/abs/#3}%
{\def\hyper@linkstart##1##2{}%
\let\hyper@linkend\@empty\citet[#1][#2]{#3}}}
\newcommandtwoopt{\citeyearads}[3][][]%
{\href{http://adsabs.harvard.edu/abs/#3}
{\def\hyper@linkstart##1##2{}%
\let\hyper@linkend\@empty\citeyear[#1][#2]{#3}}}
\makeatother

\usepackage{comment}

\usepackage{color}
\hypersetup{colorlinks=true,linkcolor=blue,citecolor=blue,urlcolor=blue}

\usepackage[breaklinks=true]{hyperref} 
\makeatletter
\renewcommand*\aa@pageof{, page \thepage{} of \pageref*{LastPage}}
\makeatother

\usepackage{xcolor}

\begin{document} 

\title{Connecting photometric and spectroscopic granulation signals with CHEOPS and ESPRESSO}
   
   \subtitle{}

\author{
S. Sulis$^{1}$, 
M. Lendl$^{2}$, 
H.~M. Cegla$^{3,4}$\fnmsep\thanks{UKRI Future Leaders Fellow}, 
L. F. Rodr\'iguez D\'iaz$^{5}$, 
L. Bigot$^{6}$, 
V. Van Grootel$^{7}$, 
A. Bekkelien$^{2}$, 
A. Collier Cameron$^{8}$, 
P. F. L. Maxted$^{9}$, 
A. E. Simon$^{10}$, 
C. Lovis$^{2}$, 
G. Scandariato$^{11}$, 
G. Bruno$^{11}$, 
D. Nardiello$^{12}$, 
A. Bonfanti$^{13}$, 
M. Fridlund$^{14,15}$, 
C. M. Persson$^{15}$, 
S. Salmon$^{2}$, 
S. G. Sousa$^{16}$, 
T. G. Wilson$^{8}$, 
A. Krenn$^{2,13}$, 
S. Hoyer$^{1}$, 
A. Santerne$^{1}$, 
D. Ehrenreich$^{2}$, 
Y. Alibert$^{10}$, 
R. Alonso$^{17,18}$, 
G. Anglada$^{19,20}$, 
T. Bárczy$^{21}$, 
D. Barrado y Navascues$^{22}$, 
S. C. C. Barros$^{16,23}$, 
W. Baumjohann$^{13}$, 
M. Beck$^{2}$, 
T. Beck$^{10}$, 
W. Benz$^{10,24}$, 
N. Billot$^{2}$, 
X. Bonfils$^{25}$,
L. Borsato$^{12}$,
A. Brandeker$^{26}$, 
C. Broeg$^{10,27}$, 
J. Cabrera$^{28}$, 
S. Charnoz$^{29}$, 
C. Corral van Damme$^{30}$, 
Sz. Csizmadia$^{28}$, 
M. B. Davies$^{31}$, 
M. Deleuil$^{1}$, 
A. Deline$^{2}$, 
L. Delrez$^{32,7}$, 
O. D. S. Demangeon$^{16,23}$, 
B.-O. Demory$^{24}$, 
A. Erikson$^{28}$, 
A. Fortier$^{10,24}$, 
L. Fossati$^{13}$, 
D. Gandolfi$^{33}$, 
M. Gillon$^{32}$, 
M. Güdel$^{34}$, 
K. Heng$^{24,35}$, 
K. G. Isaak$^{36}$, 
L. L. Kiss$^{37,38}$, 
J. Laskar$^{39}$, 
A. Lecavelier des Etangs$^{40}$, 
D. Magrin$^{12}$, 
M. Munari$^{11}$, 
V. Nascimbeni$^{12}$, 
G. Olofsson$^{26}$, 
R. Ottensamer$^{41}$, 
I. Pagano$^{11}$, 
E. Pallé$^{17}$, 
G. Peter$^{42}$, 
G. Piotto$^{12,43}$, 
D. Pollacco$^{35}$, 
D. Queloz$^{44,45}$, 
R. Ragazzoni$^{12,43}$, 
N. Rando$^{30}$, 
H. Rauer$^{28,46,47}$, 
I. Ribas$^{19,20}$, 
M. Rieder$^{10}$, 
N. C. Santos$^{16,23}$, 
D. Ségransan$^{2}$, 
A. M. S. Smith$^{28}$, 
M. Steinberger$^{13}$, 
M. Steller$^{13}$, 
Gy. M. Szabó$^{48,49}$, 
N. Thomas$^{10}$, 
S. Udry$^{2}$, 
N. A. Walton$^{50}$, 
D. Wolter$^{28}$, 
}

    \authorrunning{S. Sulis et al.}

\institute{    
$^{1}$ Aix Marseille Univ, CNRS, CNES, LAM, 38 rue Frédéric Joliot-Curie, 13388 Marseille, France\\
$^{2}$ Observatoire Astronomique de l'Université de Genève, Chemin Pegasi 51, Versoix, Switzerland\\
$^{3}$ Centre for Exoplanets and Habitability, University of Warwick, Gibbet Hill Road, Coventry CV4 7AL, UK\\
$^{4}$ Department of Physics, University of Warwick, Gibbet Hill Road, Coventry CV4 7AL, UK\\
$^{5}$ Stellar Astrophysics Centre, Department of Physics and Astronomy, Aarhus University, Ny Munkegade 120, DK-8000 Aarhus C, Denmark\\
$^{6}$ Université Nice Sophia Antipolis, Observatoire de la Côte d'Azur, Département Cassiopée CNRS/UMR 6202, BP 4229, 06304, Nice, France\\
$^{7}$ Space sciences, Technologies and Astrophysics Research (STAR) Institute, Université de Liège, Allée du 6 Août 19C, 4000 Liège, Belgium\\
$^{8}$ Centre for Exoplanet Science, SUPA School of Physics and Astronomy, University of St Andrews, North Haugh, St Andrews KY16 9SS, UK\\
$^{9}$ Astrophysics Group, Keele University, Staffordshire, ST5 5BG, United Kingdom\\
$^{10}$ Physikalisches Institut, University of Bern, Sidlerstrasse 5, 3012 Bern, Switzerland\\
$^{11}$ INAF, Osservatorio Astrofisico di Catania, Via S. Sofia 78, 95123 Catania, Italy\\
$^{12}$ INAF, Osservatorio Astronomico di Padova, Vicolo dell'Osservatorio 5, 35122 Padova, Italy\\
$^{13}$ Space Research Institute, Austrian Academy of Sciences, Schmiedlstrasse 6, A-8042 Graz, Austria\\
$^{14}$ Leiden Observatory, University of Leiden, PO Box 9513, 2300 RA Leiden, The Netherlands\\
$^{15}$ Department of Space, Earth and Environment, Chalmers University of Technology, Onsala Space Observatory, 439 92 Onsala, Sweden\\
$^{16}$ Instituto de Astrofisica e Ciencias do Espaco, Universidade do Porto, CAUP, Rua das Estrelas, 4150-762 Porto, Portugal\\
$^{17}$ Instituto de Astrofisica de Canarias, 38200 La Laguna, Tenerife, Spain\\
$^{18}$ Departamento de Astrofisica, Universidad de La Laguna, 38206 La Laguna, Tenerife, Spain\\
$^{19}$ Institut de Ciencies de l'Espai (ICE, CSIC), Campus UAB, Can Magrans s/n, 08193 Bellaterra, Spain\\
$^{20}$ Institut d'Estudis Espacials de Catalunya (IEEC), 08034 Barcelona, Spain\\
$^{21}$ Admatis, 5. Kandó Kálmán Street, 3534 Miskolc, Hungary\\
$^{22}$ Depto. de Astrofisica, Centro de Astrobiologia (CSIC-INTA), ESAC campus, 28692 Villanueva de la Cañada (Madrid), Spain\\
$^{23}$ Departamento de Fisica e Astronomia, Faculdade de Ciencias, Universidade do Porto, Rua do Campo Alegre, 4169-007 Porto, Portugal\\
$^{24}$ Center for Space and Habitability, University of Bern, Gesellschaftsstrasse 6, 3012 Bern, Switzerland\\
$^{25}$ Université Grenoble Alpes, CNRS, IPAG, 38000 Grenoble, France\\
$^{26}$ Department of Astronomy, Stockholm University, AlbaNova University Center, 10691 Stockholm, Sweden\\
$^{27}$ Center for Space and Habitability, Gesellsschaftstrasse 6, 3012 Bern, Switzerland\\
$^{28}$ Institute of Planetary Research, German Aerospace Center (DLR), Rutherfordstrasse 2, 12489 Berlin, Germany\\
$^{29}$ Université de Paris, Institut de physique du globe de Paris, CNRS, F-75005 Paris, France\\
$^{30}$ ESTEC, European Space Agency, 2201AZ, Noordwijk, NL\\
$^{31}$ Centre for Mathematical Sciences, Lund University, Box 118, 221 00 Lund, Sweden\\
$^{32}$ Astrobiology Research Unit, Université de Liège, Allée du 6 Août 19C, B-4000 Liège, Belgium\\
$^{33}$ Dipartimento di Fisica, Universita degli Studi di Torino, via Pietro Giuria 1, I-10125, Torino, Italy\\
$^{34}$ University of Vienna, Department of Astrophysics, Türkenschanzstrasse 17, 1180 Vienna, Austria\\
$^{35}$ Department of Physics, University of Warwick, Gibbet Hill Road, Coventry CV4 7AL, United Kingdom\\
$^{36}$ Science and Operations Department - Science Division (SCI-SC), Directorate of Science, European Space Agency (ESA), European Space Research and Technology Centre (ESTEC),
Keplerlaan 1, 2201-AZ Noordwijk, The Netherlands\\
$^{37}$ Konkoly Observatory, Research Centre for Astronomy and Earth Sciences, 1121 Budapest, Konkoly Thege Miklós út 15-17, Hungary\\
$^{38}$ ELTE E\"otv\"os Lor\'and University, Institute of Physics, P\'azm\'any P\'eter s\'et\'any 1/A, 1117\\
$^{39}$ IMCCE, UMR8028 CNRS, Observatoire de Paris, PSL Univ., Sorbonne Univ., 77 av. Denfert-Rochereau, 75014 Paris, France\\
$^{40}$ Institut d'astrophysique de Paris, UMR7095 CNRS, Université Pierre \& Marie Curie, 98bis blvd. Arago, 75014 Paris, France\\
$^{41}$ Department of Astrophysics, University of Vienna, Tuerkenschanzstrasse 17, 1180 Vienna, Austria\\
$^{42}$ Institute of Optical Sensor Systems, German Aerospace Center (DLR), Rutherfordstrasse 2, 12489 Berlin, Germany\\
$^{43}$ Dipartimento di Fisica e Astronomia "Galileo Galilei", Universita degli Studi di Padova, Vicolo dell'Osservatorio 3, 35122 Padova, Italy\\
$^{44}$ ETH Zurich, Department of Physics, Wolfgang-Pauli-Strasse 2, CH-8093 Zurich, Switzerland\\
$^{45}$ Cavendish Laboratory, JJ Thomson Avenue, Cambridge CB3 0HE, UK\\
$^{46}$ Zentrum für Astronomie und Astrophysik, Technische Universität Berlin, Hardenbergstr. 36, D-10623 Berlin, Germany\\
$^{47}$ Institut für Geologische Wissenschaften, Freie Universität Berlin, 12249 Berlin, Germany\\
$^{48}$ ELTE E\"otv\"os Lor\'and University, Gothard Astrophysical Observatory, 9700 Szombathely, Szent Imre h. u. 112, Hungary\\
$^{49}$ MTA-ELTE Exoplanet Research Group, 9700 Szombathely, Szent Imre h. u. 112, Hungary\\
$^{50}$ Institute of Astronomy, University of Cambridge, Madingley Road, Cambridge, CB3 0HA, United Kingdom\\
}

 \date{accepted to A\&A}

 \abstract
 {Stellar granulation generates fluctuations in photometric and spectroscopic data whose properties depend on the stellar type, composition, and evolutionary state.  Characterizing granulation is key for understanding stellar atmospheres and detecting planets.
 }
   {We aim to detect the signatures of stellar granulation, link spectroscopic and photometric signatures of convection for main-sequence stars, and test predictions from 3D hydrodynamic models.}
 {For the first time, we observed two bright stars ($T_{\mathrm{eff}} = 5833$ K and $6205$ K) with high-precision observations taken simultaneously with CHEOPS and ESPRESSO. We analyzed the properties of the stellar granulation signal in each individual dataset. We compared them to Kepler observations and 3D hydrodynamic models. While isolating the granulation-induced changes by attenuating and filtering the p-mode oscillation signals, we studied the relationship between photometric and spectroscopic observables. 
 }
 {The signature of stellar granulation is detected and precisely characterized for the hotter F star in the CHEOPS and ESPRESSO observations. For the cooler G star, we obtain a clear detection in the CHEOPS dataset only. The TESS observations are blind to this stellar signal. Based on CHEOPS observations, we show that the inferred properties of stellar granulation are in agreement with both Kepler observations and hydrodynamic models. Comparing their {periodograms}, we observe a strong link between spectroscopic and photometric observables. Correlations of this stellar signal in the time domain (flux versus radial velocities, RV) and with specific spectroscopic observables (shape of the cross-correlation functions) are however difficult to isolate due to S/N dependent variations. 
}
  {
  In the context of the upcoming PLATO mission and the extreme precision RV surveys, a thorough understanding of the properties of the stellar granulation signal is needed. The CHEOPS and ESPRESSO observations pave the way for detailed analyses of this stellar process.
  }

 \maketitle

\section{Introduction }
\label{sec1}

Stellar convection transports energy from the stellar interior to its surface in late-type stars.
The properties of this complex and multiscale plasma mixing process are key for understanding stellar structure and evolution, as the dynamics of the convective cells shape angular momentum transport within the star, impact the thermal stellar stratification, mix the chemical elements, and generate the surface acoustics modes  (see e.g., \citeads{1990A&A...228..184D}; \citeads{Stein2001}; \citeads{2014A&A...565A..42A}; \citeads{2015LRSP...12....8H}; \citeads{2017RSOS....470192S}; \citeads{2017LRSP...14....4B}; \citeads{2020A&A...644A.171P}).

This stellar phenomenon is well-studied for the Sun where it is visible in the form of granules.
For other stars, granulation is studied through indirect techniques. Two of them are photometry -- brightness fluctuations -- and spectroscopy -- radial velocity (RV) changes.

Through photometric observations, the properties of granulation as a function of stellar parameters have been revealed: CoRoT observations have shown that the granulation timescale and amplitudes decrease with the increasing characteristic frequency of the acoustic modes, the so-called frequency at maximum power $\nu_{max}$, which depends on the stellar surface gravity and temperature \citepads{2008Sci...322..558M}. 
In line with model predictions (see e.g., \citeads{2005ESASP.560..979S}; \citeads{Ludwig2006}; \citeads{Ludwig2013};  \citeads{Trampedach2013}; \citeads{Tremblay2013}; \citeads{2013A&A...558A..49B}; \citeads{2013A&A...559A..40S};  \citeads{2015A&A...581A..43B}), numerous studies based on Kepler observations (\citeads{2011ApJ...741..119M}; \citeads{2013Natur.500..427B}; \citeads{2014A&A...570A..41K}; \citeads{2016ApJ...818...43B}; \citeads{2018MNRAS.480..467P}; \citeads{2018A&A...620A..38B}; \citeads{2019ApJ...883..195T}; \citeads{2020A&A...636A..70S}; \citeads{Rodriguez2022}) have also shown that the granulation properties are dependent on the stellar fundamental parameters ($T\rm_{eff}$, ${\rm log}g$, and [Fe/H]). Granule size increases with lower stellar surface gravities and/or larger effective temperatures, and therefore with a decreasing $\nu_{max}$. The stellar photometric signal is the net contribution of all the bright granules and dark intergranular lanes on the stellar surface, which reduces the disk-integrated fluctuations (compared to the scale of the granules) and therefore depends on the stellar radius (\citeads{Trampedach1998}; \citeads{Ludwig2006}).

In spectroscopic observations, granulation produces shifts and asymmetries in the spectral lines (i.e., the line bisectors, see \citeads{1981A&A....96..345D}; \citeads{1985SoPh..100..209N}; \citeads{1987A&A...172..200D}; \citeads{2000A&A...359..729A}; \citeads{2000A&A...359..729A}; \citeads{2009ApJ...697.1032G};  \citeads{Nordlund2009}; \citeads{dravins21}). Studying RV time series of a small sample of main-sequence and subgiant stars, \citetads{2014AJ....147...29B} found  a correlation between the RV root-mean-square (RMS) evolving on timescales shorter than $8$ hours (or the ``8-hour RV scatter'')  and the stellar surface gravity. This implies a narrow relation between the RV amplitudes driven by stellar granulation signals and the associated intensities (\citeads{2013Natur.500..427B}; \citeads{Kallinger2016}).
{As a result, various empirical relations that predict the amplitude of the granulation signal in RV from photometric observations have been derived (see e.g., \citeads{2014AJ....147...29B}; \citeads{2014ApJ...780..104C}; \citeads{2017A&A...606A.107O}), but they have shown significant discrepancies in the predicted RV.}

Three-dimensional (3D) hydrodynamical (HD) and magneto-hydrodynamical (MHD) simulations of stellar convection have been developed since the 1980s (\citeads{Nordlund1982}; \citeads{1985SoPh..100..209N}; \citeads{1990A&A...228..155N}; \citeads{1990A&A...228..203D}). While computationally expensive (leading to long time series difficult to generate), they have been recently used to generate disk-integrated spectra \citepads{2019ApJ...879...55C} and RV time series \citepads{2020A&A...635A.146S} of solar granulation. They have also been used to generate synthetic brightness fluctuations whose standard deviation and autocorrelation time match those of Kepler targets from dwarfs to red giants with an overall very good agreement \citepads{Rodriguez2022}. Moreover,  \citetads{2019ApJ...879...55C} predicted correlations between the photometric and spectroscopic signals of stellar granulation, but this has yet to be confirmed with high-precision observations.  They also predict that such observed correlations may allow us to mitigate a significant fraction of the granulation variability in RV observations \citepads{2019ApJ...879...55C}.

The variability induced by granulation constitutes a significant noise source hampering the detection of smaller stellar signals, like low-amplitude acoustic or gravity modes (see e.g., \citeads{2019LRSP...16....4G} for solar-like oscillations, \citeads{2016MNRAS.457.1851R} for M-dwarfs pulsations, and \citeads{2010A&ARv..18..197A} for g-modes in the Sun) and planetary signals (\citeads{2011A&A...525A.140D}; \citeads{2015A&A...583A.118M}; \citeads{2020A&A...642A.157M}).  
From solar observations, we learn that granulation can generate variability with RMS around $40$ parts-per-million (ppm) in photometry (\citeads{dravins88}, \citeads{1997SoPh..170....1F}; \citeads{2004A&A...414.1139A}; \citeads{2020A&A...636A..70S}) and around $30$-$46$ cm/s in RVs (\citeads{10.1093/mnras/269.3.529}; \citeads{1999ASPC..173..297P}; \citeads{Garcia_2005}; \citeads{2018A&A...617A.108A}; \citeads{2019MNRAS.487.1082C}; \citeads{2021A&A...648A.103D}). This is significant compared to the signals expected from Earth-like planets around Sun-like stars: a transit depth of $84$ ppm and a Keplerian RV signal of $9$ cm/s amplitude \citepads{2018exha.book.....P}.  Since the correlation timescales of the granulation-induced noise are similar to the ingress/egress transit duration of long orbital period exoplanets, this stellar signal affects the planetary parameters inferred on the individual transits \citepads{2020A&A...636A..70S}.
Thus, the correlated noise due to stellar granulation can be seen as a source of information to study stellar physics properties but at the same time as a nuisance signal for the detection of exoplanets and stellar oscillation modes. In both cases it needs to be understood, quantified and if possible mitigated.

In this paper, we have three objectives. 
First, we aim to detect the signatures of stellar granulation based on the granulation indicators that have been recently developed, to test their predictions against high-precision CHEOPS measurements, and to compare the performances of CHEOPS for probing the photometric signature of stellar granulation to Kepler and TESS data.
Second, we aim to study the link between the spectroscopic and photometric signatures of convection for main-sequence stars with high-precision ESPRESSO observations, taken contemporaneously with CHEOPS.
Third, we aim to test the predictions from 3D hydrodynamical models of convection. \\
For these purposes, we present the analyses of high-precision CHEOPS and ESPRESSO measurements of two bright stars with different effective temperatures. 

The paper is organized as follows.
We describe the data reduction methodologies in Sec.~\ref{sec2}. We refine the stellar parameters of these two targets in Sec.~\ref{sec2b}. We analyze the stellar granulation signals for each CHEOPS, TESS and ESPRESSO observations in Sec.~\ref{sec3}. We study the links between granulation and stellar properties in Sec.~\ref{sec4}. We compare our results with the predictions from 3D hydrodynamical convection models in Sec.~\ref{sec5}. We constrain the link between spectroscopic and photometric signatures in Sec.~\ref{sec6}. We investigate the relationship between the shape of the cross-correlation functions, flux (as measured by CHEOPS), and RV in Sec.~\ref{sec7}. We conclude in Sec.~\ref{ccl}. 

\section{Observations and data reduction} 
\label{sec2}

We selected HD 67458 ($T_{\mathrm{eff}} = 5833$ K) and HD 88595 ($T_{\mathrm{eff}} = 6205$ K) as good targets based on their stellar parameters, their low apparent magnitudes, their moderate level of magnetic activity, the a priori absence of identified planets, and the excellent CHEOPS and ESPRESSO visibility windows during the time that was allocated to our program. Our dataset includes $35$ CHEOPS orbits, six dedicated nights at ESPRESSO/VLT, and five TESS sectors of observations.  We note this is the first time that high-precision spectroscopic data with ESPRESSO have been taken during several full nights for single stars.

\renewcommand{\arraystretch}{1.2}
\begin{table*}[b]\centering
\begin{tabular}{ccccccccc}
 \hline\hline
Target & Visit  & File key & Starting date & $T$ & $\tau$ & $N$ & DC & {$\sigma_{TOT}$}  \\
& ref. number&  & [UTC] & [hrs] & [s] &  & [$\%$] & [ppm] \\ 
\hline
& Visit 1 & CH\_PR100022\_TG001101\_V0200 & 2021-01-17 &8.15 &34.3 &789 & $92.5$ & 114 \\
HD 67458 & Visit 2 & CH\_PR100022\_TG001301\_V0200   &2021-01-18   &8.61 &34.3 &834 & $92.5$ & 110 \\
  & Visit 3 & CH\_PR100022\_TG001201\_V0200 & 2021-01-21  &8.15 &34.3 &807 & $94.6$ & 125 \\
\hline \hline
& Visit 1 & CH\_PR100022\_TG001601\_V0200  &2021-02-13  &8.06 &37 & 728 & 93.5 & 100 \\
HD 88595 & Visit 2 & CH\_PR100022\_TG001501\_V0200 &2021-02-14 &8.15 &37 & 698 & 88.6 & 104 \\
& Visit 3 & CH\_PR100022\_TG001401\_V0200 & 2021-02-15 & 8.14 & 37 & 718 & 91.2 & 108 \\ 
& Visit 4 & CH\_PR100022\_TG001701\_V0200  &2021-02-22   &8.14 &37 & 738 & 93.8 & 103 \\
\hline
\end{tabular}
\caption{CHEOPS observations: file keys referring to the files name in the CHEOPS database, dates (UTC), total observation duration ($T$), integration time ($\tau$), number of observations ($N$), duty cycle (DC), and standard deviation of the light curve ($\sigma_{TOT}$).}
\label{tab:table_cheops}
\end{table*}
%
%
\renewcommand{\arraystretch}{1.2}
\begin{table*}[b]\centering
\begin{tabular}{cccccccccc}
\hline\hline
Target & Night & Starting date & $T$ & Min/Max & $N$ & Min/Max  & DC & $\sigma_{TOT}$ & {${\rm log} R'_{\rm HK}$}\\ 
& ref. number & [UTC] & [hrs] & airmass &  & $\sigma_{RV}$ [m/s] & [$\%$] & [m/s] & \\ 
\hline
& Night 1 & 2021-01-17 &  $4.85$ & [1, 2.0] & {$155$} & [$0.4, 1.1$] & {$91.8$} & {$2.6$} &{-4.97} \\ 
HD 67458& Night 2 & 2021-01-18 &  $8.48$ & [1, 1.8] & $273$ & [$0.5, 2.7$] & $87.5$ & $2.0$ &{-5.20}\\ 
 & Night 3 & 2021-01-21  &  $4.75$ & [1, 1.7] & $126$ & [$0.4, 0.9$] & $72.0$ & $1.1$&{-4.96} \\ 
\hline \hline
HD 88595& Night 1  & 2021-02-14  &  $8.8$ & [1, 2.3] & $308$ &[$0.7, 2.3$] & $95.0$ & $2.54$ &{-4.96} \\
 & Night 2  & 2021-02-22 &  $9.0$  & [1, 1.9] & $313$ & [$0.7, 3.7$] & $94.0$ & $2.26$ &{-4.97} \\
\hline
\end{tabular}
\caption{ESPRESSO observations: date of observation (UTC), total observation duration ($T$), Min and Max values of airmass during the data acquisition, number of observations ($N$), Min and Max RV errorbars ($\sigma_{RV}$), duty cycle (DC), standard deviation of the detrended RV time series ($\sigma_{TOT}$), {and median value of the ${\rm log} R'_{\rm HK}$ activity indicator}.}
\label{tab:table_espresso}
\end{table*}


\subsection{CHEOPS}
\label{sec22}

CHEOPS \citepads{2021ExA....51..109B} observed the two bright stars in the visible wavelength range ($330-1100$ nm). 
HD 67458 was observed during three visits of $T\sim8.15$ hours each, with a time sampling of $\Delta t= 34.3$ seconds. Each of these measurements resulted from $7$ individual images taken with an exposure time of $\tau_{exp}=4.9$ seconds and stacked on-board by coadding them pixel-by-pixel. 
HD 88595 was observed during four visits of $T\sim8.15$ hours each, with a time sampling of $\Delta t= 37$ seconds. Each of these measurements resulted from $10$ stacked images taken with an exposure time of $\tau_{exp}=3.7$ seconds.

The duty cycle of each visits was close to $90\%$, with some gaps present within the light curves. The origin of these gaps is due to Earth occultations and the South Atlantic Anomaly (SAA) crossing of the satellite. Each series has one or two large gaps $<17.5$ min and five to ten short gaps $<2$ min. In total, the three visits of  HD 67458 contain $N=\{789, 807, 834\}$ measurements, while the four visits of HD 88595 contain $N=\{718, 698, 728, 738\}$ measurements.

All CHEOPS observations were processed with the Data Reduction Pipeline (DRP, version 13.1.0), which is described in \citetads{2020A&A...635A..24H}. 
We used the light curves provided by the DRP. We test different apertures to minimize the contribution from the background contaminants, and finally choose the default aperture of $25$ pixels radius. 

\begin{figure*}[t!]
\centering
\resizebox{\hsize}{!}{\includegraphics{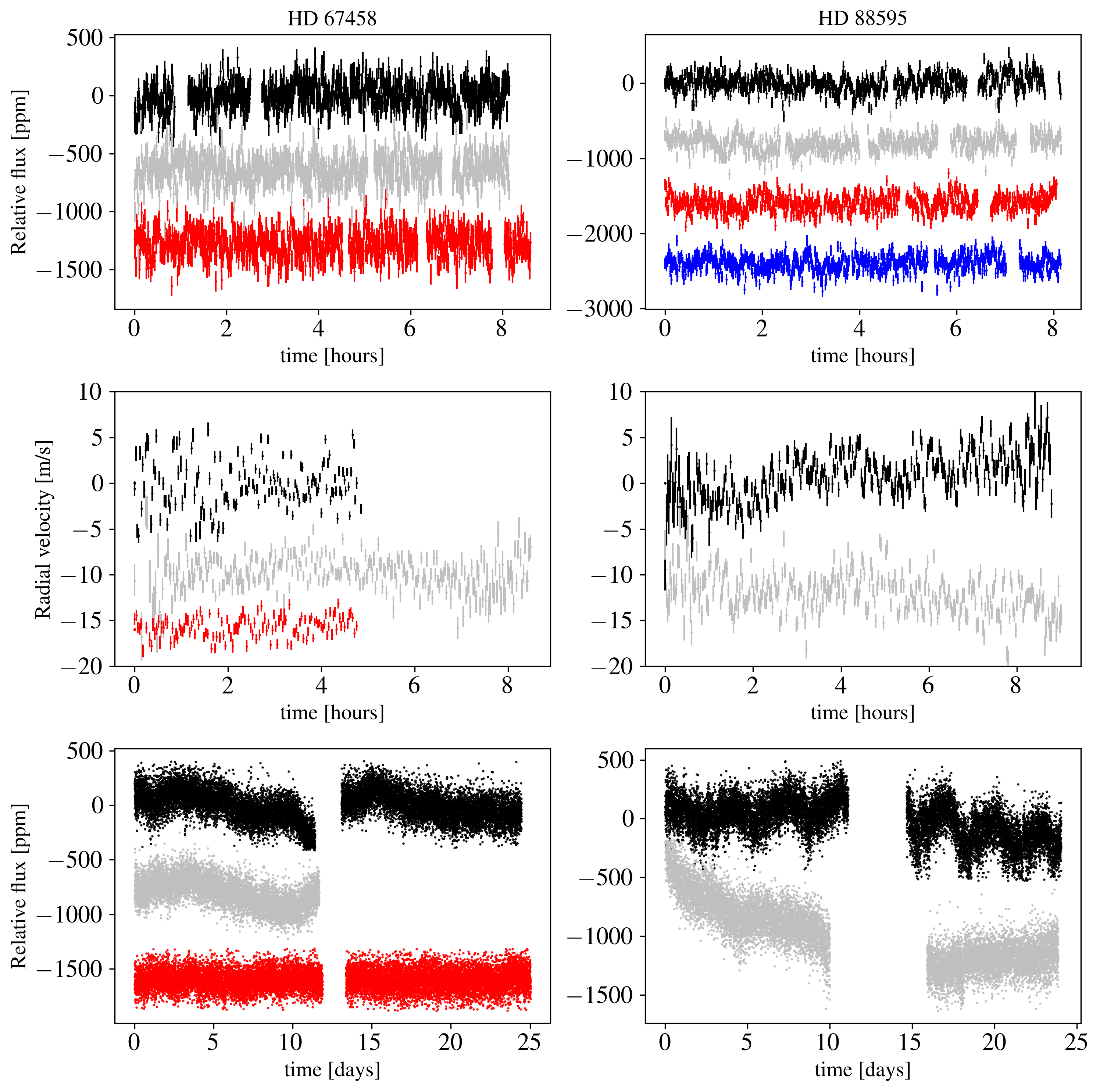}}
\caption{Detrended CHEOPS light curves (top row), ESPRESSO radial velocity time series (middle row), and TESS light curves (bottom row) for HD 67458 (left panels) and HD 88595 (right panels). In each panel, the colors indicate a different set of observations (CHEOPS visits, ESPRESSO nights, TESS sectors). Observations are y-shifted for visibility, and the dates of each time series are normalized to start at $0$ (with a display that goes from the earliest date on the top, to the latest one on the bottom).
}
\label{fig_data}  
\end{figure*}

We detrended each CHEOPS visit from the satellite systematics using the \texttt{Pycheops} software\footnote{\url{https://github.com/pmaxted/pycheops}} (version 1.0.6, \citeads{2022MNRAS.514...77M}).
We first corrected for contaminating flux from background sources, with the contamination estimation in Pycheops based on the Gaia DR2 catalog. We then performed a $5\sigma$ clipping to each visit, before detrending from the x and y centroid variations, roll angle, background, and smear systematics.
The main characteristics of these observations are listed in Table~\ref{tab:table_cheops}, and the detrended light curves are shown in the top rows of Fig.~\ref{fig_data}.

%
%
\renewcommand{\arraystretch}{1.2}
\begin{table*}[t]\centering
\caption{{TESS observations: date of observation (UTC), total observation duration ($T$), integration time ($\tau$), number of observations ($N$), duty cycle (DC), and standard deviation of the light curve ($\sigma_{TOT}$).}}
\begin{tabular}{cccccccc}
\hline\hline
Target & Sector & Starting date & $T$ & $\tau$ & $N$ & DC & $\sigma_{TOT}$ \\ 
       & number & [UTC]         & [days] & [s]  &     & [$\%$] & [ppm] \\ 
\hline
          & 7 & 2019-01-08 &  24.4  &  120  &  16283  &  92.5  &  134 \\ 
HD 67458  & 8 & 2019-02-02   &  24.6  &  120  &  13369  &  75.5  &  275 \\
          & 34 & 2021-01-14 &  25.0  &  120  &  16802  &  93.2 & 92\\
\hline \hline
HD 88595  & 9 & 2019-02-28  &  24.0  &  120  &  15164  &  87.6  &  175 \\
          & 35 & 2021-02-09 & 23.8 & 120 & 12866 & 74.9 & 269 \\
\hline
\end{tabular}
\label{tab:table_tess}
\end{table*}


\subsection{ESPRESSO}
\label{sec23}

\begin{figure}[t!]
\centering
\resizebox{\hsize}{!}{\includegraphics{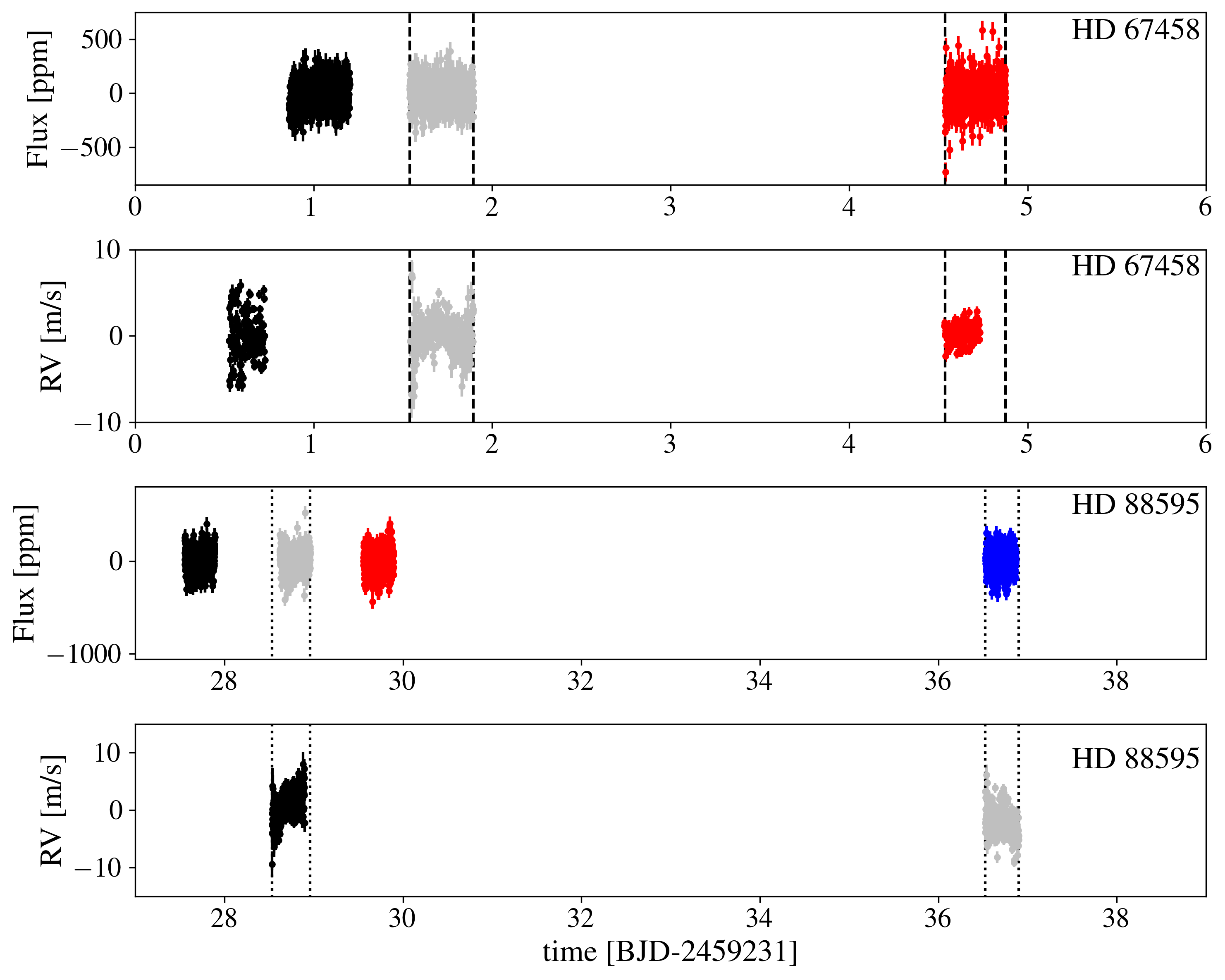}}
\caption{Contemporaneous observations of HD 67458 (two first rows) and HD 88595 (two last rows) taken with CHEOPS and ESPRESSO (same color code as in Fig.~\ref{fig_data}: black, gray, red and blue dots). For each star, there are two simultaneous sets of CHEOPS/ESPRESSO observations {shown with the dashed and dotted vertical lines.}
}
\label{fig_dates}  
\end{figure}

ESPRESSO \citepads{2013Msngr.153....6P} observed the two targets in high-resolution spectroscopy ($R\sim  140 000$) in the wavelength range $\lambda \in [380,780]$ nm, during six full nights\footnote{Program IDs: 106.2186.001 {to 106.2186.006.}}. Each measurement resulted from an exposure time of $\tau_{exp}=60$ seconds. 
The ESPRESSO observations have been taken simultaneously or contemporaneously with the CHEOPS observations (see Fig.~\ref{fig_dates}).
They are the first ESPRESSO observations taken during full nights for such a program. 

Among these six nights of observations at the VLT, one\footnote{Program ID 106.2186.005 {(UTC time: 2021/02/15)}.} encountered significant technical problems (PLC-ADC communication issues). We remove this RV time series from this study and are left with three RV time series for the solar-like star HD 67458, and two for the hotter star HD 88595. 

All ESPRESSO data were processed with the Data Reduction Software (DRS) provided with the instrument and publicly available from the ESO pipeline repository\footnote{\url{www.eso.org/sci/software/pipelines/}}. We refer to \citetads{2021A&A...645A..96P} for a short description of the main processing steps. Radial velocities were obtained by cross-correlation with binary {F9 and F7 masks for HD 67458 and HD 88595, respectively. We note that} we had to process the HD 88595 data enforcing an F7 spectral type to override the (wrong) F5 spectral type provided in the OCS.OBJ.SP.TYPE FITS keyword. Given the exquisite short-term RV precision required to characterize stellar granulation, we paid extra attention to the instrumental drift measured on fiber B and the blue-to-red flux balance in the CCF computation. Both effects may introduce instrumental RV systematics within the night if not properly calibrated out. 

From the processed RV time series of the solar-like star HD 67458, we removed the mean value and we clipped out the {$2\sigma$ outliers for all nights.} 
The standard deviation of the three final RV time series are {$2.6$}, $2.0$ and $1.1$ m/s, respectively. The errorbars are between $0.4$ and $2.7$ m/s. 
{We find that the RV dispersion changed significantly between the different nights, but also during a given night. In particular, during the first night we found an RMS of $\sim 3.4$ m/s during the first two hours of acquisition and $\sim 2.0$ m/s during the rest of the night. As we detail in Sec.~\ref{sec31}, this discrepancies are certainly of instrumental origin.}

From the processed RV time series of the F star HD 88595, we also removed the mean value and we clipped out the $5\sigma$ outliers for both nights.
The standard deviations of the two final RV time series are $2.54$ and $2.26$ m/s respectively. The errorbars are between $0.7$ and $3.7$ m/s. \\
The characteristics of these observations are listed in Table~\ref{tab:table_espresso}, and the RV time series are shown in the middle rows of Fig.~\ref{fig_data}. {For both targets, we also report in Table~\ref{tab:table_espresso} the median value of the chromospheric activity indicator ${\rm log} R'_{\rm HK}$ \citepads{1984ApJ...279..763N}, which is based on the intensity of the Ca II H\&K reemission lines. For both stars, we found ${\rm log} R'_{\rm HK}\sim-4.9$, which is consistent with relatively inactive stars \citepads{2008LRSP....5....2H}. Moreover, as no significant variation of this activity indicator is observed during the nights, the stellar magnetic activity is probably not at the origin of the variability observed in the data of HD 67458.}


\subsection{TESS}
\label{sec24}

{TESS \citepads{2015JATIS...1a4003R} observed the two stars in the red-optical bandpass ($600-1100$ nm). HD 67458 was observed in sectors \{7, 8, 34\}, and HD 88595 in sectors \{9, 35\}.}

In this work we made use of the short-cadence light curves released by the TESS team. However, we do not use the Presearch Data Conditioning Simple Aperture Photometry (PDCSAP) flux because sometimes they are affected by some systematic errors due to over-corrections and/or injection of spurious signals. The light curves we used were obtained applying Cotrending Basis Vectors (CBVs) to the Simple Aperture Photometry (SAP) flux as done in \citet{2021MNRAS.505.3767N}. The CBV were obtained by using the SAP light curves of the stars in the same Camera/CCD in which the targets are located and following the procedure described in detail in \citet{2019MNRAS.490.3806N,2020MNRAS.495.4924N}. For the analysis of the light curves, we rejected the points with \texttt{DQUALITY>0}, as recommended by the {\it TESS} Science Data Products Description Document\footnote{\url{https://archive.stsci.edu/files/live/sites/mast/files/home/missions-and-data/active-missions/tess/_documents/EXP-TESS-ARC-ICD-TM-0014-Rev-F.pdf}}.
Additionally, we clipped out the points corresponding to a sky background value $>3040$ e$^{-}$/s in sector 35 (since this sector was affected by known technical issues such as thermal stability and telescope pointing). We finally clipped out the $3\sigma$ outliers of each sectors for both targets.
The main characteristics of these observations are listed in Table~\ref{tab:table_tess}, and the lightcurves are shown in the bottom rows of Fig.~\ref{fig_data}. 

{For comparison with the CHEOPS observations in the following, we extracted subseries of $8$-hours duration from each TESS sector. Then, we removed the subseries affected by large gaps and kept the subseries for which the duty cycle was $>90\%$. We obtained $L=172$ subseries for HD 67458, and $L=113$ for HD 88595.}


\subsection{The Sun as a reference star}
\label{sec25}

As a reference star, we refer throughout the paper to the Sun that has been observed continuously from space with the \textit{Solar and Heliospheric Observatory} (SoHO) since 1996. 
On board SoHO, the  VIRGO instrument measures the solar spectral irradiance with the three channels sun photometer (SPM) at wavelength of 402 (blue), 500 (green) and 862 (red) nm (\citeads{1995SoPh..162..101F}; \citeads{1997SoPh..170....1F}; \citeads{2002SoPh..209..247J}). In parallel, the GOLF spectrophotometer observes the solar disk-integrated position of the Sodium doublet lines at $\lambda=5895.9$ and $5889.9~\AA$, from which it extracts the projected radial velocities (\citeads{1993ExA.....4...87B}; \citeads{1995SoPh..162...61G}; \citeads{Garcia_2005}; \citeads{2018A&A...617A.108A}). 
Together, VIRGO and GOLF observations form a unique set of high-precision solar observations with an excellent duty cycle of almost $96\%$ over the past $26$ years. In the following, we use both datasets sampled at one point per minute, taken out of the initial higher cadence sequences.

For GOLF observations, we used the level-2 GOLF data\footnote{\url{www.ias.u-psud.fr/golf/templates/access.html}} calibrated as described in \citetads{2018A&A...617A.108A}. To be conservative, we only selected the year 1996 from this dataset since, after that, the detector was affected by an instrumental failure. This year corresponds to a solar cycle minima, leading then to a negligible impact of the magnetic regions (which is out-of-scope of the present study). We divided this time series into subseries of $8$-hours duration to compare them with our ESPRESSO set of observations. From the total number of available $8$-hours subseries, we selected $75$ subseries with a condition of regular sampling (i.e., we only took the subseries that avoid gaps).

For VIRGO observations, we also selected the observations taken in 1996 to be consistent with the selected GOLF data. We corrected them from the instrumental degradation over time following the recipe described in Sec. 2 of \citetads{2020A&A...636A..70S}. We then divided them into $8$-hours subseries to mimic the duration of CHEOPS observations, and we picked $265$ of these subseries with a condition of regular sampling. 
Throughout the paper, we use these VIRGO and GOLF subseries as references for comparing the properties of solar granulation with our two main-sequence targets. 

In addition, we also compare our set of observations in Sec.~\ref{sec31} with ground-based solar observations taken by the HARPS-N spectrograph, for which the first three years of observations (2015-2018) have been made recently available\footnote{\href{https://dace.unige.ch/sun/}{https://dace.unige.ch/sun/}} (\citeads{2019MNRAS.487.1082C}, \citeads{2021A&A...648A.103D}). We extracted $298$ daily subseries from this released dataset. The HARPS-N subseries have a median duration $T\sim 6.75$ hours and a sampling rate $\Delta t \sim 5.4$ min.
In Appendix~\ref{App_solar_obs}, we compare these different sets of solar observations. 
 
\begin{table*}[t] \centering
\caption{Properties of the stars HD 67458 and HD 88595. }
\vspace{-0.1cm}
\label{tab_params} 
\begin{tabular}{*4c}
\noalign{\smallskip}\hline\noalign{\smallskip}
Parameters & \multicolumn{2}{c}{Values} & Source \\
\noalign{\smallskip}\hline\hline\noalign{\smallskip}
  & HD 67458 & HD 88595 & \\
Target names  & HIP 39710 & HIP 50013 & Simbad$^a$ \\
  & TIC 154234114 & TIC 5506063 &\\
  & {\small Gaia EDR3 5596797039059063680} & {\small Gaia EDR3 5669349344592951936} & \\
\noalign{\smallskip}\hline\noalign{\smallskip}
Spectral type & G0V &  F7V & Simbad \\ 
\noalign{\smallskip} \noalign{\smallskip}
Right Ascension (ep=J2000) & 08 07 00.522 & 10 12 37.96 & Simbad \\ 
Declination (ep=J2000) & -29 24 10.57 &  -19 09 10.94 & Simbad \\ 
\noalign{\smallskip} \noalign{\smallskip}
Gaia G-band magnitude & 6.64 & 6.34 & Gaia archive$^b$\\ 
Distance [pc] &  	25.68 & 	42.01 & This work$^c$ (IRFM) \\ 
\noalign{\smallskip}\hline\noalign{\smallskip}
Effective Temperature (K) & $5833 \pm 62$ &  $6205\pm35$ & This work (spectroscopy) \\
\noalign{\smallskip} \noalign{\smallskip}
Metallicity [Fe/H] & $-0.18\pm0.04$ & $0.17\pm0.03$ & This work (spectroscopy) \\ 
\noalign{\smallskip} \noalign{\smallskip}
$\rm [Mg/H]$ & $-0.14\pm0.06$ & $0.10\pm0.07$ & This work (spectroscopy)   \\ 
\noalign{\smallskip} \noalign{\smallskip}
$\rm [Si/H]$ & $-0.20\pm0.04$ & $0.12\pm0.05$ & This work (spectroscopy)  \\ 
\noalign{\smallskip} \noalign{\smallskip}
$\rm [Ti/H]$ & $-0.17\pm0.06$ & $-0.01\pm0.05$ & This work (spectroscopy)  \\ 
\noalign{\smallskip} \noalign{\smallskip}
 ${\rm log}g$ [cgs] & $4.37\pm0.10$ & $3.99\pm0.06$  & \ This work (spectroscopy)   \\ \noalign{\smallskip} \noalign{\smallskip}
 & 4.390 $\pm$ 0.026 & 4.148 $\pm$ 0.017 & This work (from $R_*$ and $M_*$)\\
\noalign{\smallskip} \noalign{\smallskip}
Radius ($R_\odot$) & $1.021\pm0.020$ & $1.616\pm0.017$ & This work (IRFM)  \\ 
\noalign{\smallskip} \noalign{\smallskip}
Mass ($M_\odot$) & $0.935^{+0.042}_{-0.043}$ & $1.351^{+0.031}_{-0.058}$ & This work (isochrones) \\ 
\noalign{\smallskip} \noalign{\smallskip}
Age [Gyr]& $7.7\pm2.1$ &  $2.8\pm0.4$ &  This work (isochrones) \\ 
$\nu_{\rm max}$ [$\mu$Hz] & $2756 \pm 169$ &  $1534 \pm 85$ &  Eq. (\ref{numax}), This work (ESPRESSO)\\
\noalign{\smallskip}\hline\noalign{\smallskip}
Mean ${\rm log} R'_{\rm HK}$& $-4.99\pm0.14$ & $-4.96\pm0.15$ &  This work (spectroscopy)\\ 
Rotation period [days] & $10.57\pm0.06$  & $3.1151\pm0.0003$ & This work (TESS photometry)\\ 
$v\sin{i}$ [km/s] & $ 2.179\pm0.2$ & $7.24\pm0.35$ & HARPS$^d$,  This work   \\ 
Stellar inclination [degrees] & {$\sim 26$} & {$\sim 16$} & This work \\ \noalign{\smallskip}\hline\noalign{\smallskip}
\end{tabular}
\parbox{7in}{
\footnotesize Notes. $^a$SIMBAD astronomical database from the Centre de Données astronomiques de Strasbourg (\href{http://simbad.u-strasbg.fr/simbad/}{http://simbad.u-strasbg.fr/simbad/}). $^b$Archive of the Gaia mission of the European Space Agency (\href{https://gea.esac.esa.int/archive/}{https://gea.esac.esa.int/archive/}).
$^c$ Values are in agreement with the Gaia Early Data Release 3  (catalog \href{http://vizier.u-strasbg.fr/viz-bin/VizieR?-source=I/352}{http://vizier.u-strasbg.fr/viz-bin/VizieR?-source=I/352} from \citeads{2021AJ....161..147B}).
$^d$ HARPS data initially published in \citetads{2018A&A...615A..76S} and reanalyzed in this work.

}
\vspace{-0.1cm}
\end{table*}

\section{Derivation of stellar parameters}
\label{sec2b}

First concerning the stellar atmospheric parameters, we coadded the individual exposures taken with ESPRESSO (see Sec. \ref{sec23}) 
{after correcting for each individual radial velocity shift. This was done for each }target in order to create a combined spectra with higher signal-to-noise ratio. We used each ESPRESSO master spectrum to derive the stellar spectroscopic parameters ($T_{\mathrm{eff}}$, ${\rm \log}g$, micro-turbulence, [Fe/H]) and the respective uncertainties following the ARES+MOOG methodology as described in \citetads{Sousa-21}; \citetads{Sousa-14}; \citetads{Santos-13}. The \texttt{ARES} code\footnote{\url{https://github.com/sousasag/ARES}} (\citeads{Sousa-07}; \citeads{Sousa-15}) was used to measure in a consistent way the equivalent widths of iron lines included in the line list presented in \citetads{Sousa-08}. Briefly, ARES+MOOG performs a minimization process looking for the ionization and excitation equilibrium to find convergence on the best set of spectroscopic parameters. For the computation of the iron abundances we make use of a grid of Kurucz model atmospheres \citepads{Kurucz-93} and the radiative transfer code \texttt{MOOG} (v2019) \citepads{Sneden-73}.
In addition we have also used the IDL package \texttt{Spectroscopy Made Easy} (SME) to do the spectral analysis for these stars \citep{Valenti1996, Piskunov2017}. This code utilizes an input of stellar parameters to perform radiative transit calculations in order to synthesize models, that through an iterative minimizing procedure with the observed spectrum as a template, arrive at a set of final stellar parameters. In this process one varies one parameter while keeping the other fixed and works with several different atmospheric models  and atomic and molecular line lists from \texttt{VALD} \citep{Piskunov95}. In this case we utilized again a grid of Kurucz model atmospheres \citepads{Kurucz-93}.

For HD 67458 both spectral analyses provided completely consistent parameters and we selected the values given by ARES+MOOG. For HD88595 we rely on the parameters derived by SME, mostly because of the higher $v\sin{i}$ for this star which degrade a bit the precise measurements of the equivalent widths in the ARES+MOOG method. The adopted spectroscopic parameters are listed in Table \ref{tab_params}.

We determined the stellar radii of HD\,67458 and HD\,88595 using a modified IRFM method in a Markov-Chain Monte Carlo (MCMC) approach (IRFM; \citeads{Blackwell1977}; \citeads{Schanche2020}). This was done by computing the bolometric fluxes for the targets by fitting stellar atmospheric models to broadband photometry that are converted to effective temperatures and angular diameters using the physical relationships between these parameters. Utilizing the target's parallaxes, we subsequently determined the radii from the angular diameters. For HD\,67458 and HD\,88595, we use \textit{Gaia}, 2MASS, and \textit{WISE} broadband photometry (\citeads{Skrutskie2006}; \citeads{Wright2010}; \citeads{GaiaCollaboration2021}) with the ATLAS catalog of stellar atmospheric models \citepads{Castelli2003}, and the offset-corrected \textit{Gaia} EDR3 parallax \citepads{Lindegren2021} and find $R_\star=1.021\pm0.020 ~R_\odot$  and $R_\star=1.616\pm0.017~R_\odot$, respectively. These radii are reported in Table~\ref{tab_params}.

Adopting $T_{\mathrm{eff}}$, [Fe/H], and $R_{\star}$ as basic input set, we then derived the isochronal mass $M_{\star}$ and age $t_{\star}$ of each star from two different stellar evolutionary models. In detail, we used the isochrone placement algorithm (\citeads{bonfanti15}; \citeads{bonfanti16}), which interpolates the input parameters within precomputed grids of \texttt{PARSEC}\footnote{\textit{PA}dova \& T\textit{R}ieste \textit{S}tellar \textit{E}volutionary \textit{C}ode: \url{http://stev.oapd.inaf.it/cgi-bin/cmd}} \citepads{marigo17} isochrones and tracks, to retrieve a first pair of mass and age values. To improve the convergence, we also accounted for the stellar $v\sin{i}$, coupling the isochronal interpolation scheme with gyrochronology as outlined in \citetads{bonfanti16}. The second pair of mass and age values, instead, was computed through \texttt{CLES} (Code Liégeois d'Évolution Stellaire; \citeads{scuflaire08}), which builds the best-fit stellar track according to the input parameters following the Levenberg-Marquadt minimization scheme as explained in \citetads{salmon21}.
Finally, for each star and for each output parameter, we merged the two respective distributions derived from \texttt{PARSEC} and \texttt{CLES}, after checking their mutual consistency through a $\chi^2$-based criterion (see \citeads{bonfanti21} for further details). We obtained $M_{\star}=0.935_{-0.043}^{+0.042}\,M_{\odot}$ (resp. $M_{\star}=1.351_{-0.058}^{+0.031}\,M_{\odot}$) and $t_{\star}=7.7\pm2.1$ Gyr (resp. $t_{\star}=2.8\pm0.4$ Gyr) for HD\,67458 (resp. HD\,88595). The computed stellar parameters are listed in Table~\ref{tab_params}. 

The uncertainties associated with each of the stellar parameters are those obtained by the above procedure, which is the one applied to all CHEOPS exoplanet host targets (see again \citeads{bonfanti21}, for full details).  This procedure is performed here on two bright main-sequence stars with high quality spectra, detailed abundances, and numerous and accurate broadband photometric measurements.
Concerning the stellar surface gravity, we want to recall that its spectroscopic determination is prone to several problems that directly affect the accuracy of the derived value. Problems such as the assumption of plane parallel stellar atmosphere models, or the use of the ionization balance where few optimal lines of ionized iron is present, can affect strongly the accuracy of the spectroscopic analysis for the ${\rm log}g$. Fortunately these do not strongly affect the determination of other atmospheric parameters. For example, it was demonstrated that when using equivalent-width methods the other atmospheric parameters are mostly independent from the surface gravity (e.g., \citeads{Torres-2012}). In our procedure to derive the stellar radius, mass, and age, the spectroscopic surface gravity is only marginally used as a prior for the radius determination (where it is allowed to vary within the MCMC), but is not used anymore in deriving mass and age. Given these considerations, for HD 88595 the spectroscopic ${\rm log}g$ is reasonably consistent with the one computed from the derived mass and radius values (see Table~\ref{tab_params}), which in turn very well agree with the asteroseismic $\nu_{\rm max}$ measured for this star (see Sec. \ref{sec33}). Concerning the solar-like star HD 67458, the spectroscopic ${\rm log}g$ and the one derived from mass and radius are in perfect agreement (see Table~\ref{tab_params}).

Finally, analyzing the TESS observations based on the generalized Lomb-Scargle periodograms \citepads{2009A&A...496..577Z}, we were able to constrain the rotation period of the F star ($P_{\rm rot} \simeq 3.1151\pm0.0003$~days) and the G star ($P_{\rm rot} \simeq 10.57\pm0.06$~days). We report these rotation periods in Table~\ref{tab_params} and the details in Appendix.~\ref{App_Prot}. Using the known $v\sin{i}$ and $R_\star$ from Table~\ref{tab_params}, this gives a stellar inclination of {$i\simeq16^\circ$} for the F star, and {$i\simeq26^\circ$} for the G star. Both stars are then seen nearly pole-on.

\section{Granulation signals in high-precision photometric and spectroscopic observations}
\label{sec3}
 
Stellar {granulation} generates stochastic fluctuations in both photometric and spectroscopic observations. These fluctuations are correlated over timescales from some minutes to several hours, depending on the stellar parameters. {The objective of this section is to identify the contribution of the stellar granulation among the various sources of noise present in the CHEOPS, TESS and ESPRESSO datasets.}

{We first look at the behavior of these fluctuations in the time domain, and how their amplitudes evolve over different timescales (Sec.~\ref{sec31}). We then analyze how they behave when using the common observational strategies to mitigate them (i.e., long exposure time for RV data acquisition or light curve binning over short timescales), and we discuss these behaviors in the context of small exoplanets detection (Sec.~ \ref{sec32}). We conclude this section by analyzing the observations in the frequency domain (periodograms), and in particular we show how the instrumental noise (dominated by photon noise) impacts the characterization of the stellar granulation signal (Sec.~\ref{sec33}). The main conclusions of this section are summarized in Table.~\ref{tab:table_summary}.}
 
\subsection{Amplitude of the granulation signal}
\label{sec31}

\begin{figure*}[t]
\centering
\resizebox{\hsize}{!}{\includegraphics{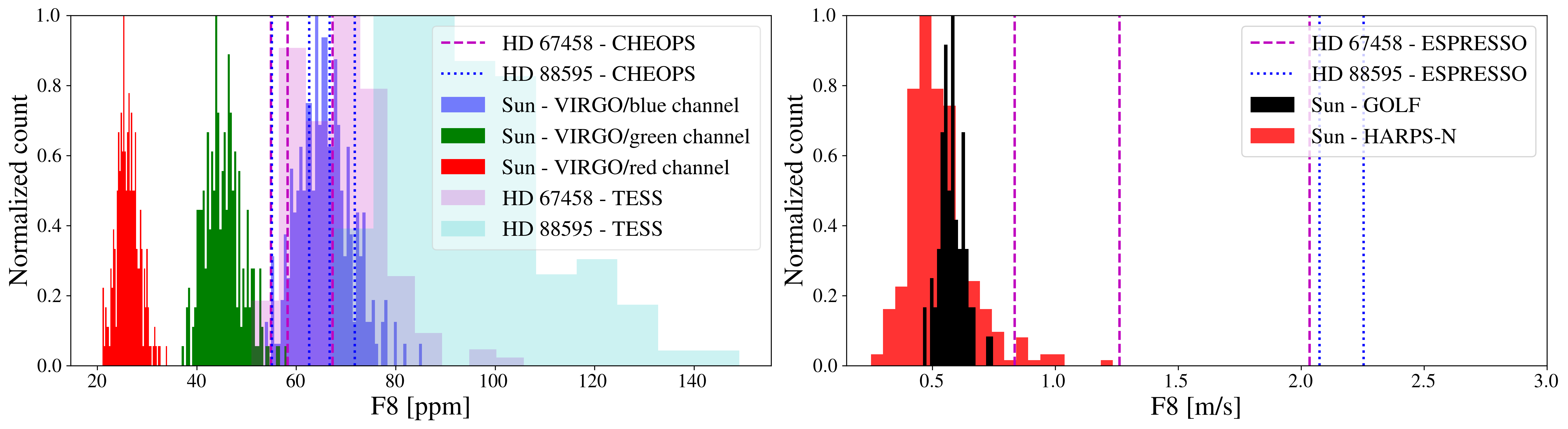}}
\caption{Comparison of the stellar signal amplitude at short time scale for different stars (the Sun, HD 67458, and HD 88595). 
\textit{Left:} {Comparison of the F8 metric computed on the $8$-hours subseries of VIRGO observations (red, green, and blue histograms) and CHEOPS observations of HD 67458 and HD 88595 (colored dashed and dotted vertical lines, respectively). The F8 metric has been slightly modified from the initial definition of \citetads{2013Natur.500..427B} to include fluctuations evolving on timescales from $5$ min to $8$ hours. F8 values computed on each TESS $8$-hour subseries are shown for comparison (magenta and cyan histograms). \textit{Right:} Same for spectroscopic solar observations with GOLF (black histogram) and HARPS-N (red histogram) compared with ESPRESSO observations of HD 67458 and HD 88595 (colored dashed and dotted vertical lines, respectively).}
}
\label{fig_f8}  
\end{figure*}

{We assume that each time series results from the contribution of three phenomena: instrumental noise, stellar oscillations, and stellar granulation. 
We neglect stellar magnetic activity because i) the chromospheric activity indicator ${\rm log} R'_{\rm HK}$ indicates that the two stars are relatively inactive (${\rm log} R'_{\rm HK}~\sim$ -4.9), ii) no correlation between ${\rm log} R'_{\rm HK}$ and RV is observed, and iii) our observations are very close in time so any signature of magnetic activity (spot or plages) should act as a trend over the nightly datasets.
The instrumental noises are dominated here by the photon noise and affect differently the CHEOPS, TESS and ESPRESSO observations. The stellar oscillations (or p-modes) evolve on short timescales for both stars, typically over $10$ min or less. The stellar granulation, which we seek to identify here, evolves on longer timescales. We expect the amplitudes of the stochastic fluctuations generated by the stellar granulation to follow a Gaussian distribution (see Sec. 2.3 of \citeads{2020A&A...636A..70S}), and to decrease with increasing stellar surface gravity \citepads{2013Natur.500..427B}. We therefore expect the overall amplitude of the granulation signal to be comparable to the solar values extracted from the visible wavelengths for HD 67458, and to be larger for HD 88595. In addition, since the granulation signal amplitude decreases at redder wavelengths (Planck's law), we expect the amplitude of the granulation signal to be smaller in the TESS passband than in the CHEOPS one. }\\

{We start by extracting the contribution of photon noise, which dominates the high-frequency part of our time series. To this end, we filter each time series with a high-passband filter with cut-off frequencies $\nu_{cut}$. We visually determine these cut-off frequencies $\nu_{cut}$ for each dataset by examining the flat regions of the different periodograms (see details in Sec.~\ref{sec33}). 
We then measure the standard deviation of each filtered time series ($\sigma_w$) that corresponds to one visit (CHEOPS), one night (ESPRESSO), or one sector (TESS).  For TESS, we report the median value of the standard deviations calculated on each $8$-hour subseries contained in a given sector.
For HD 67458, we obtain: $\sigma_w=\{79, 81, 93\}$ ppm (CHEOPS), $\sigma_w=\{97, 93, 85\}$ ppm (TESS), and $\sigma_w = \{2.1, 1.6, 0.95\}$ m/s (ESPRESSO). For HD 88595, we obtain: $\sigma_w=\{77, 75, 77, 78\}$ ppm (CHEOPS), $\sigma_w=\{181, 177\}$ ppm (TESS), and $\sigma_w = \{1.37, 1.49\}$ m/s (ESPRESSO). \\
For CHEOPS observations, the observed white Gaussian noise (WGN) amplitudes are in remarkably good agreement with the predictions from the \texttt{CHEOPS Exposure Time Calculator}\footnote{ \url{https://cheops.unige.ch/pht2/exposure-time-calculator}}. Indeed, including contributions of instrumental (readout, smearing, quantization and dark current), background (sky and straylight) and photon noise (which is by far the dominating factor), the ETC predicts white noise levels at the cadence of the downloaded images of $85$ ppm and $71$ ppm for HD 67458 and HD 88595, respectively.
We observe larger white noise levels for TESS observations (especially for HD 88595) than for CHEOPS. This is also in agreement with expectations based on the characteristics of the two satellites (equivalent collective area, photometric performances at a given stellar magnitude, see e.g., \citeads{Futyan2020}).
For ESPRESSO observations, the observed white noise amplitudes do vary significantly between the different nights of HD 67458. This would mean than the global dispersion $\sigma_{TOT}$ (given in Table~\ref{tab:table_espresso}) is driven by short term instrumental noise. In order to identify whether these variations were of instrumental origin, we looked at many indicators such as atmospheric conditions, signal-to-noise ratio, instrumental drifts, or pipeline quality controls. Unfortunately, we could not clearly identify a source for this high-frequency variability. However, this excess white noise with no apparent structure will not affect the conclusions presented in the rest of this study.\\
}

We continue with the extraction of the granulation signal. To this end, 
we use a slight modification of the $8$-hour flicker (or F8) metric, defined in \citetads{2013Natur.500..427B}. 
We start by binning each time series into $5$-minute intervals (and not $30$-minutes as originally defined in \citetads{2013Natur.500..427B}, since we would miss the typical timescales of stellar granulation). We then use a $8$-hours length boxcar filter\footnote{We use the function \texttt{convolution.Box1DKernel} available from the Python package \url{www.astropy.org}.} to remove the long term stellar activity (we note the small impact of this additional step since the length of our subseries are short). Results are shown and compared to the Sun in photometry (right) and spectroscopy (left) in Fig.~\ref{fig_f8}.

\begin{figure}[t!]
\centering
\resizebox{\hsize}{!}{\includegraphics{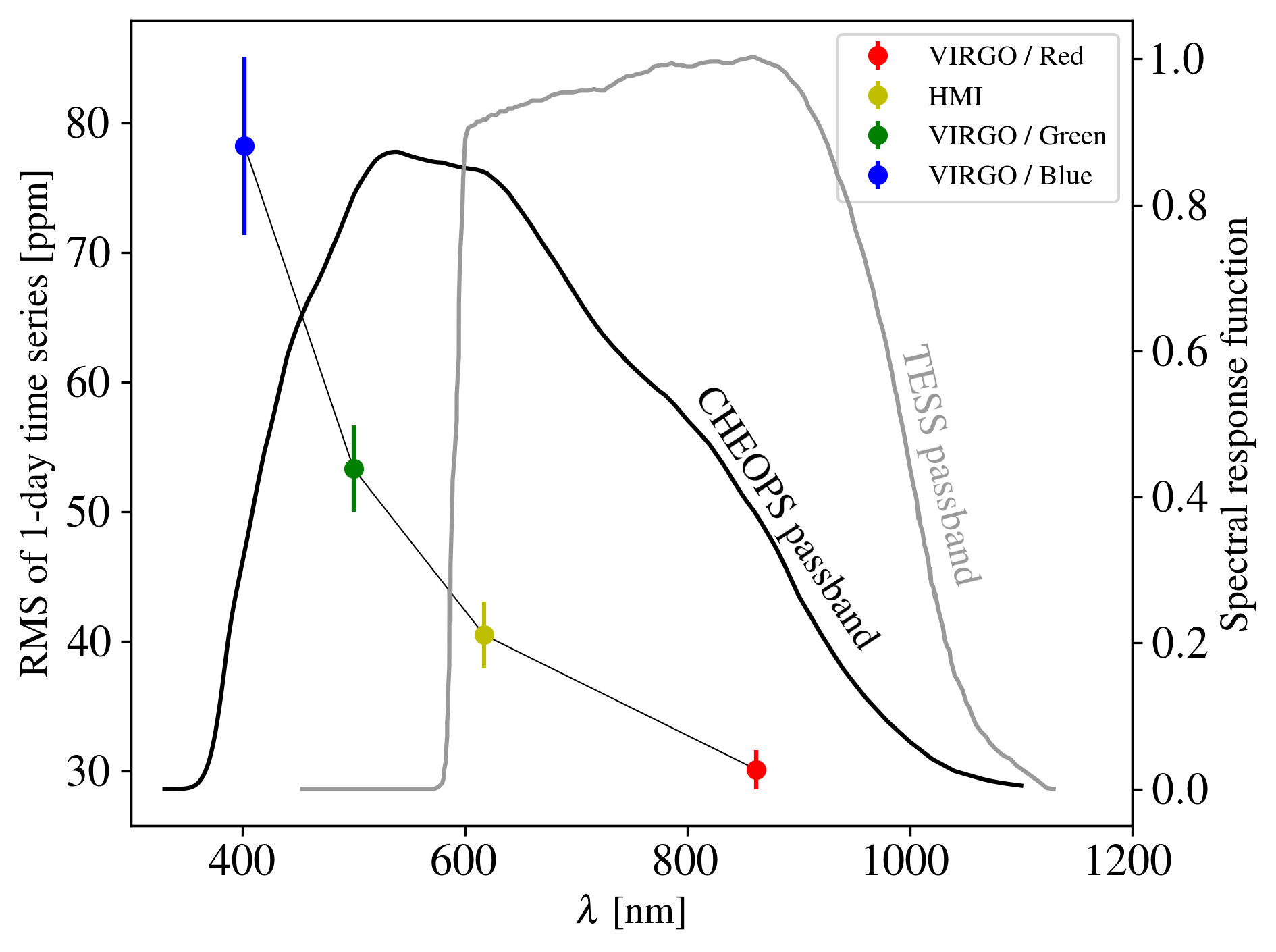}}
\caption{{Decrease in the amplitude of the granulation signal as a function of wavelength. The granulation amplitude is estimated by the RMS of 265 solar $1$-day subseries (VIRGO), since the level of white noise is negligible in solar observations. The different solar values are extracted from Fig. 4 of \citetads{2020A&A...636A..70S}. On the right y-axis, we show the spectral response function of the CHEOPS (black) and TESS (gray) telescopes. }}
\label{Fig_passband}  
\end{figure}

The solar photometric values are extracted from the narrow passbands of the VIRGO red, green and blue SPM channels. To compare CHEOPS and VIRGO observations, as done in \citetads{2010ApJ...713L.155B} and \citetads{2016A&A...596A..31S} for Kepler, we therefore need to consider a combination of the red ($862$ nm) and green ($500$ nm) channels. The CHEOPS values for HD 67458 are then expected to fall in the interval defined by the red and green histograms of Fig.~\ref{fig_f8}. The TESS passband being redder, the TESS values for HD 67458 are expected to fall closer to the red histogram.  
This is however not what we observe, with F8 values of $\{67, 55, 58\}$ ppm for the three CHEOPS visits, and F8$\in [50,106]$ ppm for the whole set of $8$-hours subseries from TESS observations.
We explain these discrepancies by two reasons. First, the level of white noise is significantly larger in both CHEOPS and TESS observations than in solar observations. Then, the discrepancy with solar observations is larger for TESS observations since the granulation amplitude decrease with increasing wavelengths (see Fig.~\ref{Fig_passband}).  In TESS data, the high-frequency noise masks the granulation signal and does not allow to identify it. 

Computing now the F8 metric on the four CHEOPS visits of the F star HD 88595, we obtain  F8$=\{63, 67, 72, 55\}$ ppm.  As expected from a star with a lower surface gravity, these values are above the solar ones (at all wavelengths).

On the right panel of Fig.~\ref{fig_f8}, we show the histogram of the F8 metric computed on each $8$-hour solar GOLF (black) and HARPS-N (red) subseries. We observe F8 values in the interval $[0.46,0.75]$ m/s for GOLF and $[0.25,1.2]$ m/s for HARPS-N.  Although these solar values are in agreement, we are not surprised by the larger RV dispersion of HARPS-N data because GOLF data are obtained using one line (Sodium Doublet), that is one height in the atmosphere,  whereas HARPS-N uses a series of lines in the visible range, that is an average of various contributions at different heights. 

Computing the F8 metric over the three ESPRESSO observation series for the G star, we obtain F8$=\{2.0,1.2,0.8\}$ m/s; which are larger but still in agreement with the right tail of the HARPS-N solar data distribution. We recall that the first night of observations was affected by large variations during the night (see Sec.~\ref{sec22}), certainly of instrumental origin, and should be considered with caution.
On the other hand, we measure stable F8 values for the two nights of the F star with, as expected, values larger than the solar ones (F8$=\{2.07, 2.25\}$ m/s).

Finally, we note that, if we drastically filter out the long periods ($> 3$ hours instead of $>8$ hours), the F8 values do not change significantly. Indeed, we observe a decrease of only $6-10$ ppm for the CHEOPS observations of HD 88595, which is the star with the fastest rotation rate. This sanity check supports the assumption that the impact of long-term stellar magnetic activity is negligible in our analyses.

\subsection{Mitigation of the granulation signal}
\label{sec32}

{To reduce the contribution of the stellar oscillations and granulation for the detection and characterization of exoplanets, the common strategy is to average the observations. In radial velocity, this results in the use of a longer exposure time than the stellar p-modes timescales (see \citeads{2019AJ....157..163C} for a recent optimization of this exposure time), followed by a binning of data points taken over the course of a night to reduce stellar granulation signals (\citeads{2011ApJ...743...75H}; \citeads{2011A&A...525A.140D}). In photometry, this results in binning the data points over some minutes, depending on the characteristics of the studied planetary transit.
Below, we first study how the high precision ESPRESSO observations behave when using this mitigation strategy, then we turn to the CHEOPS photometric observations. In both cases, we compare with the expected signal amplitude of an Earth-like planet that would orbit in the habitable zone (HZ) of the two stars.
}\\

\begin{figure}[t!]
\centering
\resizebox{\hsize}{!}{\includegraphics{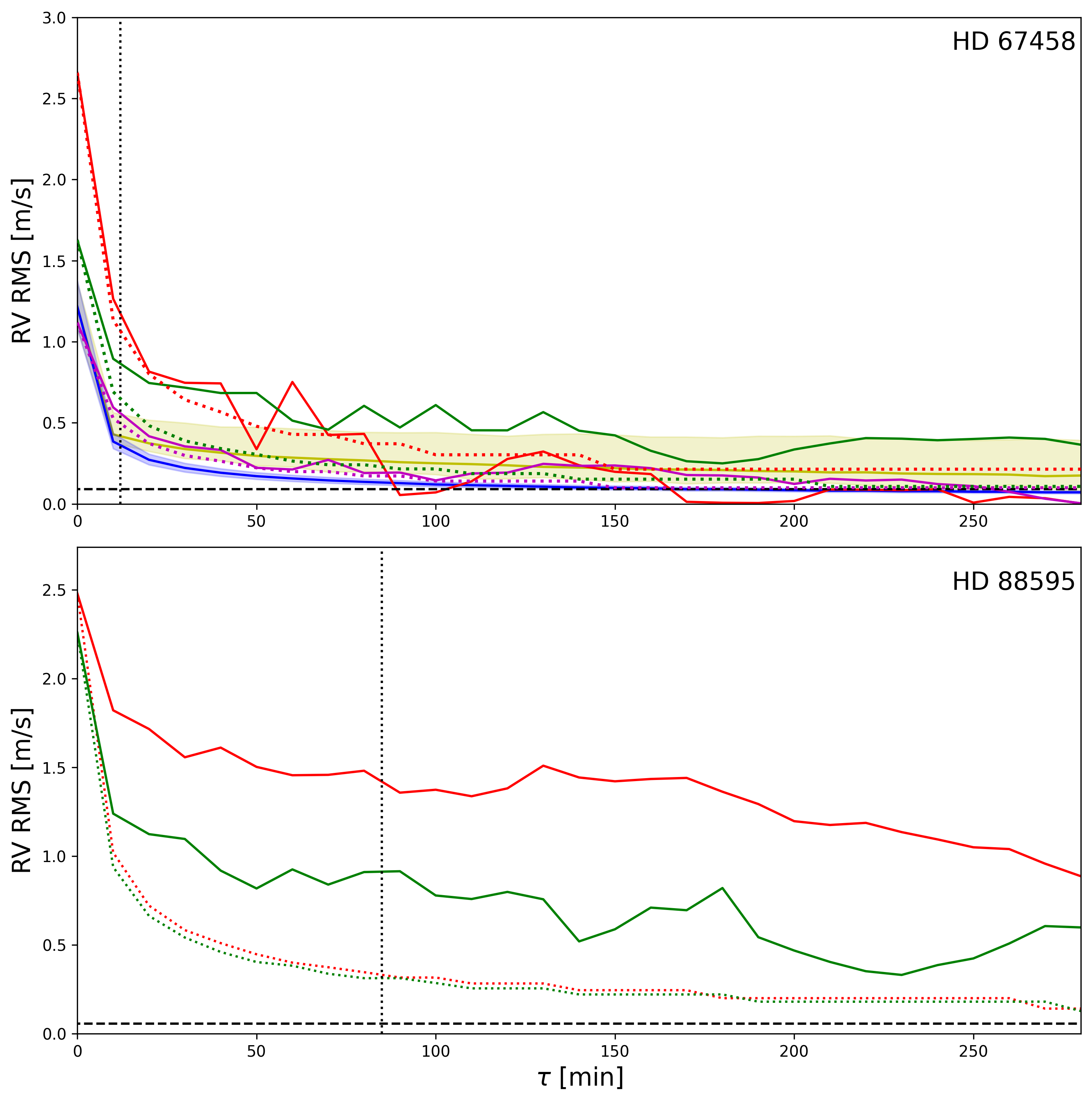}}
\caption{
Effect of temporal binning on the RV RMS of HD 67458 (top) and HD 88595 (bottom). Each color (red, green and magenta) represents one night of ESPRESSO observations. The behavior of WGNs having the same variance as the considered dataset is shown by the colored dotted lines.  RMS values obtained on the $75$ solar GOLF subseries are shown in yellow in the top panel, and the corresponding WGNs in blue (median values are shown by the yellow and blue solid lines).
The RV semi-amplitude of an Earth-like planet orbiting in the HZ of each star are shown by the dashed horizontal lines. The exposure times needed to mitigate the p-modes oscillations down to the $\sim 10$ cm/s level are shown by the dotted vertical lines.
}
\label{fig_RVbin}  
\end{figure}

{In Fig.~\ref{fig_RVbin}, we show the decreases of the RV  amplitudes as a function of the binning timescale $\tau$. We compare them with the decrease expected for WGNs of the same variance, that we take as a reference to determine if the RV RMS at a given $\tau$ is dominated by white noise or by stellar signals. The decrease of WGNs behaves as $\sigma_{TOT}\sqrt{N_{\rm bin}/N}$, with $N_{\rm bin}$ the number of binned data points.}

For the solar-like star HD 67458 (top panel), we observe a different behavior for the three nights: one is consistent with the behavior of a white noise, while the two others are not. At $\tau=20$ min we get an RV RMS of $\sim0.4$ m/s for the last night (magenta line) and the WGNs, while we get an RV RMS of $0.7-0.8$ m/s for the other nights (red and green lines). 
To determine if these two behaviors are consistent with solar observations, we compute the RV RMS as a function of $\tau$ for each of the $75$ GOLF solar $8$-hours subseries and for each corresponding WGN. 
We see that solar RV RMS measurements\footnote{We note that the reported RV RMS are also consistent with the predictions based on solar-like RV simulations from  \citetads{2015A&A...583A.118M}, with values around $30$ to $40$ cm/s for a solar-like star at $\tau=1$ hour.} show a large dispersion but do not match the decrease observed for a WGN. The RV RMS derived on HD 67458 dataset are statistically compatible with solar data at large $\tau$.
Comparing with the RV semi-amplitude of an Earth-mass planet in the habitable-zone ($K\sim9$ cm/s for HD 67458, see Appendix.~\ref{App_pl}), this demonstrates that this strategy is not robust to mitigate enough the short-timescale stellar signal. This is in total agreement with \citetads{2015A&A...583A.118M}, who found that we need to bin over $\tau>8$-hours to get down to the level of an Earth-like RV signature.
For the F star HD 88595 (bottom panel), we observe for both nights a behavior inconsistent with a WGN. At $\tau=20$ min, we read RV RMS values of $1.7$ and $1.1$ m/s for the first and second nights, while we read an RMS of $0.7$ m/s for the WGN. This demonstrates the failure of this observational strategy for a slightly evolved star as HD 88595, with an RV dispersion remaining above $50$ cm/s even after a binning of $\tau=150$ min. For comparison the RV semi-amplitude of an Earth-mass planet in the habitable-zone of this star would be $K\sim5.6$ cm/s (see Appendix.~\ref{App_pl}). 
 
{While it is clear that the RV RMS at large $\tau$ is not driven by WGN, it remains to estimate the contribution of the stellar oscillation modes. }
Following \citetads{2019AJ....157..163C}, we estimate the exposure time needed to mitigate the contribution of these  modes down to the $10$ cm/s level for our two stars\footnote{\url{https://github.com/grd349/ChaplinFilter}}. We find $\tau_{osc}\sim12$ min and $\sim85$ min for HD 67458 and HD 88595, respectively (see dotted vertical lines in Fig.~\ref{fig_RVbin}). 
Since \citetads{2019AJ....157..163C}'s methodology does not include the contribution of the stellar granulation signal (stochastic correlated noise) but is designed for the p-modes mitigation (which behaves at first approximation as pure sines for oscillations), we conclude that the remaining signal contribution at timescales $\tau>\tau_{osc}$ is dominated by stellar granulation.\\

\begin{figure}[t]
\centering
\resizebox{\hsize}{!}{\includegraphics{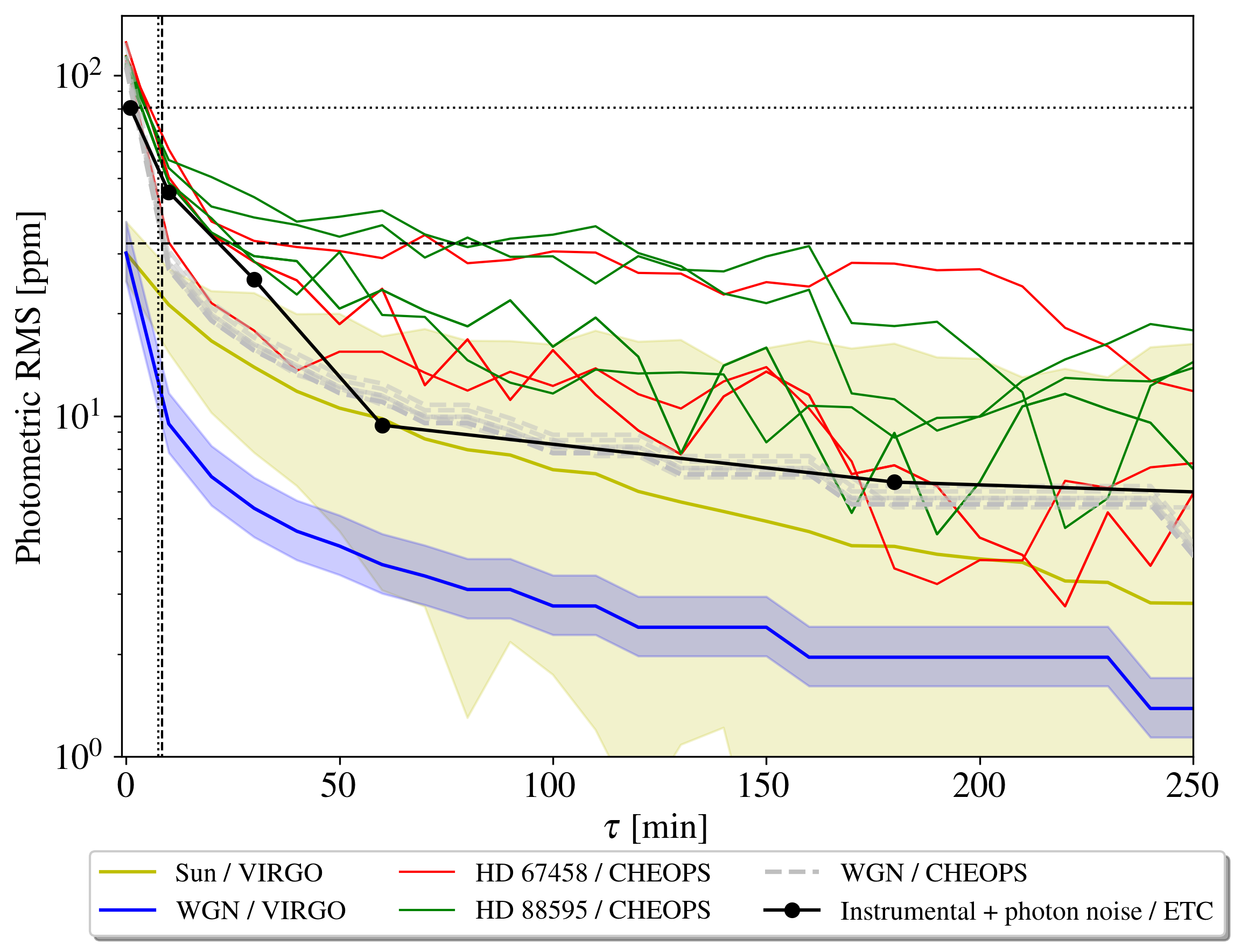}}
\caption{
Effect of temporal binning on the RMS of the photometric data of HD 67458 (three red solid lines for the three CHEOPS visits) and HD 88595 (four green solid lines for the four CHEOPS visits).  The behavior of WGNs of the same variance than each CHEOPS visit are shown with the dashed gray lines.
RMS values obtained on the $265$ solar VIRGO subseries are shown with the yellow beam (median value in solid), and the corresponding behavior of  WGNs in blue (median value in solid). Prediction from the CHEOPS ETC are shown with the black thin lines with dots.
The transit depth of an Earth-size planet orbiting in the HZ of each star are shown by the dashed horizontal lines. Their typical transit ingress duration are shown by the vertical dotted and dashed lines.  
}
\label{fig_Fluxbin}  
\end{figure}

In Fig.~\ref{fig_Fluxbin}, we show the decreases of the CHEOPS photometric amplitudes as a function of the binning timescale $\tau$.
Compared to WGNs of same input variance, these observations show larger amplitude at $\tau<150$ min. This indicates that the signal remaining at large $\tau$ comes from stellar variability.  

It can be argued that the instrumental noises contain not only white noise components but also red noises.  To evaluate this, we computed from the CHEOPS ETC the full expected noise taking into account the impact of both red and white noise components, and we check how it behaves when we integrate observations on timescales up to $6$ hours. We note that in addition to the main contributions already listed in Sec.~\ref{sec31}, these new estimates include also variable dark current, instabilities in gain, quantum efficiency and analog electronics, as well as timing errors, and flat field homogeneities. Comparing these estimates with CHEOPS observations in Fig.~\ref{fig_Fluxbin}, we still observe discrepancies indicating that the excess signal at $\tau<150$ min is driven by stellar variability. 

Finally, since we expect the p-modes photometric contribution to be smaller than in spectroscopic observations, we can conclude that the significant difference between our observations and WGNs is driven by stellar granulation. 

This stellar signal may bias the inferred parameters of long-period transiting exoplanets (from which few transits may be observable, see e.g., \citeads{2020A&A...636A..70S}). Indeed, their amplitudes are comparable to the transit depth of Earth-size planets,  which are $80.6$ ppm for HD 67458 and $32.2$ ppm HD 88595 (see Appendix.~\ref{App_pl}). Moreover, these signals' amplitudes remain significant (brightness RMS $\in [32-76]$ ppm) even after a binning of $\tau\sim7.5-8.4$ min, which are the typical durations of the transit ingresses of Earth-size planets in the HZ (see Appendix.~\ref{App_pl}). Accessing how this stellar signal is correlated is {therefore important} to infer accurate and precise exoplanet parameters.

\subsection{{Periodograms} of the granulation signal}
\label{sec33}

\begin{figure*}[t!]
\centering
\resizebox{\hsize}{!}{\includegraphics{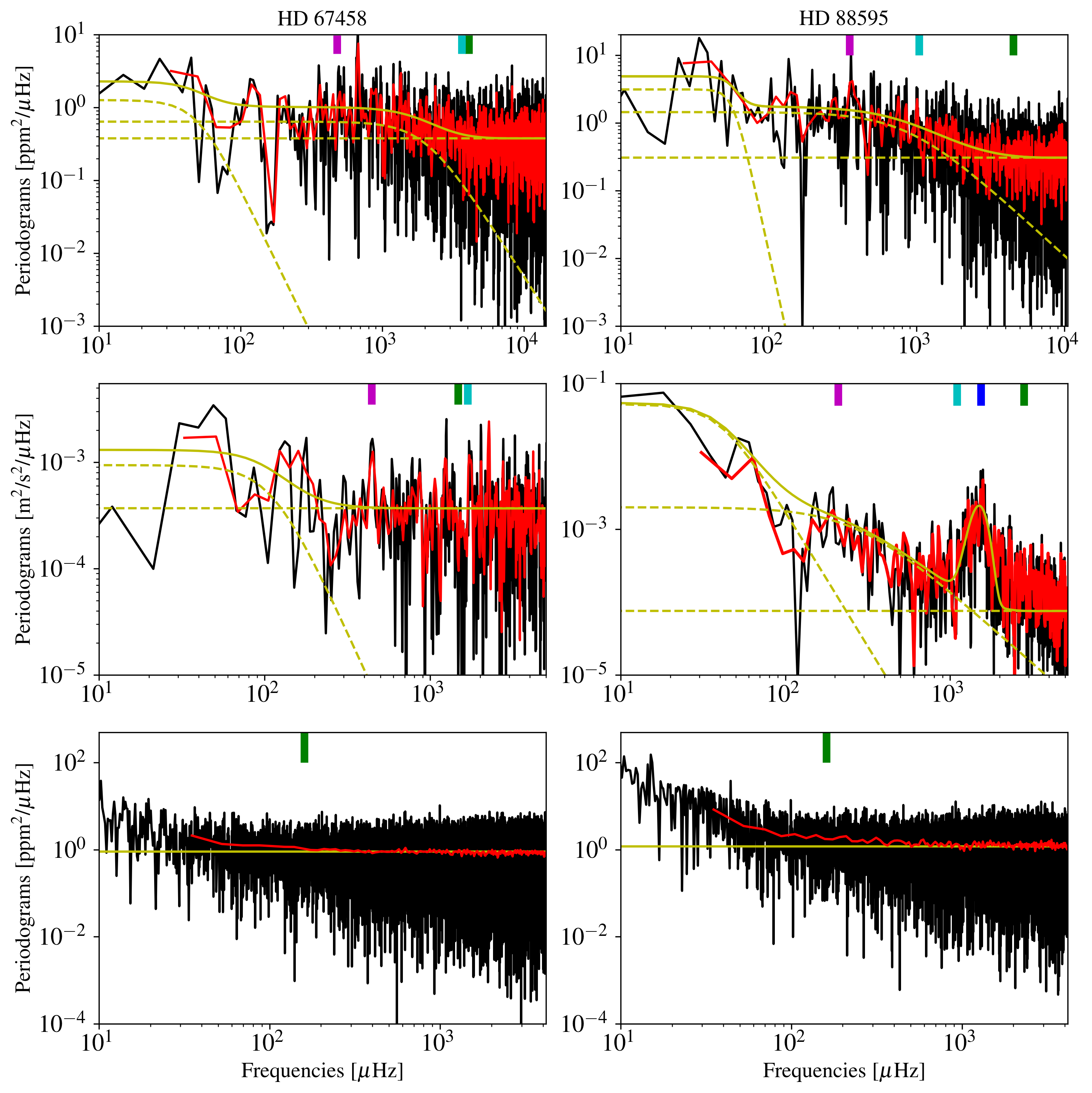}}
\caption{{Periodograms of HD 67458 (left) and HD 88595 (right). 
\textit{From top to bottom}: Lomb-Scargle periodograms of CHEOPS, ESPRESSO and TESS observations computed on the whole dataset are shown in black. 
Their respective averaged periodograms, resulting from the average of the periodograms of each CHEOPS visits, ESPRESSO nights, or TESS $8$-hours subseries, are shown in red.
Yellow lines represent the best-fitting Harvey models, given in  Eq.\eqref{eq_harvey}, that fits the Lomb-Scargle periodograms of each datasets. The different components of this model (white noise, oscillation modes, granulation, supergranulation) are represented in dotted yellow. 
When relevant, the different cut-off frequencies are represented by vertical lines: 
$\nu_{cut}$ in green (used to infer the level of white noise), $\nu_{max}$ in blue (frequency only relevant for ESPRESSO data of HD 88595), $f_H$ in cyan (frequency generally close to $\nu_{cut}$), and $f_g$ in magenta. 
}
}
\label{fig_per}  
\end{figure*}

\setlength{\tabcolsep}{2pt}
\begin{table}[t]\centering
\caption{Best-fitting parameters $\{a_1, b_1, c_1\}$ associated to the  stellar granulation component in the Harvey functions given in Eq.~\eqref{eq_harvey} and found for CHEOPS, ESPRESSO and TESS datasets of HD 67458 and HD 88595. The best-fitting parameters found for the solar VIRGO data without and with {white} noise added are shown in the two last rows. Symbol $\emptyset$ highlights the nondetection cases (i.e., parameters $\{a_1, b_1, c_1\}$ consistent with zero within their $1\sigma$ uncertainties). Parameter $a_1$ is in units of ppm$^2/\mu$Hz or m$^2$/s$^2/\mu$Hz, and parameter $b_1$ in units of $\mu$Hz.
}
\label{tab:harvey}
\begin{tabular}{ccccc}
\hline\hline
Target & Instrument & $a_1$ & $b_1$ & $c_1$ \\ 
\hline 
 HD     & CHEOPS & $0.64 \pm 0.01$ & $2005 \pm 44$ & $3.04 \pm 0.19$\\
67458   & ESPRESSO & $938 \pm 15$ & $110 \pm 12$ & $3.44 \pm 0.02$\\
        & TESS & $\emptyset$ & $\emptyset$ & $\emptyset$\\
\hline 
HD      & CHEOPS & $1.44 \pm 0.02$ & $875 \pm 18$ & $2 \pm 0.01$\\
88595   & ESPRESSO & $2000 \pm 3$& $270.0 \pm 0.3$ & $2.0\pm0.01$ \\
        & TESS & $\emptyset$ & $\emptyset$ & $\emptyset$\\
\hline
 Sun & VIRGO & $0.86  \pm 0.08$ & $2065.9 \pm 147 $ & $3.8 \pm 0.66 $\\
     & VIRGO + WGN  &$0.31  \pm 0.05$ & $3405 \pm 332$ & $3.3 \pm 1.1 $\\
\hline
\end{tabular}
\end{table}

The power spectral density of stellar granulation is well-known to act as a red-like noise, that is a power increase in a given frequency range. For main-sequence stars, this frequency range correspond {to $\nu >10~\mu$Hz.}

In Fig.~\ref{fig_per}, we show the Lomb-Scargle periodograms \citep{1982ApJ...263..835S} of each CHEOPS, ESPRESSO and TESS datasets.
To fit these periodograms, classical models are Harvey-functions \citepads{1988IAUS..123..497H}.  These functions are defined as the sum of Lorentzian functions parameterized by different timescales and amplitudes to distinguish the stellar activity components that dominate different frequency regions in the periodogram (generally attributed from high to low frequencies to:  {instrumental noise, stellar oscillation modes,} granulation, supergranulation, and active regions). 
Some debates about the exact shape of these Lorentzian functions and the number of free parameters to use exist (see e.g., \citeads{2011ApJ...741..119M}; \citeads{2014A&A...570A..41K}).
 In this work, we chose to model each periodogram shown in Fig.~\ref{fig_per} with a Harvey-function of the form \citepads{2014A&A...570A..41K}:
\begin{equation}
P_H(\nu_k^{+}) \!:= \eta^2(\nu_k^{+}) \left(\sum_{i=1}^{2} \frac{a_i}{1+\Big(\frac{\nu_k^{+}}{b_i}\Big)^{c_i}} \!\! + \!\! P_{osc} \exp\left({\frac{-(\nu_k^{+}-\nu_{max})^2}{2\sigma_{osc}^2}} \right)\right)+ \sigma_H^2,
\label{eq_harvey}
\end{equation}
where the set of parameters $\{a_i, b_i, c_i\}_{i=1,2}$ collects the amplitude, characteristic frequency and power of the Harvey functions for the stellar granulation signal ($i=1$) and the low frequency region ($i=2$). Parameters $\{P_{osc}, \nu_{max}, \sigma_{osc}\}$ refer to the oscillation p-modes signals, which are only clearly detected in the ESPRESSO observations of HD 88595 (see the middle right panel of Fig.~\ref{fig_per}). Parameter $\sigma_H^2$ refers to the variance of the high-frequency noise component, assumed to be a WGN in model \eqref{eq_harvey}. Notation $\nu_k^{+}$ means that positive Fourier frequencies are considered, and $\eta := {\rm sinc}(\frac{\pi}{2}\frac{\nu_k^{+}}{\nu_{Ny}})$ is an attenuation factor based on the Nyquist frequency (defined for a regular sampling with time step $\Delta t$ as $\nu_{Ny}=1/2\Delta t$). %
For each star and each periodogram, we infer the parameters of model \eqref{eq_harvey} using first a nonlinear least-squares minimization\footnote{We use the \texttt{LMFIT} Python package, see \url{https://lmfit.github.io/lmfit-py/} \citep{newville_matthew_2014_11813}.}.
Then, we use the \texttt{EMCEE} package \citepads{emcee} to calculate with MCMC the probability distribution for each parameters, from which we take the median values and the $1\sigma$ uncertainties.
We note that large uniform priors are used to fit all the parameters of the model except the power index $\{c_i\}$ which we consider $\le10$ to avoid being too sensitive to large local variations between two peaks of the periodogram.  
Estimated values that are relevant to the granulation signals $\{a_1, b_1, c_1\}$ are reported in Table~\ref{tab:harvey}. 
The best-fitting models are shown in Fig.~\ref{fig_per} (yellow lines).  
From these fits, we conclude that the signature of granulation is evident from both CHEOPS and ESPRESSO observations of HD 88595. For HD 67458, only the CHEOPS observations show a clear granulation signal. {We note an increase of the periodogram at frequency $\nu<200~\mu$Hz for both stars that could be the signature of supergranulation in the ESPRESSO data, but the length of our observations are too short to conclude on the nature of this signal.} On the contrary, TESS is blind to the {granulation} signal due to the high level of WGN {and the small amplitude of this signal in this passband (see Fig.~\ref{Fig_passband})}.

We also find an oscillation frequency at maximum power $\nu_{\rm max} = 1534 \pm 85$ $\mu$Hz from ESPRESSO observations of HD 88595, which is consistent with the prediction from the asteroseismic scaling relation \citep{1995A&A...293...87K}:
\begin{equation}
\label{numax}
    \nu_{\rm max}=\nu_{\rm max,\odot}\;\Big(\frac{M_*}{M_{\odot}}\Big)
     \;\Big(\frac{R_*}{R_{\odot}}\Big)^{-2} \Big(\frac{T_{\rm eff}}{T_{\rm_{eff},\odot}}\Big)^{-1/2},
\end{equation}
that gives $\nu_{\rm max}=$ 1542.5$\pm$60 $\mu$Hz, with ${T_{\rm_{eff},\odot}}=5777$ K and $\nu_{\rm max,\odot}=3090$ $\mu$Hz \citep{2011ApJ...743..143H} and stellar parameters from Table \ref{tab_params}.
%

%
\renewcommand{\tabcolsep}{8pt}
\begin{table*}[t]\centering
\caption{Summary of the data analyses: target name, instrument, reference of the observations, total dispersion of the time series ($\sigma_{TOT}$), estimated level of white noise$^\star$ from the high frequency noise filtering ($\sigma_w$, see Sec.~\ref{sec33}), estimated granulation amplitude from F8 (F8, see Sec.~\ref{sec31}), and from predictions from HD simulations (F8$_{\rm HD}$, see Sec.~\ref{sec5}). Units are in ppm or m/s. The last column indicates if the granulation signal is detected (yes/no/marginal). 
}
\begin{tabular}{cccccccc}
\hline
Target & Instrument & Reference & $\sigma_{TOT}$ & $\sigma_w$ & F8 & F8$_{\rm HD}$ & Detection ?\\ 
\hline
HD 67458 & CHEOPS & visit 1 & 114 & 79 & 67 & 48 &marginal\\
         & CHEOPS & visit 2 & 110 & 81 & 55 & 48 &marginal\\
         & CHEOPS & visit 3 & 125 & 93 & 58 & 48 &marginal\\
         & TESS   &  sector 7&  101 - [91, 116] & 97 - [88, 106] & 73 - [54, 106] & $<48$ & no\\
         & TESS   &  sector 8& 98 - [93, 213] & 93 - [86, 122] &  72 - [59,80] & $<48$ & no\\
         & TESS   & sector 34 &  89 - [81, 101] & 85 - [79, 99]&  61 - [51,76] & $<48$ & no\\
         & ESPRESSO & night 1 & 2.6 & 2.1 & 2.0 & - & no\\ 
         & ESPRESSO & night 2 & 2.0 & 1.6 & 1.2 & - & no\\ 
         & ESPRESSO & night 3 & 1.1 & 0.95 & 0.8 & - & no\\ 
\hline
HD 88595 & CHEOPS & visit 1 & 100 & 77 & 63 & 66 & yes\\ 
         & CHEOPS & visit 2 & 104 & 75 & 67 & 66 & yes\\
         & CHEOPS & visit 3 & 108 & 77 & 72 & 66 & yes\\
         & CHEOPS & visit 4 & 103 & 78 & 55 & 66 & yes\\
         & TESS & sector 9 & 199 - [166, 258] & 181 - [156, 223] &  87 - [67,128] & $<66$ & no \\ 
         & TESS & sector 35 & 190 - [163, 293] & 177 - [159, 204]& 98 - [70,149] &  $<66$ & no \\
         & ESPRESSO & night 1 & 2.54 & 1.37 & 2.07  & - & yes\\ 
         & ESPRESSO & night 2 & 2.26 & 1.49 & 2.25 & - & yes\\ 
\hline
\end{tabular}
\vspace{0.1cm}
\parbox{7in}{
\footnotesize $^\star$We note that the analyses of TESS data are based on a large number of $8$-hours subseries, we then indicate the median values, and the min/max values under intervals.

}
\label{tab:table_summary}
\end{table*}

\begin{figure}[th!]
\centering
\resizebox{\hsize}{!}{\includegraphics{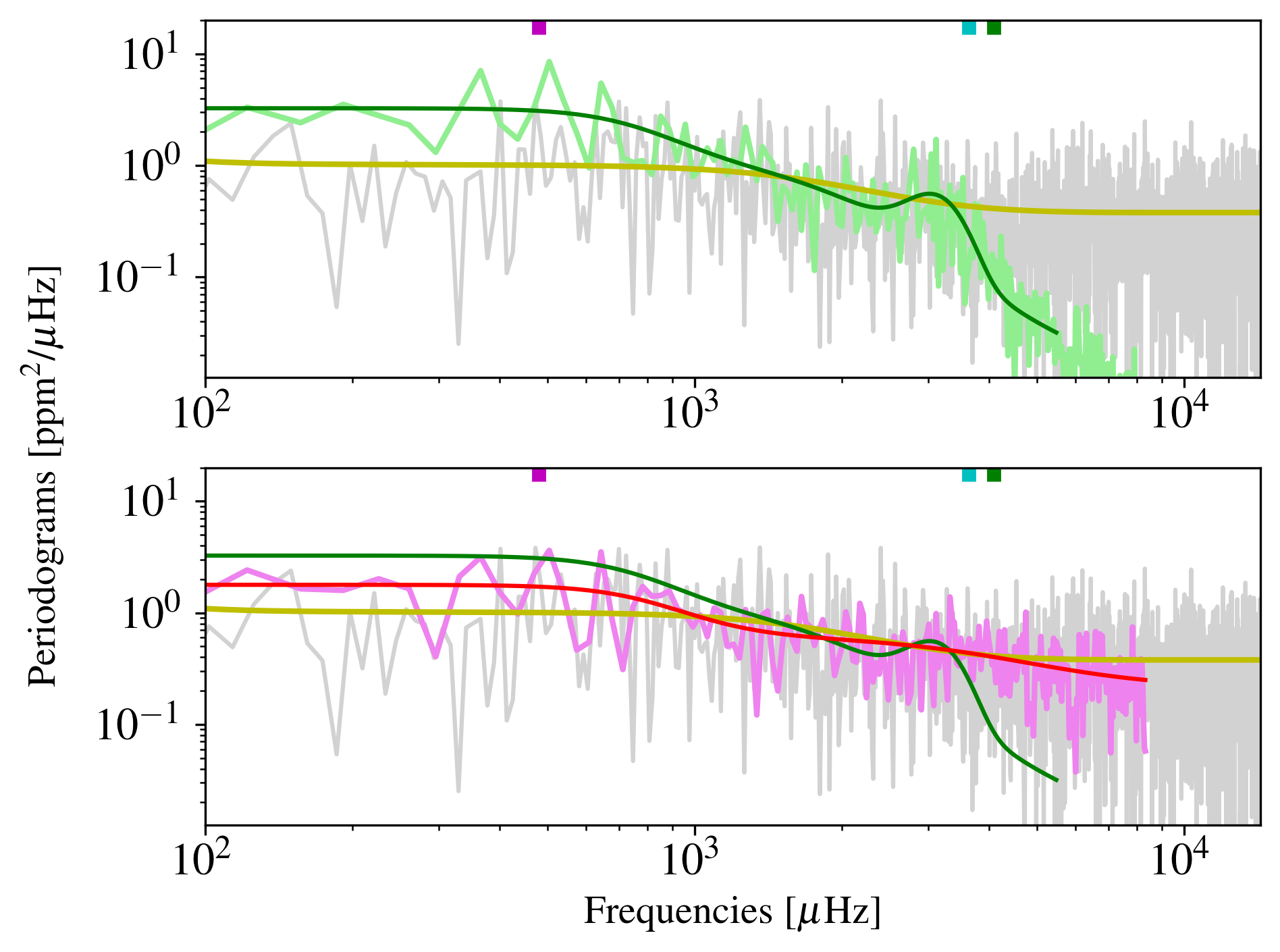}} 
\caption{Impact of the WGN on the periodograms. Top:  Comparison of the periodogram of HD 67458 (gray) and the averaged periodograms of three solar $8$-hour subseries (green).  Best-fitting models resulting from Harvey-functions (Eq.\eqref{eq_harvey}) fits on these CHEOPS and VIRGO periodograms are shown in yellow and green, respectively. Bottom: Same comparison but with the averaged periodogram of three solar subseries with the WGN added (pink).  The associated best-fitting Harvey function is shown in red. We see how the WGN impacts the periodogram shape (and so the extraction of the granulation properties) by comparing the green and red solid lines. In both panels, the different cut-off frequencies  $\nu_{cut}$, $f_H$, and $f_g$ inferred on CHEOPS observations are represented by the green, cyan and magenta vertical lines, respectively.
}
\label{fig_perSun}  
\end{figure}

{In each panel of Fig.~\ref{fig_per}, we have indicated the cut-off frequencies $\nu_{cut}$ that delineate the white noise and granulation regimes (green vertical lines). In the CHEOPS and ESPRESSO datasets, $\nu_{cut}$ correspond to periods in the interval $[244, 810]$ s. In the TESS dataset, $\nu_{cut}$ corresponds to periods around $100$ min. We note that, in the dataset of ESPRESSO observations of HD 88595, the  approximation of white noise is only partially true (the high frequency part of the periodogram is not perfectly flat).}
These levels of {white} noise impact the detection of the granulation signal, with the periodogram slope in the frequency region of stellar granulation that decreases with the increase of WGN. Without correction of this WGN, the inferred Harvey parameters from model \eqref{eq_harvey} may show discrepancies around their expected values in the Hertzsprung-Russell (HR) diagram. 
To illustrate this effect, we compare the periodograms of CHEOPS observations of the solar-like star HD 67458 and of the solar VIRGO observations with WGN added (because intrinsically the WGN {component} is very low in solar data). 
For this comparison, we compute the averaged periodogram defined as in \citetads{2017ITSP...65.2136S}:
\begin{equation} 
	P_L(\nu_k): =  \frac{1}{L} \sum_{\ell=1}^{L}  P^\ell(\nu_k), 
    \label{eq_PL}
\end{equation}
with $P^\ell$ { the periodogram computed on $L=3$ solar $8$-hours subseries taken randomly in the VIRGO sample.  
For the case with WGN, we added to each $L$ solae subseries a WGN with standard deviation $\sigma_w$. The value of $\sigma_w$ has been scaled until the high-frequency region of the VIRGO periodograms matches the one of CHEOPS observations. This corresponds to $\sigma_w \approx 30$ ppm. 
The averaged periodograms of the solar data (with and without WGN added) as well as the best-fitting Harvey functions are shown in Fig.~\ref{fig_perSun}. The best-fitting parameters related to the solar granulation signal component in Eq.~\eqref{eq_harvey} are given in Table~\ref{tab:harvey}. 
We see the impact of WGN on the Harvey parameters and particularly the decrease of the power index value ($c_1$ parameter) with the increase of $\sigma_w$.}
When parameter $c_1$ is fixed in Eq.~\eqref{eq_harvey} (for example $c_1=2$ corresponds to a standard Harvey model), this can bias the comparison of the best-fitting Harvey parameters derived for stars with different apparent magnitude (i.e., different levels of {white} noise).
Since Harvey functions are classical empirical functions used to model the stars' power spectral density, when we compare the inferred Harvey parameters for stars of different apparent magnitude, one needs to be careful about the impact of {this white} noise on the fitted parameters. The interpolation of the inferred parameters to their value at a ``reference level'' of white noise is necessary to avoid any bias.

\section{Relationship between granulation and stellar properties}
\label{sec4}

\subsection{Computing the flicker index}
\label{sec41}

In this section, we made use of the flicker index \citepads{2020A&A...636A..70S}. This index is a granulation indicator that has been defined as the slope of the averaged periodogram $P_L$ in Eq.~\eqref{eq_PL} in the frequency region where the granulation signal dominates. Related to the parameter $c_1$ in Harvey functions (Eq.~\eqref{eq_harvey}), this index has been shown to be correlated to the stellar fundamental parameters once it is corrected from the influence of the high-frequency white noise component.

\begin{figure}[t]
\centering
\resizebox{\hsize}{!}{\includegraphics{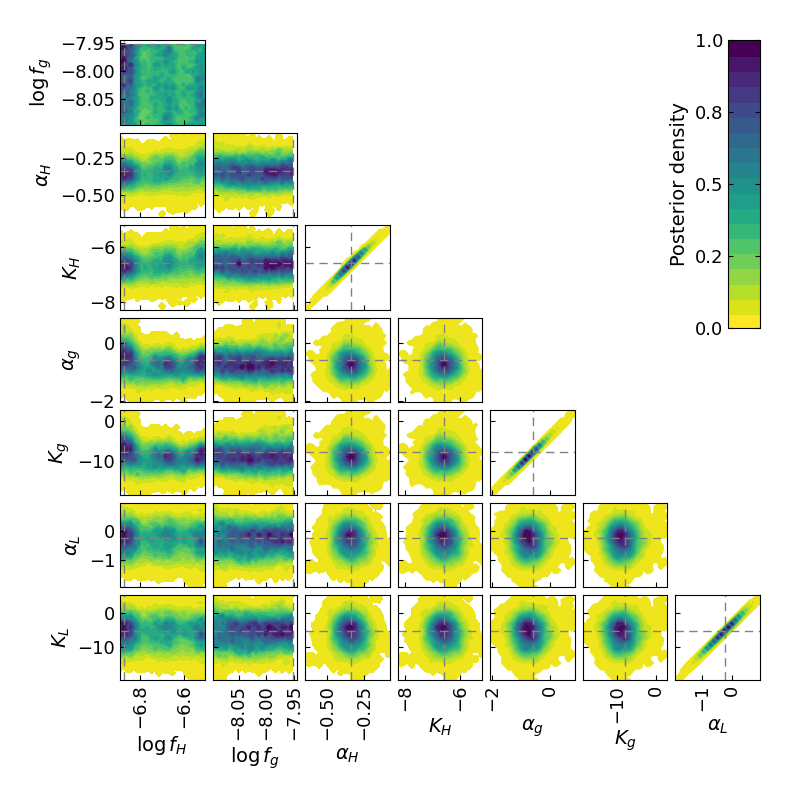}}
\vspace{-0.2cm}
\caption{Posterior distribution and correlation between all free parameters involved in model \eqref{eq_model} fitted to the averaged periodogram of HD 88595 CHEOPS observations. Frequency $f_g$ is not optimally constrained since the duration of the time series is relatively short ($<8$ hours).
}
\label{fig_mcmc}  
\end{figure}

To compute the averaged periodograms of TESS observations, we consider all the available $8$-hour subseries that have a duty cycle $\ge 90 \%$. This corresponds to $L=172$ subseries for HD 67458, and  $L=113$ for HD 88595.
Since the duty cycle of the individual CHEOPS and ESPRESSO visits of both stars is excellent (see Tables.~\ref{tab:table_cheops} and \ref{tab:table_espresso}), we use our whole set of observations to compute $P_L$.
The resulting averaged periodograms are shown in red in Fig.~\ref{fig_per}.

We then fit each averaged periodogram with a model defined as a sum of power laws functions of the form:
\begin{equation}   
    \log P_L(\nu_k^{+}) = \sum_{i=1}^{3} -\alpha_i \log (\nu_k^{+}) + \beta_i,
    \label{eq_model}
\end{equation}
with $\alpha_i = \{\alpha_{H}, \alpha_g, \alpha_{L}\}$ the periodogram slopes in three particular frequency regimes split by two cut-off frequencies $\{f_g, f_H\}$, and $\beta_i = \{\beta_{H}, \beta_g, \beta_{L}\}$ the corresponding amplitudes. {Model \eqref{eq_model} takes the form of straight lines in the log-log space.}
Parameters $\{\alpha_{H}, \beta_{H}\}$ represent the periodogram in the high frequency regime (dominated by the photon/instrumental noises, and also stellar oscillations for ESPRESSO observations of HD 88595), parameters $\{\alpha_g, \beta_g\}$ in the  regime dominated by the granulation signal, and parameters $\{\alpha_{L}, \beta_{L}\}$ in the regime dominated by low-frequency stellar signal. We note that photometric and RV observations are expected to be sensitive to different noise sources in the low frequency region (e.g.,  the signal of supergranulation is large in RV observations but negligible in photometric observations, see Sec.~\ref{sec61}). 

Therefore, model \eqref{eq_model} has eight free parameters (indices $\alpha$ and amplitudes $\beta$ for each three PSD regimes, plus two cut-off frequencies $\{f_g, f_{H}\}$ that mark these three regimes out). Using the MCMC scheme described in Sec. 4.2 of  \citetads{2020A&A...636A..70S}, we fit model \eqref{eq_model} to each averaged periodogram. An example  MCMC posterior is shown in Fig.~\ref{fig_mcmc}. The best fitting parameters for the flicker indexes ($\alpha_g$), the flicker power amplitudes ($\beta_g$) and the two cut-off frequencies are listed in Table~\ref{tab:flicker}. 

For reference, the flicker index inferred from the VIRGO averaged periodogram in the frequency range $\nu \in [550,2083]$ $\mu$Hz is $\alpha_g =1.33\pm 0.03$ (for the red SPM channel). We note that consistent indexes are found for all SPM channels, as shown in \citeads{2020A&A...636A..70S}).
The flicker index inferred from GOLF averaged periodogram in the frequency range $\nu \in [550,1236]$ $\mu$Hz is $\alpha_g =1.34 \pm 0.14$, {consistent with VIRGO observations}. Note the difference in the cut-off frequency $f_H$ between the two instruments.

From Table~\ref{tab:flicker}, we read a flicker index for the solar-like star HD 67458 smaller than the one found on solar VIRGO data. As discussed in Sec.~\ref{sec32}, this is due to the {large level of white noise} in CHEOPS data. 
For comparing different stellar observations, we need to correct the flicker index values for this effect. This will be done in Sec.~\ref{sec42}.
On the other hand, the flicker index deduced from the ESPRESSO periodogram of HD 67458 is consistent with zero, {leading us to the conclusion that the granulation signal is this dataset is not clearly identified}.
%
On the contrary, indexes inferred from  CHEOPS and ESPRESSO averaged periodograms of HD 88595 are both large (i.e., $\alpha_g >>0$), and are consistent within their $1\sigma$ errorbars. 

Finally, the flicker indexes inferred from TESS averaged periodograms of HD 67458 and HD 88595 are $\alpha_g <0.1$, both consistent with a nondetection. Analyzing the two TESS sectors of HD 88595 allows, however, a marginal detection of the granulation signal with $\alpha_g = 0.26 \pm 0.36$ for sector 9 (non detection), and $\alpha_g=0.71 \pm 0.39$ for sector 35 (marginal detection). We investigate the influence of the high-frequency noise level, the duration of the subseries, and the temporal sampling on the inferred flicker index in Appendix.~\ref{App_params}.

\renewcommand{\arraystretch}{1.4}
\begin{table}[t]
\caption{Best-fitting parameters of model \eqref{eq_model} fitted to CHEOPS and ESPRESSO averaged periodograms of HD 67458 and HD 88595. Frequencies $f_{H}$ and $f_{g}$ are expressed in $\mu$Hz. The last row indicates the flicker index based on solar VIRGO and GOLF averaged periodograms.}
\vspace{-0.3cm}
\label{tab:flicker}
\begin{center}
\begin{tabular}{ cccc }
\hline\hline
 Target & Parameter & Photometry &  Spectroscopy \\ 
\hline
         & $\alpha_g$ & $0.52^{+0.12}_{-0.13}$ & {$0.02^{+0.43}_{-0.28}$}   \\  
HD 67458 & $\beta_g$ & $-4.88^{+0.74}_{-0.42}$ & $-12.60^{+2.82}_{-2.08}$   \\  
         & $f_{H}$  & $3636^{+1645}_{-577}$  & $1680^{+711}_{-125}$   \\  
         & $f_g$ & $480^{+26}_{-132}$ & $ 442^{+189}_{-10}$   \\ 
\hline 
         & $\alpha_g$ &  $0.61^{+0.49}_{-0.25}$ & $0.77^{+38}_{-0.28}$   \\  
HD 88595 & $\beta_g$ &  $-8.08^{+3.59}_{-1.80}$ & $-18.49^{+2.75}_{-2.15}$   \\  
         & $f_{H}$ & $1036^{+450}_{-29}$ & $1104^{+7}_{-8}$   \\  
         & $f_g$ & $351^{+2}_{-37}$ & $209^{+111}_{-5}$  \\
\hline 
Sun & $\alpha_g$ & $1.33\pm 0.03$ & $1.34 \pm 0.14$   \\  
\hline
\end{tabular}
\end{center}
\vspace{-0.3cm}
\end{table}

\subsection{Comparison with Kepler bright stars}
\label{sec42}

\begin{figure}[t!]
\centering
\resizebox{\hsize}{!}{\includegraphics{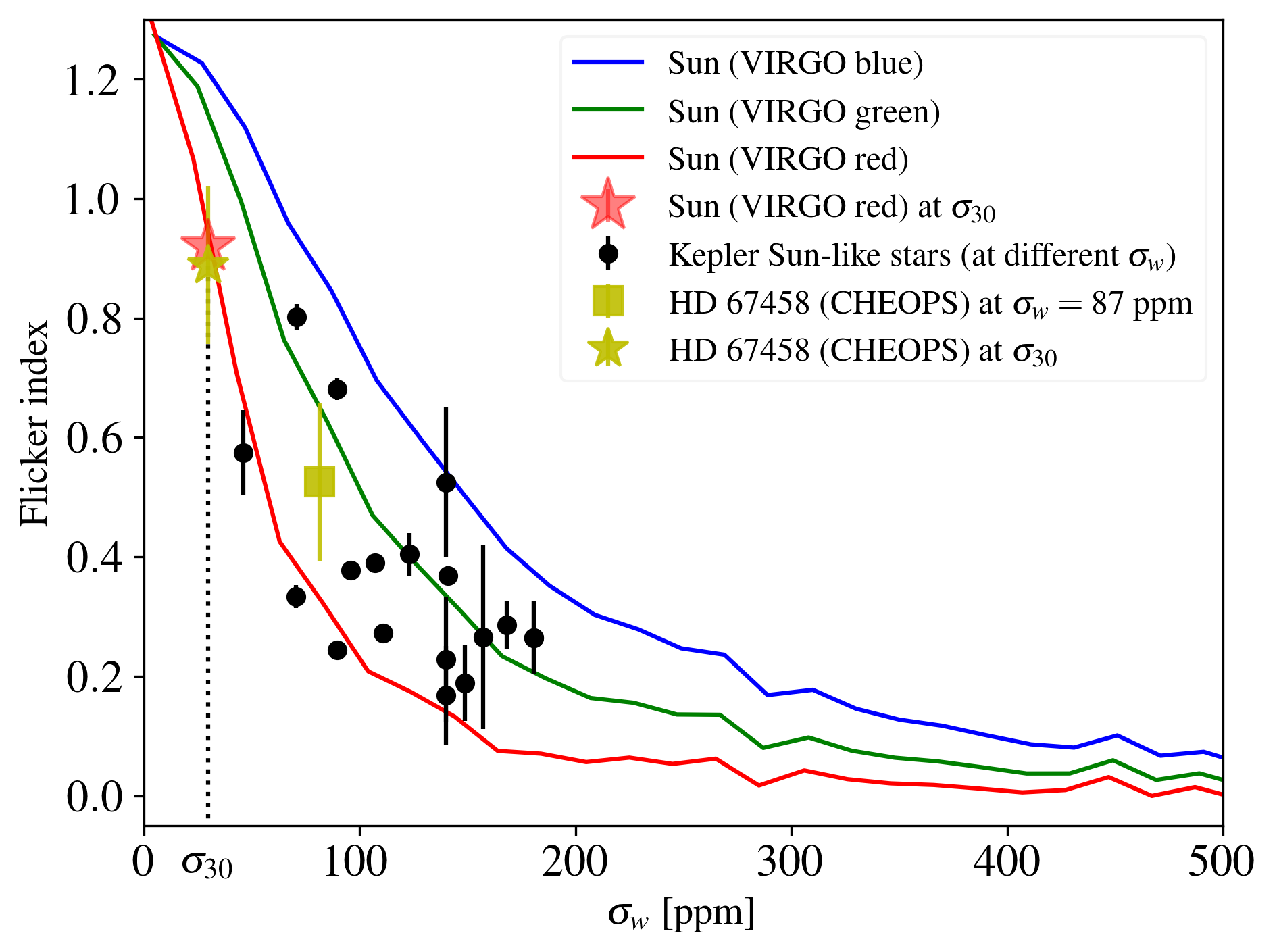}}
\caption{Estimated values of the flicker index ($\alpha_g$) as a function of the white noise level ($\sigma_w$) added in the VIRGO time series (red, blue, and green SPM channels, see Fig. 19 of \citeads{2020A&A...636A..70S}). The red star corresponds to the flicker index of the red channel measured at $\sigma_w=\sigma_{30}= 30$ ppm. Black dots show the flicker indexes inferred from the averaged periodograms of Kepler Sun-like stars. The yellow square shows the flicker index inferred from the averaged periodogram of HD 67458 (CHEOPS), and the yellow star shows its interpolation at $\sigma_{30}$. }\label{fig_interp}  
\end{figure}

To compare the flicker indexes for stars of different apparent magnitudes (i.e., different {white} noise levels), we have to interpolate the behavior of the periodogram's slope at a constant {white} noise level. Following  \citetads{2020A&A...636A..70S}, we target the level $\sigma_{w} = 30$ ppm (hereafter $\sigma_{30}$), {which was arbitrarily chosen as a good compromise for all studied stars (since all estimates of $\sigma_{w}$ are above $30$ ppm).} 
{For each photometric dataset (VIRGO's three SPM channels, CHEOPS observations), we then applied the following procedure. 
\begin{enumerate}[label=(\alph*)]
    \item We added a synthetic white Gaussian noise of standard deviation $\sigma_{w}^{(i)}$ to each $L$ available light curves.
    \item We compute the averaged periodogram $P_L^{(i)}$ with Eq.\eqref{eq_PL} using these $L$ new light curves.
    \item We evaluate the flicker index $\widehat{\alpha}_g(\sigma_{w}^{(i)})$ by fitting model \eqref{eq_model} to $P_L^{(i)}$.
    \item We performed steps (a) to (c) for $\sigma_{w}^{(i)} = 0$ (initial conditions) to $\sigma_{w}^{(i)} =1000$ ppm. 
\end{enumerate}
}
\noindent {This gives us the empirical curve $\widehat{\alpha}_g(\sigma_{w}^{(i)})$, which behaves as a decreasing exponential function of the form} (see Eq. (5) in \citeads{2020A&A...636A..70S}):
\begin{equation}
\widehat{\alpha_g}(\sigma_{w}^{(i)}) = a_e ~ e^{-b_e~\sigma_{w}^{(i)}} + c_e,
\label{eq_expo}
\end{equation}
with $\{a_e,b_e,c_e\} \in \mathbb{R}$, that are found correlated with the stellar parameters\footnote{We note that we do not find coefficients $\{a_e,b_e,c_e\}$ consistent with the predictions from Eq.(6) of \citetads{2020A&A...636A..70S}. This may be due to the small number ($<5$) of targets with mass $< 1$ $M_\odot$ and/or $>1.5 M_\odot$ that were used to derive the Eq.(6) of \citetads{2020A&A...636A..70S}. This may also be due to uncertainties on the stellar parameters of the Kepler sample or to the few number of CHEOPS visits that leads to an approximate interpolation of the curve $\widehat{\alpha}_g(\sigma_i)$ involved in Eq.\eqref{eq_expo}.}.
This behavior is clearly shown in Fig.~\ref{fig_interp} for the three VIRGO SPM channels. For the red channel, we read  $\widehat{\alpha}_g(\sigma_{30}) = 0.92 \pm 0.14$ (see yellow star).
The flicker index $\alpha_g = 0.52^{+0.12}_{-0.13}$ inferred on the CHEOPS data of HD 67458 (see Table \ref{tab:flicker}) is shown at $\sigma_w\approx87$ ppm (value computed using the three visits, see yellow square symbol). 
Based on the fit of model \eqref{eq_expo} to the empirical curve $\widehat{\alpha}_g(\sigma_{w}^{(i)})$ of HD 67458, we found $\widehat{\alpha_g}(\sigma_{30}) = 0.89^{+0.12}_{-0.13}$ (see yellow star symbol). The flicker index of HD 67458 is therefore in complete agreement with the solar flicker index at comparable white noise levels.
For HD 88595, we found $\alpha_g =0.61^{+0.49}_{-0.25}$ at $\sigma_w=77$ ppm (see Table \ref{tab:flicker}) and $\widehat{\alpha_g}(\sigma_{30}) = 2.08^{+0.49}_{-0.25}$ based on the interpolation.

\begin{figure*}[t!]
\centering
\resizebox{\hsize}{!}{\includegraphics{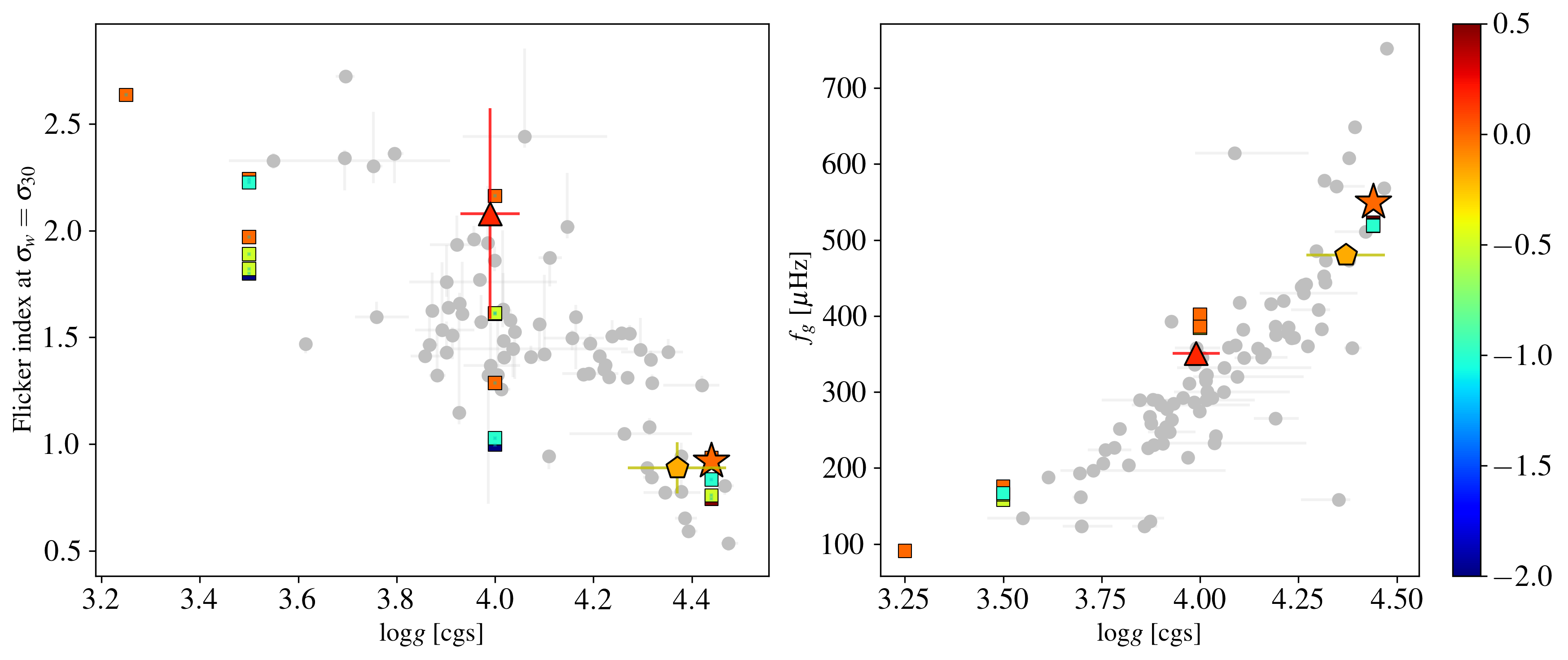}}
\caption{Flicker parameters as a function of the stellar surface gravity.  The flicker indexes (left) have been interpolated to a fixed level of white noise ($\sigma_w=30$ ppm) for this comparison. The flicker frequencies are shown in the right panel. In both panels, the gray dots indicate values obtained for the sample of Kepler bright targets (magnitude $<10$) involved in \citetads{2020A&A...636A..70S}. The pentagon symbols indicate values obtained for HD 67458 and the triangle symbols for HD 88595 based on CHEOPS high-precision observations. The star symbols indicate the value obtained for the Sun (VIRGO observations). The square symbols indicate the values obtained from HD simulations of $17$ stars (see Sec.~\ref{sec5}). The color code indicates the stellar metallicity [Fe/H] (not shown for Kepler targets). 
}
\label{fig_flicker}  
\end{figure*}

We compare these values with the brightest stars (magnitude $<10$) in the Kepler sample studied in \citetads{2020A&A...636A..70S}. This sample includes $87$ stars (among which the $17$ solar-like stars, with inferred flicker indexes added to Fig.~\ref{fig_interp}), from which the flicker indexes have been inferred from averaged periodograms computed based on hundreds of $1$-day subseries and then interpolated at the {white} noise level $\sigma_{30}$ using Eq.\eqref{eq_expo}. 
Parameters $\widehat{\alpha}_g(\sigma_{30})$ and the flicker cut-off frequency $f_g$ are shown as a function of the stellar surface gravity\footnote{The ${\rm log}g$ for the Kepler targets was obtained by different techniques, from the highest priority to the lowest ones, depending on availability: asteroseismology, high-resolution spectroscopy, low-resolution spectroscopy, flicker, photometric observations, and finally the KIC (see \href{https://exoplanetarchive.ipac.caltech.edu/docs/KSCI-19097-004.pdf}{https://exoplanetarchive.ipac.caltech.edu/docs/KSCI-19097-004.pdf}). } in  Fig.~\ref{fig_flicker}. 
The two bright stars observed by CHEOPS (pentagon and triangle symbols for HD 67458 and HD 88595, respectively) are in complete agreement with Kepler (gray dots) and solar (star symbol) predictions.
The inferred parameters are listed in Appendix.~\ref{App_flicker} (see Table~\ref{tab:table_flicker}, which is also available online). 

We note that the frequency $f_g$ behaves similarly to the convective characteristic timescale ($1/b_1$) that is fitted empirically by the usual Harvey functions in Eq.~\eqref{eq_harvey} assuming an exponential decay with time.
They are both strongly correlated with the stellar maximum oscillation frequency $\nu_{max}$. The frequency $f_g$ can be interpreted as the upper tail of the distribution of the granule cells' lifetime  \citepads{2011A&A...532A.108S}.  
From  Fig.~\ref{fig_flicker}, we read that the maximum correlation timescale for our sample stars ranges between $\sim22$ min ($750 ~\mu$Hz) and $\sim2.7$ hours ($100 ~\mu$Hz). 


\section{Predictions from 3D hydrodynamic models of convection}
\label{sec5}

\subsection{Description and predictions for HD 67458 and HD 88595}
\label{sec51}

{ The minimum level of stellar activity of a late type star is due to surface convective motions that produces stochastic variations of the light curves. The amplitude and time scale of these fluctuations depend on the spectral type and generally increase toward red giant type or earlier type (F star) due to more vigorous convection in these stars. In this work we aim at reproducing this stellar noise with the use of state-of-the-art 3D hydrodynamical simulations. These simulations do not account for magnetic fields and therefore no plage or spot but they showed a remarkable agreement with bolometric light variations of KEPLER targets \citepads{Rodriguez2022}. Previous studies have also used these kind of models to study the granulation signal (see e.g., \citeads{Ludwig2006}; \citeads{Ludwig2013};  \citeads{Trampedach2013}; \citeads{Tremblay2013};  \citeads{2013A&A...559A..40S}), and compared their results with observations. However, those studies could not reproduce the observational trends with high accuracy or without introducing the Mach number, a quantity that is very difficult to compute for observational data.}

Therefore, we generated long time series of box-in-a-star type 3D hydrodynamical simulations across the HR diagram using the STAGGER-code (\citeads{Nordlund1995}; \citeads{Magic2013}). This code solves the equations for the conservation of mass, momentum, and energy, as well as the radiative transfer equation assuming local thermodynamic equilibrium (LTE). For more information about the code we refer to \citetads{Rodriguez2022} and references therein. The box-in-a-star type means that the 3D models are centered around the stellar photospheres. That is, they cover the photosphere, the superadiabatic region, and the quasi-adiabatic deeper convective layers, where a flat entropy profile is ensured at the bottom boundary. These layers are distributed in a specific 3D Cartesian geometry. 

The 3D models are defined by three stellar parameters: the effective temperature $T\rm_{eff}$, the surface gravity ${\rm \log}g$, and the metallicity [Fe/H]. $T\rm_{eff}$ is defined by the entropy value at the bottom boundary of the models, while [Fe/H] is defined by the abundance of chemical elements present in the models. 

Each model contains typically $10$ {granules}, whose sizes are a few tens of pressure scale heights. This means that the sizes of the granular cells are bigger for models representing early-type stars or evolved stars.

Realistic radiative transfer is performed using long characteristic along several rays at different inclinations across the simulation domain in order to account for heating and cooling in the energy equation (e.g., \citeads{2003ASPC..288..519S}). These radiative intensities are integrated in wavelengths and inclinations to give bolometric fluxes which can be compared to observations. 
To compare with observations, this quantity has to be rescaled by the number of granules visible on the disk, as \citetads{Trampedach1998} and \citetads{Ludwig2006} proposed. With this scaling, \citetads{Rodriguez2022} was able to determine the standard deviation of the stellar disk-integrated intensity from the small box models. We refer the reader to \citetads{Rodriguez2022} for a more detailed description of the method. 

\citetads{Rodriguez2022} found that at solar metallicity the standard deviation of the flux scales like $\sigma_{HD} \sim \nu_{\rm max}^{-0.567}$, and the characteristic timescale follows $\tau_{ACF} \sim \nu_{\rm max}^{-0.997}$ (see Table 4. in their paper). 
We note that the characteristic timescale is defined as the autocorrelation time of the entire time series. It is a different timescale than the one related to the flicker frequency $f_g$ (measured on the observed periodogram). {For consistency with how we evaluate the granulation amplitude in Sec.~\ref{sec31}, we have also determined a relation for F8 based on these 3D HD models (see App.~\ref{appD_F8} for details). We found  F8$_{HD} \propto \nu_{\rm max}^{-0.530}$ .}

Using such scaling and the parameters given in Table~\ref{tab_params}, the two {values predicted by the 3D simulations are F8$_{HD}  \approx 48 \, {\rm ppm}$ (HD 67458) and F8$_{HD}  \approx 66 \, {\rm ppm}$ (HD 88595)}. The predicted characteristic timescales are $\tau_{ACF}  \approx 203.2 \, {\rm s}$ (HD 67458) and $\tau_{ACF}  \approx 369.3 \, {\rm s}$ (HD 88595). {For HD 67458, the F8 values inferred from CHEOPS dataset are larger than the predictions from HD simulations, indicating that white noises are dominating the dataset. For HD 88595, the inferred and predicted F8 are in very good agreement (see Sec.~\ref{sec31} and Table \ref{tab:table_summary}).} 

\subsection{Flicker indexes and relationship with the stellar properties }
\label{sec52}

From the total sample of $27$ stars simulated with the 3D models in \citetads{Rodriguez2022}, we select the $17$ main-sequence stars. The selected stars have $T\rm_{eff} \in [4727,6485]$ K, ${\rm log}g \in [3.25,4.44]$ cgs, and [Fe/H]$ \in [-2,0.5]$. 

The length of all the synthetic time series correspond to at least $1000$ convective turnover times, with these turnover times that are adapted to the stellar properties of the synthetic target stars. For all of these target stars, we compute the classical periodogram and scale the two cut-off frequencies $f_g$ and $f_c\ge f_H$ with the stellar parameters, with $f_c$ the corner frequency that marks the p-modes frequency domination regime out (see \citeads{2020A&A...636A..70S}). 
We finally evaluate the flicker indexes as in Sec.~\ref{sec41} based on these synthetic periodograms of stellar granulation. 
Results are shown with the square symbols in Fig.~\ref{fig_flicker}. The color code indicates the stellar metallicity (not given for Kepler stars). We note that the HD simulated time series do not contain white noise, but only granulation signals and are therefore used as reference.

The flicker index indicators $\alpha_g$ derived from the HD convection models are in good agreement with the bright Kepler and CHEOPS targets. In line with \citetads{Rodriguez2022}, this demonstrates the success of these 3D models in reproducing realistic photometric time series of stellar granulation. We however note a larger dispersion of the simulations compared to observations for stars with ${\rm log}g=4.0$. This dispersion seems to be strongly correlated with the stellar metallicity: low metallicity shows smaller flicker index. While the results are different at other  ${\rm log}g$ values, this needs to be investigated further based on a larger synthetic stellar population.  

The flicker frequency $f_g$ derived from the HD convection models are also in good agreement with the values inferred from Kepler and CHEOPS observations. We note however a slight shift compared to Kepler values at ${\rm log}g=4.0$.

\section{Link between the spectroscopic and photometric signatures}
\label{sec6}

Only a few variations of the physical granulation  properties were observed during the solar magnetic cycles \citepads{Garcia_2005}, with $2\%$ variation in the density and mean granules' area observed during the solar cycle \citepads{2021A&A...652A.103B}. This cannot be directly verified on other stars since their surfaces cannot be resolved on the scale of the granulation cells. However, this can also be confirmed indirectly by studying variations in their power spectra  \citepads{2011A&A...532A.108S,2018A&A...616A..87M, 2020A&A...636A..70S}. A comparison of the current {periodogram} of HD 88595 with future observations taken in a few years would allow us to confirm if the granulation signal is indeed stationary with the stellar magnetic cycle. Without having any dataset -- to our knowledge -- to verify this statement, we assume in this section that photometric observations of the granulation signal can allow us to predict its signature (amplitude and timescales) in spectroscopic observations (e.g., in line with techniques that have been developed for magnetic activity, such the FF' technique described in \citetads{2012MNRAS.419.3147A}). This is important since high-precision photometric surveys allow the accumulation of ten to hundred of continuous $1$-day observation of stars from space missions, while RV ground-based surveys typically have poor sampling to characterize this noise source for exoplanet detection (typically: one to three data points per night spread over long term campaigns).

\subsection{Comparison of the {periodograms}}
\label{sec61}

\begin{figure*}[t!]
\centering
\resizebox{\hsize}{!}{\includegraphics{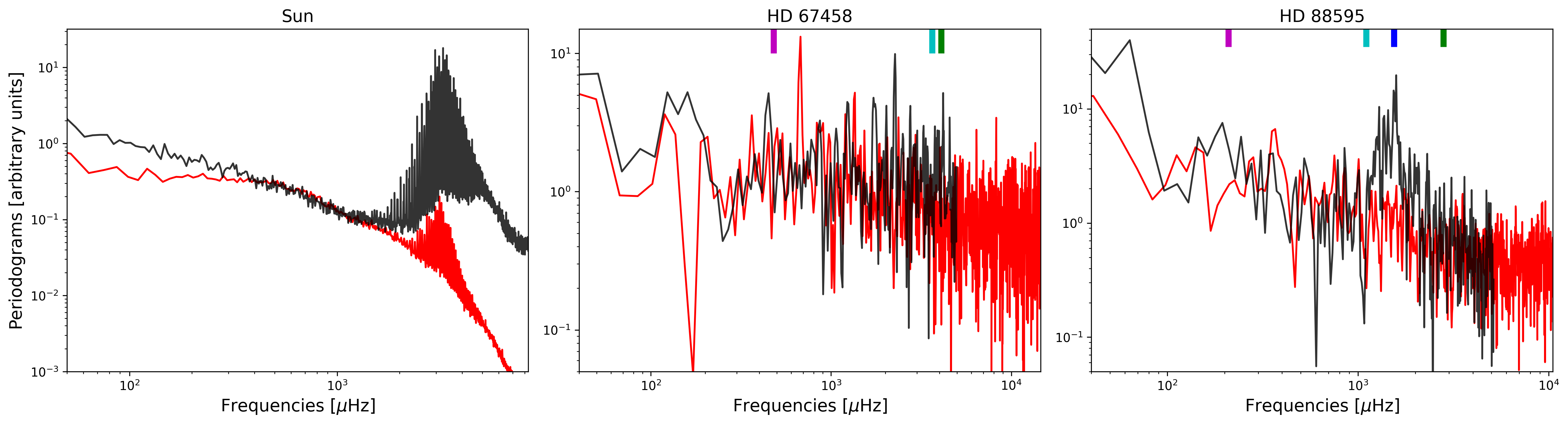}}
\caption{Comparison of the averaged periodograms of photometric (red) and spectroscopic (black) observations. Left: Solar VIRGO (red SPM channel) and GOLF observations. Middle: CHEOPS and ESPRESSO observations of HD 67458. Right: Same for HD 88595. The periodograms have been arbitrarily normalized to compare their slope in the frequency regime that is dominated by stellar granulation. When relevant, the different cut-off frequencies $\nu_{cut}$, $\nu_{max}$, $f_H$, and $f_g$ represented by the green, blue, cyan and magenta vertical lines, respectively. 
}
\label{fig_sun}  
\end{figure*}

In the left panel of Fig.~\ref{fig_sun}, we compare the (arbitrarily normalized\footnote{We note that this normalization factor has no impact on the inferred flicker index values, which are based on the periodograms' slope and not their amplitudes.}) averaged periodograms of solar VIRGO (red) and GOLF (black) observations. We observe comparable periodogram' slopes on data taken in photometry and spectrophotometry with flicker indexes $\alpha_g =1.33 \pm 0.03$ and $\alpha_g =1.34 \pm 0.14$, respectively (see Sec.~\ref{sec41}). In both dataset, the level of {white} noise is low and does not affect the characterization of the stellar granulation signal. We note that, in the frequency region where the supergranulation starts to dominate ($\nu<200~\mu$Hz), the {periodogram} of brightness becomes flat while it still increases in the {periodogram} of RV.

{In the two other panels, we show the averaged periodograms of CHEOPS and ESPRESSO observations of HD 67 458 and HD 88595. The flicker index deduced from the ESPRESSO periodogram of HD 67458 is consistent with zero (see Table~\ref{tab:flicker}), but its value remain consistent within their $1\sigma$ errorbars with the flicker index inferred from CHEOPS observations. While we assert a marginal detection of the granulation signal for this star, this could explain the visual good match of the two periodograms; the white noise being too large to infer a precise flicker index value. More subseries ($>>3$) would be needed to improve the precision on this parameter. On the other hand,} 
the flicker indexes of HD 88595 inferred on CHEOPS ($\alpha_g = 0.61^{+0.49}_{-0.25}$) and ESPRESSO data ($\alpha_g = 0.77^{+38}_{-0.28}$; see Table~\ref{tab:flicker}) are in complete agreement within their $1\sigma$ errorbars. 

This indicates that the properties (i.e., correlation timescales where this signal dominates, amplitudes distribution, and type of correlation with the flicker index) of the stellar granulation signal in RV data could be predicted from high-precision photometric observations (CHEOPS and PLATO) of bright stars.

\subsection{Prediction from empirical laws}
\label{sec62}

Some relations between the amplitudes of the stellar granulation signal in spectroscopy (RV RMS) and photometry (F8 metric) have been derived in the literature. We aim to test these predictions with the CHEOPS and ESPRESSO datasets of HD 67458 and HD 88595.

A first relation has been derived in \citetads{2014AJ....147...29B}, based on a small sample of $12$ stars observed first by Kepler and later  by RV ground-based surveys at the Keck and Lick observatories. The stars in this sample have been characterized as ``chromospherically quiet stars'' since they showed low-amplitude photometric variability ($\le 3$ ppt) in their Kepler light curves. However, the RV precision was limited to $\ge 4$ m/s leading to a difficult characterization of the stellar granulation signal. Fitting a linear law between F8 and the corresponding RV RMS for each stars,  the relation: 
\begin{equation}
{\rm RV~RMS} = (31.99 \pm 3.95) \times F8 + (3.46 \pm 1.19)
\label{eq_bastien14}
\end{equation}
has been derived with RV RMS in m/s and F8 in ppt (see also Sec.3.5 of \citeads{2019ApJ...883..195T}). 

A second relation has been derived in \citetads{2014ApJ...780..104C}, but based on indirect RV measurements. This study used a statistically more robust stellar sample ($944$ stars) observed by Kepler in photometry and with the GALEX survey \citepads{Martin_2005} in far ultraviolet (FUV). From the FUV observations, the authors converted the data to chromospheric activity proxies (${\rm log} R'_{\rm HK}$) using the conversion extracted from \citetads{Findeisen_2011}, and then converted again to an RV RMS using the relation given in \citetads{2003csss...12..694S}\footnote{We note that they found similar RV RMS based on another work of \citetads{2005PASP..117..657W}, but higher RV RMS values when using predictions from \citetads{2000A&A...361..265S}.}.  This work led to the linear relations: 
\begin{equation}
\begin{split}
{\rm RV~RMS} &=18.04  \times F8 + 0.98 ~~\text{if $T\rm_{eff} < 6000$ K},\\
{\rm RV~RMS} &=84.23  \times F8 -3.35 ~~\text{if $T\rm_{eff} \ge 6000$ K}.
\end{split}
\label{eq_cegla14}
\end{equation}
These relations were derived from targets deemed to be on the so called flicker floor in \citetads{2014ApJ...780..104C} and therefore convection dominated; stars with large amplitude photometric variations did not show a clear correlation between F8 and RV RMS.

Finally,  \citetads{2017A&A...606A.107O} extended the work of \citetads{2014AJ....147...29B} with $9$ new targets observed with K2 and HARPS. However, their HARPS RV measurements contained only few data points ($<12$) and the RV RMS remained large ($>10$ m/s) meaning the RV  signal was dominated by other noise sources than granulation (e.g., active regions). We then do not consider this study in the present work.

\begin{table}[t] \centering
\caption{Radial velocity RMS for HD 67458, HD 88595, and the Sun as predicted from Eqs.\eqref{eq_bastien14} and \eqref{eq_cegla14} based on the F8 metric computed on our CHEOPS and VIRGO observations. The last column shows the RV RMS measured on ESPRESSO observations {(see Table~\ref{tab:table_espresso})}. Radial velocity RMS for the Sun are based on the $75$ GOLF subseries. All RV units are in m/s. }
\vspace{-0.1cm}
\label{tab:r8_rv} 
\begin{tabular}{c|cc|c}
\hline\hline
Target &\multicolumn{2}{c}{RV RMS predicted from } & Measured\\
 & Eq.\eqref{eq_bastien14} & Eq.\eqref{eq_cegla14} &  RV RMS \\
\hline
HD 67458  & {$4.31 \pm 1.19$} & {$1.46$} &  $1.1, {2.6}$\\
HD 88595  & $5.59 \pm 1.22$ &  $2.27$ &  $2.26, 2.54$\\
Sun  & $4.10 \pm 1.19$  & $1.34$ &  $[0.87, 1.63]$ \\
\hline
\end{tabular}
\end{table}

{We note that in Eqs.\eqref{eq_bastien14} and \eqref{eq_cegla14} the F8 metric has been derived based on the initial definition, that is using data binned into $30$-min intervals ($\tau$). However, this $\tau$ is not optimal for main-sequence stars since the granulation timescales are shorter than $30$ minutes.
We therefore investigated different binning intervals. 
To reduce the contribution of the high-frequency noises, we found that a data binning of $\tau=15$ minutes and $\tau=5$ minutes (as in Sec.~\ref{sec31}) was sufficient to extract the signal of granulation in the CHEOPS data of HD 67458 and HD 88595, respectively. In particular, since ESPRESSO data of HD 67458 are dominated by photon noise we have to use a more drastic data binning than in Sec.~\ref{sec31} to reduce its contribution in the measured RV data and makes an approximation comparison with the literature possible.}
We then derive the RV RMS following the relations given in Eqs.\eqref{eq_bastien14} and \eqref{eq_cegla14}. {Results are shown in Table \ref{tab:r8_rv}.} Clearly, comparing these predictions with the RV RMS found on our ESPRESSO time series (see last columns) are in favor of Eq.\eqref{eq_cegla14} derived by \citetads{2014ApJ...780..104C}. 

We note however the high sensitivity of this relation to the length of the temporal binning $\tau$ used to compute Eq.\eqref{eq_cegla14}. 
Indeed, computing the F8 metric with $\tau=\{5, 15, 30\}$ min in Eq.\eqref{eq_cegla14} leads to a predicted RV RMS of $\{2.0, 1.61, 1.4\}$ m/s for HD 67458, $\{1.46, 1.34, 1.26\}$ m/s for the Sun, and $\{2.27, 1.05, -0.10\}$ m/s for HD 88595 (the last value representing a limit of Eq.\eqref{eq_cegla14} for F8 values $<39.77$ ppm). The choice of $\tau$ for computing the F8 metric does not significantly affect the predictions for the Sun (where the level of high-frequency noise in VIRGO data is negligible). For HD 67458, using a too short temporal binning for computing the F8 metric of the solar-like star HD 67458 makes the time series completely dominated by the high-frequency noise. Using a temporal binning of $30$-minutes, as in \citetads{2014ApJ...780..104C}, leads to predicted RV RMS of similar order of magnitude. For the F star HD 88595, the relation in  Eq.\eqref{eq_cegla14} is not adapted to $\tau > 15$ minutes. 
 Computing the F8 metric based on a temporal binning that is adapted to the granulation timescales, as proposed in \citetads{2018A&A...620A..38B}, may allow the photometric versus RV relation derived in \citetads{2014ApJ...780..104C} to be refined.

\begin{figure}[h!]
\centering
\resizebox{\hsize}{!}{\includegraphics{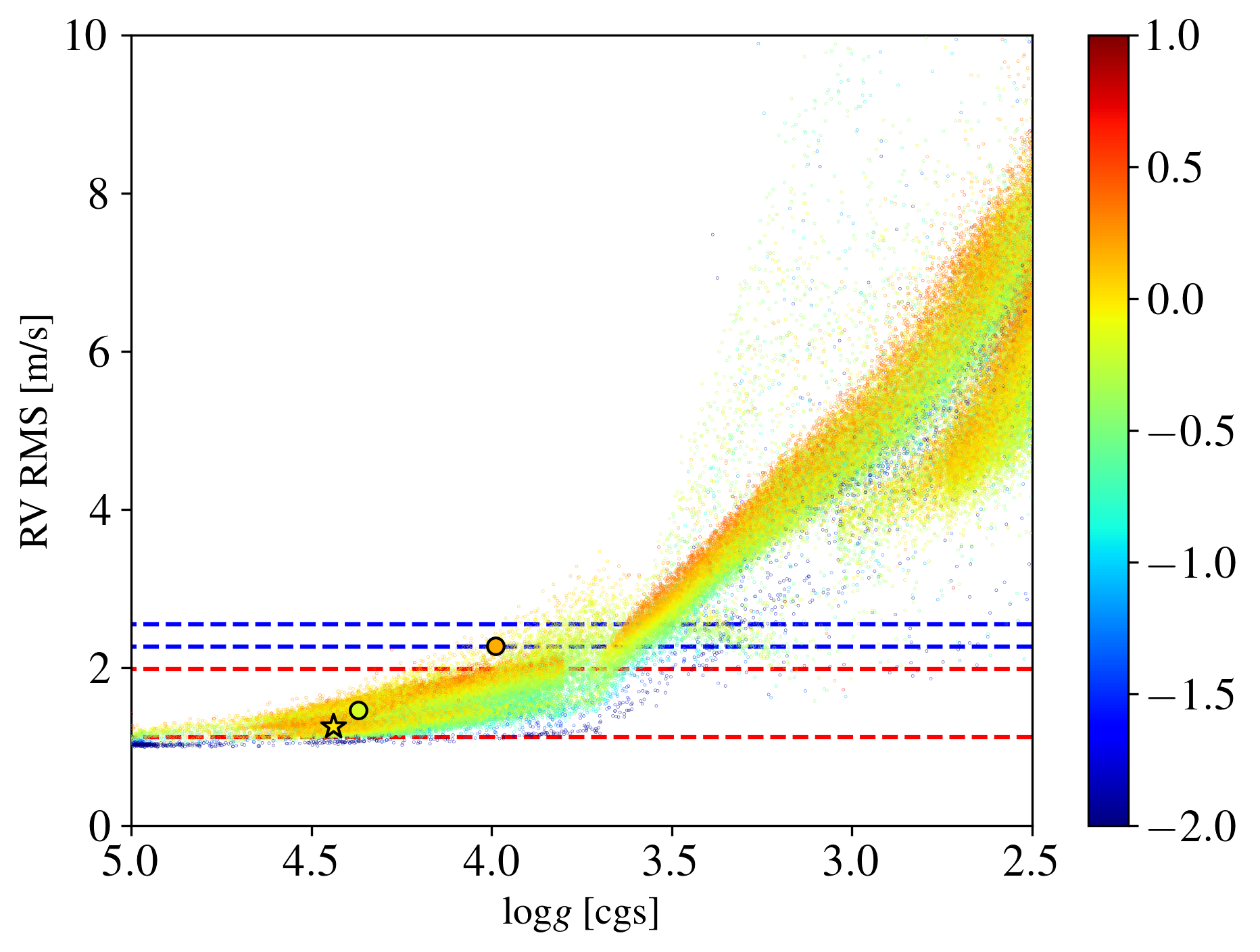}}
\caption{Radial velocity RMS as a function of the stellar surface gravity. 
Individual small dots represent the values for the $129~055$ stars of the APOGEE sample studied in \citetads{2019ApJ...883..195T}. For this stellar sample, the RV RMS was computed based on Eq.\eqref{eq_cegla14} \citepads{2014ApJ...780..104C}. 
The two big dots represent the RV RMS predicted from  Eq.\eqref{eq_cegla14} for HD 67458 and HD 88595. The star symbol represents the RV RMS predicted from  Eq.\eqref{eq_cegla14} for the Sun.
The color code for all targets represents the stellar metallicity [Fe/H].
The blue and red horizontal dashed lines show the RV RMS values inferred from ESPRESSO observations of HD 88595 and HD 67458, respectively. }
\label{fig_compa}  
\end{figure}

In addition, we compare the measured RV RMS of our two targets and the Sun with the stellar surface gravity and metallicity in Fig.~\ref{fig_compa}. The dots represent the predicted values from Eq.~\eqref{eq_cegla14}, both are in agreement with ESPRESSO observations (see the interval delimited by the dashed horizontal lines), as shown in Table~\ref{tab:r8_rv}. In this figure, we also represent the targets involved in the study of \citetads{2019ApJ...883..195T}, who computed the RV RMS with Eq.~\eqref{eq_cegla14} using an F8 metric coming from the stellar parameters. 
{This leads us to the conclusion that, although the predictions of \citetads{2014ApJ...780..104C} are rough estimates (i.e., no confidence interval is given while the granulation signal behaves -- a priori -- as Gaussian colored noise with amplitudes that should be defined with mean and standard deviation parameters, see \citeads{2020A&A...636A..70S}), they lead to values that are quite consistent with CHEOPS and ESPRESSO observations.}

\section{Analysis of CCF shape} 
\label{sec7}

 \renewcommand{\tabcolsep}{10pt}
 \renewcommand{\arraystretch}{1.2}
\begin{table*}[t!]
\caption{Correlations with RV as assessed by Spearman's Rank ($\rho_S$) for the binned datasets. Also listed in brackets is the average S/N from echelle order 10 per night; strong correlations with low S/N may be spurious (HD 67458), see for example Fig.~\ref{fig:shape_rv_hd67}. ``N.O'' means there was no overlap between CHEOPS and ESPRESSO observation for this night (see Fig.~\ref{fig_dates})}.
\vspace{-0.5cm}
\begin{center}
\begin{tabular}{c|ccc|cc}
    \hline
     & \multicolumn{3}{c|}{HD 67458} & \multicolumn{2}{c}{HD 88595}  \\
     Indicator & Night 1 {(88)} & Night 2 {(68)} & Night 3 {(87)} & Night 1 {(263)} & Night 2 {(285)}\\
    \hline  
    \hline  
	FWHM & $-0.75$ & -0.92 & $-0.77$ & $-0.31$ & 0.71 \\
    Contrast & $-0.68$ & -0.86 & -0.83 & $-0.26$ & $-0.14$\\
    EW & $-0.67$ & $-0.89$ & $-0.82$ & $-0.26$ & 0.26\\
    BIS & 0.35 & 0.05 & 0.56 & -0.77 & 0.31\\
    Curvature & -0.11 & 0.1 & 0.82 & -0.77 & -0.77\\
    A$_{\rm{b}}$ & -0.28 & 0.01 & -0.15 & 0.6 & 0.94\\
    Flux & N.O & $-0.56$ & 0.73 & 0.0 & 0.03 \\
    
    \hline
  \end{tabular}
\end{center}
\label{tab:ccf_corr}
\vspace{-0.5cm}
\end{table*}

 We know from solar observations that granulation gives rise to a ``C''-shaped bisector in most stellar absorption lines; the uprising/blue-shifted granules contribute more overall to the stellar lines due to their larger brightness and surface area, while the dark downfalling and red-shifted intergranular lanes serve to depress the redward wing. As the granules evolve over time, and the ratio of granule to intergranular lane changes, we expect to see a corresponding change in line shape, asymmetry and overall brightness. Such changes in line shape may then manifest as RV shifts. Both 3D MHD and HD simulations predict correlations between granulation-induced line shape and both RV and brightness (\citeads{2019ApJ...879...55C}; \citeads{dravins21}). Consequently, in this section we explore variations in the shape of the cross-correlation functions (CCFs) and RVs, as measured from the ESPRESSO observations, and brightness as measured from CHEOPS. We examine CCFs rather than individual stellar lines due to the increase in S/N; we note that due to the weighting in the CCF template mask, the `C'-shaped bisector of the individual lines is not preserved in the CCF, but the overall shape should respond to the corresponding temporal variations experienced by the individual lines. 
 
Before we can explore the impact of granulation, we must exclude the impact of the p-mode oscillations; this is particularly important in RV-space where they may dominate over the granulation-induced shifts. We take two approaches to disentangling the granulation and p-mode effects.

\subsection{Data binning}
\label{sec71}
First, we bin the individual CCFs following \cite{2019AJ....157..163C}, to optimally average out the p-modes. Using the \texttt{OscFilter} script provided by \cite{2019AJ....157..163C}, we predicted in Sec.~\ref{sec32} that an exposure time of $\sim$12 min should result in a remaining RMS (due to the p-modes) of $10$~cm/s for HD~67458. However, to reach a similar (p-mode) RMS for the hotter and slightly evolved HD~88595 would require binning over $\sim$85 min.  We note that we have not propagated the uncertainties in the stellar parameters through \texttt{OscFilter} nor accounted for the variability in mode amplitudes. From Table~1 of \cite{2019AJ....157..163C}, we expect these uncertainties to be of order a few minutes for HD~67458, but they may be as large as $50$ minutes or more for HD~88595. For HD~67458, we account for this uncertainty by averaging over $20$~minute time bins; unfortunately, with $85$~minutes time bins for HD~88595 we are left with only $6$ points per night so we do not attempt to account for this uncertainty here. 

For each time bin, we sum all the CCFs that fall within the bin and then continuum normalize; uncertainties on the CCF flux are propagated accordingly. We then fit a Gaussian to the binned CCFs to determine the RV, full width at half maximum (FWHM), and contrast. To investigate any potential trends with S/N, we add the individual S/N, provided by the DRS for each order, in quadrature. We follow \cite{2019ApJ...879...55C} and measure a variety of shape diagnostics for the binned CCFs, including the FWHM, contrast, equivalent width (EW), bisector inverse span (BIS), bisector curvature, and bisector amplitude (A$_{\rm{b}}$). Both the BIS and bisector curvature split the CCF bisector into discrete regions. Here we use the standard definitions for BIS and curvature (BIS: top region is bounded by 10-40\% of the line depth, bottom region by 55-90\% of the line depth; curvature: top region is 20-30\%, middle is 40-55\%, 75-95\% of the line depth); we explore optimizing these definitions in Appendix~\ref{App_bis_curve_opt}\footnote{We are able to improve the correlation between BIS or curvature and RV by tweaking the top, bottom, and middle definitions, which may indicate we are identifying regions more sensitive to granulation. However, these regions change across nights and further exploration is needed to determine the root cause; see Appendix~\ref{App_bis_curve_opt} for more details.}. For nights with overlapping simultaneous CHEOPS and ESPRESSO observations (see Fig.~\ref{fig_dates}), we average the CHEOPS light curves over the same time bins as the ESPRESSO data. 
 
 In each instance, we explore the relationship between the shape indicator or brightness and RV on a night by night basis for each target. If we are able to detect the impact of granulation on the CCF shape, we expect to see correlations between the shape indicators, flux and RV that are consistent across the various observing nights and independent of S/N. Irregardless of the observations here, the CCF contrast, and to a lesser extent FWHM, is often correlated with S/N, especially at low S/N due to difficulties in accurately subtracting the background (bias level, dark current, diffuse inter-order background, sky background, etc.). Hence, any correlation between shape indicator and RV that is S/N-dependent should be taken with caution. 

We use the Spearman's Rank correlation coefficient\footnote{{We use the \texttt{spearmanr} function from the \texttt{scipy} Python library.}} ($\rho_S$) to assess, in a nonparametric way, the strength 
of any potential correlations between the shape indicator and the measured RV shift. The results are shown in Table~\ref{tab:ccf_corr}, and a subset of these are shown in Figs.~\ref{fig:shape_rv_hd67} and \ref{fig:shape_rv_hd88} for HD~67458 and HD~88595, respectively. 

For HD~67458, the FWHM, contrast, and EW show clear correlations with RV that are consistent across all three nights (see Fig.~\ref{fig:shape_rv_hd67}); however, these correlations are dependent on the S/N, as shown by the data color-coded, with the S/N from the 10th echelle order (there does not appear to be a dependence on
 \begin{center}
 \begin{figure}[t!] \centering
\includegraphics[trim=0.2cm .3cm 0cm 0cm, clip, scale=0.48]{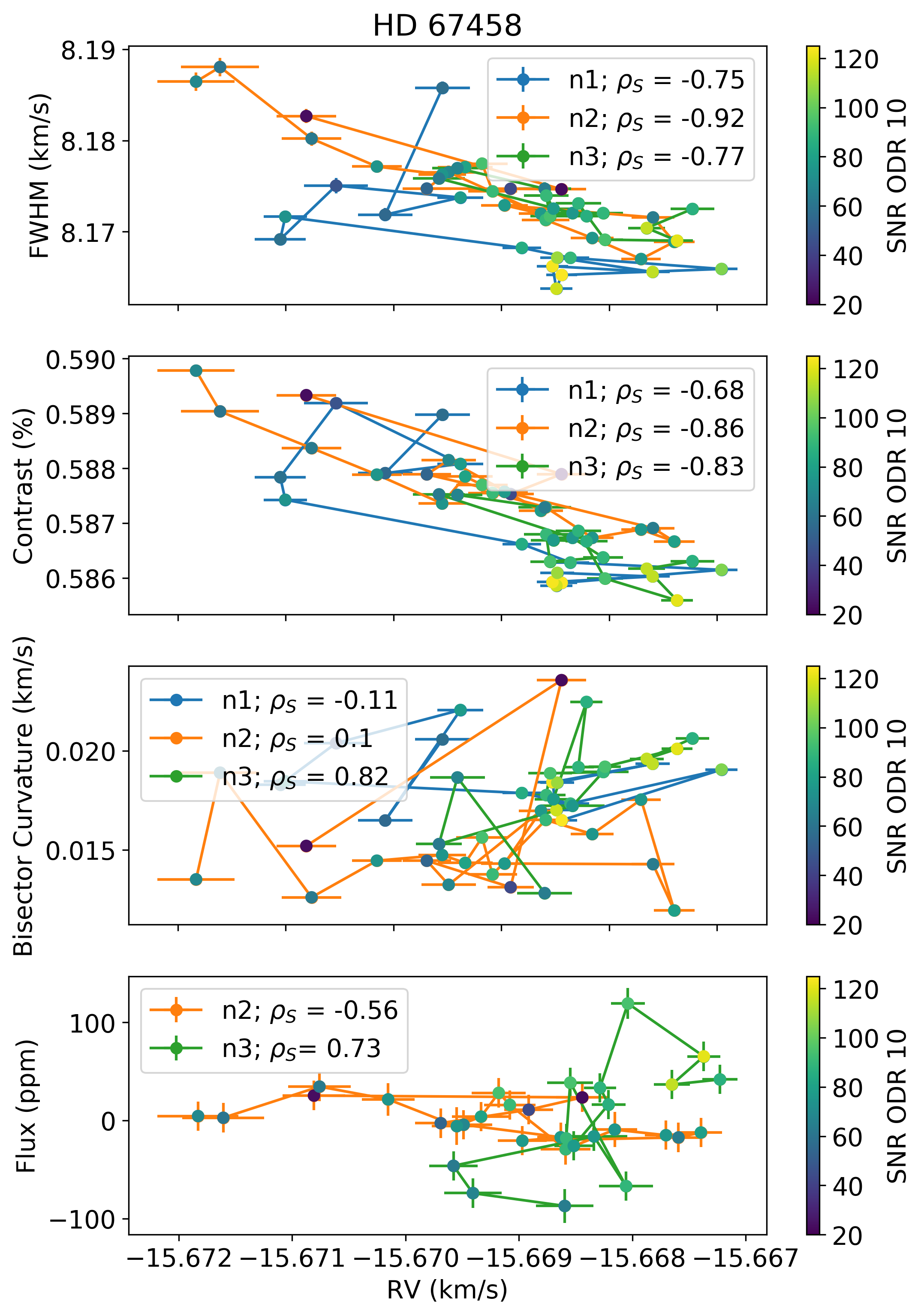}
\caption{Relationship between various shape indicators and brightness (y-axis) vs RV for HD~67458 shown as individual data points for CCFs binned in 20~minute time intervals, where errorbar color, and lines connecting data in time, indicates the night of the observation {($n1, n2, n3$)}. Each point is color-coded by the average S/N of the 10th echelle order. {The Spearman's correlation coefficients is indicated by $\rho_s$ for each night.}
}
\label{fig:shape_rv_hd67}  
\end{figure}
\vspace{-1cm}
\end{center}
 \begin{center}
 \begin{figure}[t!]
\centering
\includegraphics[trim=0.2cm .3cm 0cm 0cm, clip, scale=0.48]{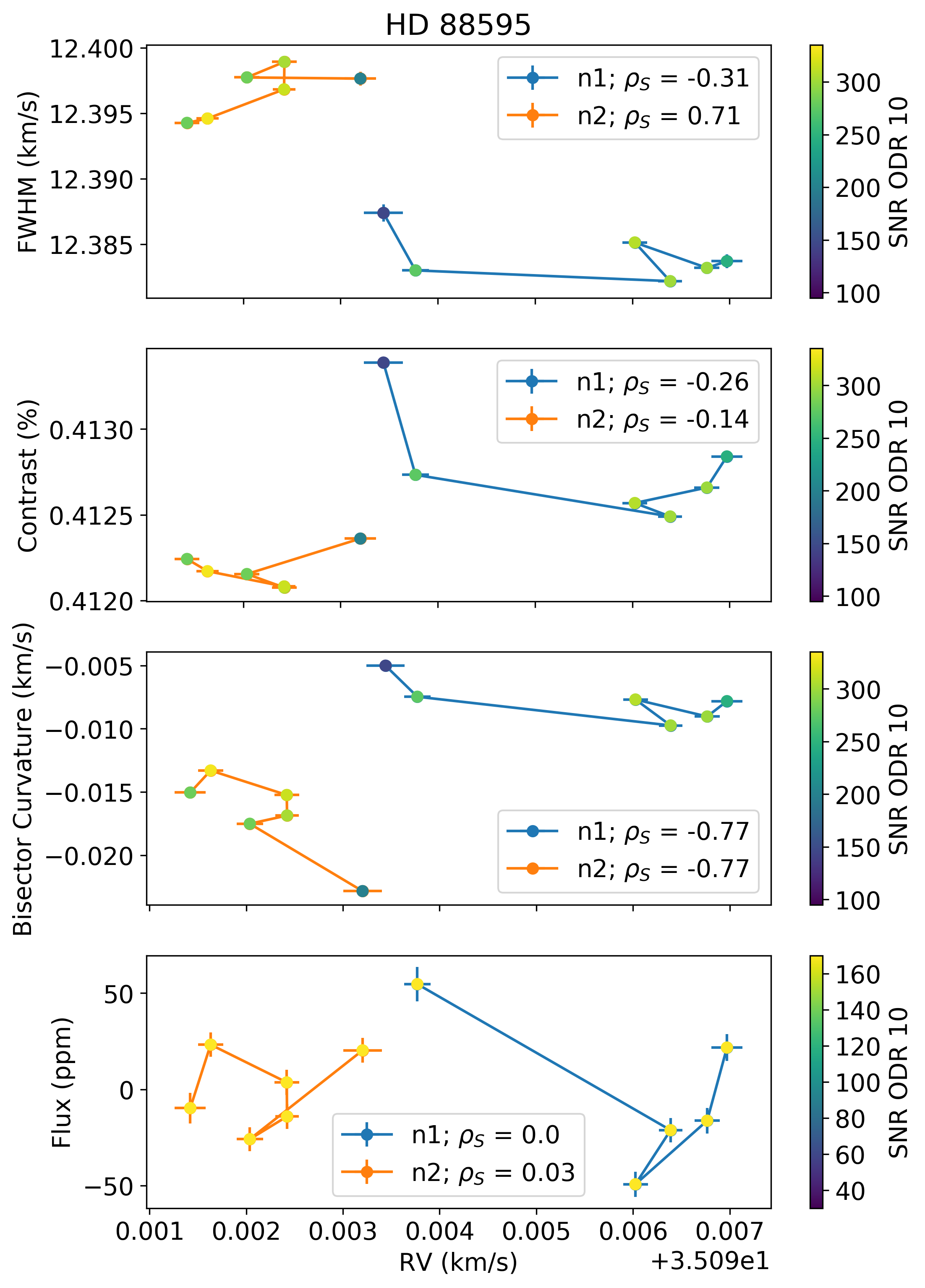}
\caption{Same as Fig.~\ref{fig:shape_rv_hd67}, but for HD 88595. For this star, the CCFs are binned over $85$ minutes.}
\label{fig:shape_rv_hd88}  
\end{figure}
\end{center}
airmass). The remaining shape indicators do not show any consistent or strong correlations with RV. The flux appears to correlate with RV, but changes from anticorrelated in the second night to positively correlated in the third night. The peak-to-peak RV variation of $\sim$5~m/s is also potentially larger than we might expect based on solar observations of granulation. For all these reasons, the correlations we do see are likely due to difficulties in accurately subtracting the sky background, rather than driven by true convection changes. Future observations under more ideal observing conditions may be key to confirm the nature of the observed correlations; alternatively, a more refined treatment of the p-modes may also help.

For HD~88595, with the exception of the bisector curvature, we do not see any strong correlations that are consistent across nights (see Fig.~\ref{fig:shape_rv_hd88}). Even with the bisector curvature there are still some inconsistencies; although the strength (and sign) of the correlation is similar, there are offsets in the actual values of the curvature and RV. We find that the bisectors between the two nights show a different shape, indicating the overall shape of the CCF is different between nights (see Fig.~\ref{fig:ccf_bisectors}). We also note that for the flux-RV exploration in the first night, there is a lack of CHEOPS observations at the start of the night that means there is no overlap in the first time bin and limited overlap in the second bin (the latter corresponding to the data point in bottom right subplot of Fig.~\ref{fig:shape_rv_hd88} with the highest flux).

At this stage, it is not clear if these differences in behavior between nights are driven by changes in the instrumental PSF or stellar changes happening on longer timescales, such as supergranulation. The two nights are separated in time by $\sim$8~days and although we would expect the supergranular lifetime to be longer than on the Sun (which is just under 2 days) this seems likely too long for the supergranulation to remain correlated; that said, there is not a strong enough understanding of stellar supergranulation to completely rule this out. If supergranulation were the dominant effect here (rather than the short-term granulation), then there might be tentative evidence for correlations across the nights (that often differ in sign from the nightly behavior). Unfortunately, the long duration required to mitigate the p-modes significantly reduces our sampling and prevents us from making strong conclusions here.

\subsection{Data filtering}
\label{sec72}

 \begin{figure*}[t!]
\centering
\includegraphics[trim=4.5cm 1.5cm 1.cm 1cm, clip, scale=0.4]{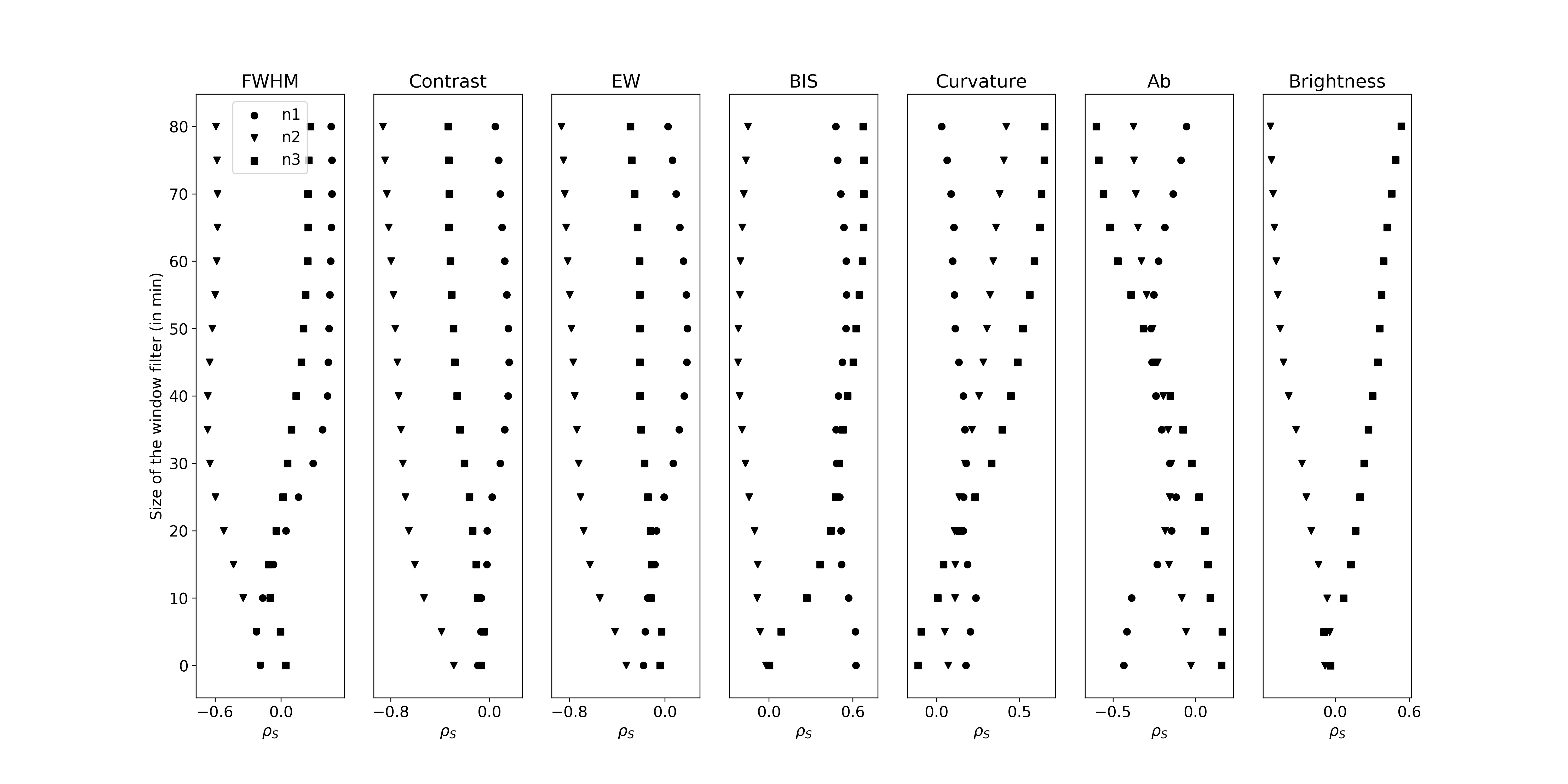}
\caption{HD~67458 correlation strengths for the relationship between (ESPRESSO) CCF shape indicator (FWHM, contrast, equivalent width [EW], BIS, bisector curvature, bisector amplitude [A$_{\rm{b}}$]) or (CHEOPS) flux and RV, as determined by a Spearman's rank correlation test, for various Butterworth filter cutoff frequencies for each night (n1, n2, and n3) as indicated by marker style. CHEOPS data (last column) was interpolated onto the ESPRESSO timestamps prior to filtering; no simultaneous observations were performed during n1.}
\label{fig:filt_shape_rv_hd67}  
\end{figure*}

 Motivated by the large timespans required to bin out the p-modes for HD~88595 (and likelihood for a large uncertainty on the optimal binning), we also attempt to filter out the p-mode impact in the RVs and shape indicators. For this, we apply a Butterworth filter provided by the Python library \texttt{scipy}. Here, we use the individual RVs, the FWHM and contrast provided by the ESPRESSO DRS. We fit each of the remaining shape indicators for each individual CCF. For nights where there is an overlap between CHEOPS and ESPRESSO observations, we (linearly) interpolate the CHEOPS lightcurves onto the ESPRESSO timestamps. Then, on a night by night basis, we apply the Butterworth filter to the RV and shape indicator time-series with a cutoff frequency corresponding to 5 min and increasing in 5 min intervals up to 80 and 255 min for HD~67458 and HD~88595, respectively (with the latter being approximately three times the estimated time to bin the p-modes down to $10$~cm/s RMS for this target). For each cutoff frequency, we asses the correlation strength between the filter RV and indicator time-series following the Spearman's Rank correlation coefficient. We repeat this analysis on the unfiltered (unbinned) data. 
 
 We examine the behavior of each indicator for all observed nights for each target, as well as the behavior of all indicators across a given night. If we are filtering out the p-modes with increasing cutoff frequency, then a given indicator should increase in correlation strength in a consistent manner across the nights. However, we do run the risk of over-filtering and flattening the signal, and therefore artificially creating a correlation (e.g., if both the RV and shape indicator or flux are constant over time then they will have a perfect correlation). If all indicators grow in correlation strength in a similar way with cutoff frequency, but behave differently on different nights, this may be a sign we are over-filtering the data. 
 
\begin{figure*}[t!]
\centering
\includegraphics[trim=4.5cm 1.5cm 1.cm 1cm, clip, scale=0.4]{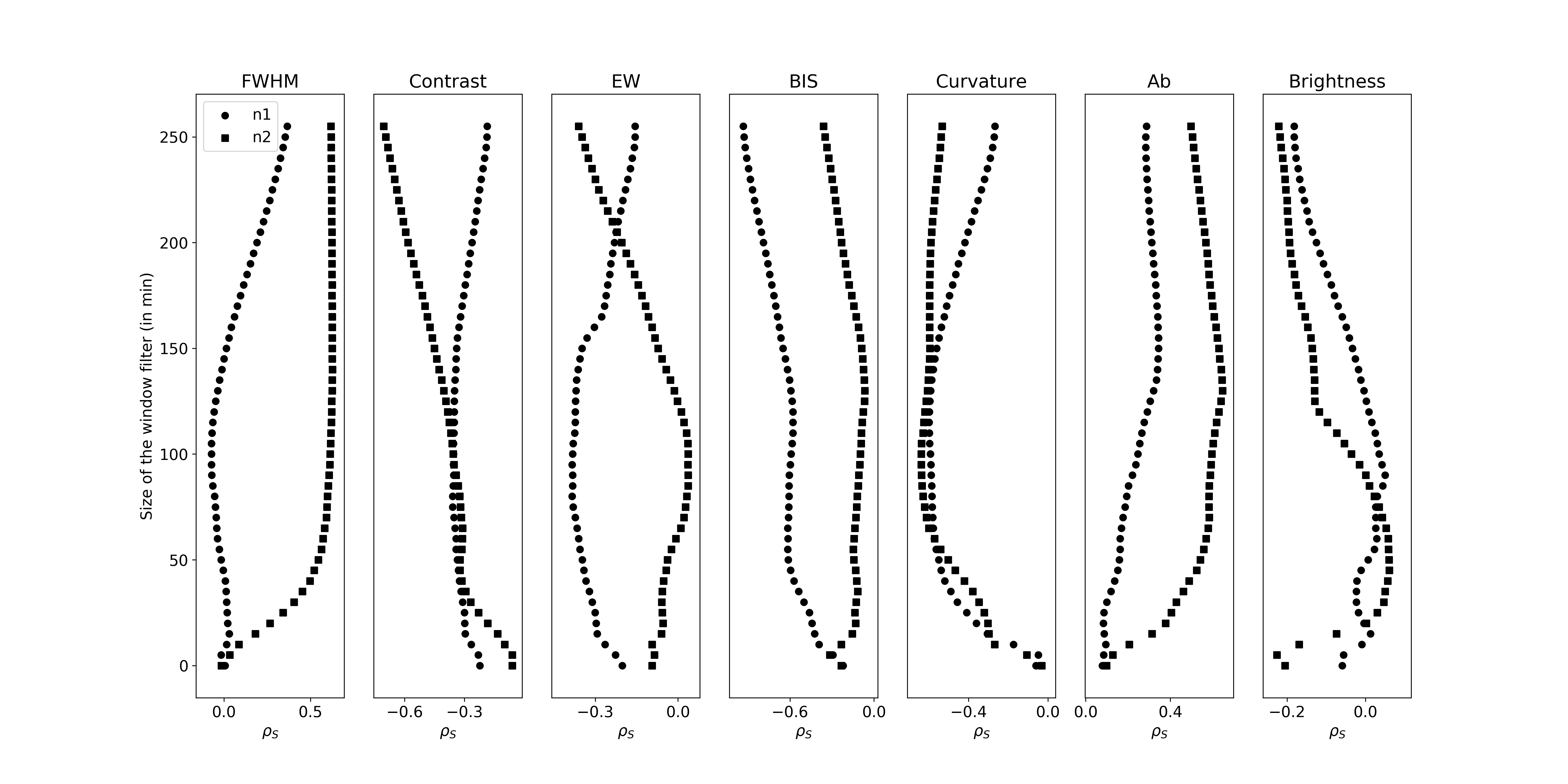}
\caption{Same as Fig.~\ref{fig:filt_shape_rv_hd67}, but for HD~88595.}
\label{fig:filt_shape_rv_hd88}  
\end{figure*}

 In general we find, for both targets and all nights, the correlation strength between RV and most shape indicators or flux tends to grow as a function of cutoff frequency,  as shown in Figs.~\ref{fig:filt_shape_rv_hd67} and \ref{fig:filt_shape_rv_hd88}. However, the behavior of the correlation strength as a function of cutoff frequency is largely inconsistent across nights for most indicators, for both targets. This is consistent with the binned analysis in Section~\ref{sec71}, where the sign of the correlations which did not trend with S/N changed across nights. The inconsistencies across nights here casts doubt on the detection of granulation-induced shape changes and the increase in correlation strength with cutoff frequency may indeed be due to simply flattening both time-series. The exception to this is the relationships with contrast and bisector curvature for HD~88595, where the correlation with RV increases in strength for cutoff frequencies corresponding to longer timescales in a similar manner across both nights. This also adds further evidence that the strong correlation between bisector curvature and RV in the binned data {in Section~\ref{sec71}} (Fig.~\ref{fig:shape_rv_hd88}) might be indeed be due to granulation.
 
 The relationship between flux and RV (as a function of cutoff frequency) is also somewhat consistent across nights; we note that the p-values associated with the Spearman rank correlations strengths indicate that a window size of $100-150$ minutes or more may be required to reject the null hypothesis, that means we may only be able to trust the correlations with a large filtering window. This behavior might explain why we were unable to detect an RV-flux correlation in the $80$-minute binned data in Section~\ref{sec71}.
 Interestingly, the difference in strength of the p-values associated with the Spearman rank correlation may further indicate the difference in data quality between nights (e.g., Fig.~\ref{fig_RVbin} shows how the RV RMS of the first nights is consistently larger than for the second night). 
 
 Nonetheless, a further refinement of the treatment of the p-modes is required to confidently confirm the nature of this behavior; this is beyond the scope of this paper and will be the subject of a future study. Additionally, future observations would also be useful to confirm the origin of the CCF shape differences between nights and the relationship between granulation-induced RV shifts and flux. 
 
\section{Conclusions}
\label{ccl}

The first goal of this paper was to detect the stellar granulation signal of two bright stars for the first time with high-precision photometric CHEOPS observations taken simultaneously or contemporaneously with high-resolution spectroscopic ESPRESSO observations. 
We detected the stellar granulation signal for the F star HD 88595 in both datasets, but only in the CHEOPS dataset for the solar-like star HD 67458 (Sec.~\ref{sec3}). {In particular, we observed significant variabilities between the three nights of observations of HD 67458 with ESPRESSO. Although we could not clearly identify the origin of these variabilities, we showed that they are induced by an excess of white noise in the observations masking the signatures of the stellar granulation (Sec. \ref{sec31}).}
We demonstrated the sensitivity of high-precision photometric CHEOPS observations to this stellar signal, which was not detected on TESS observations ({large photon noise and small amplitude of the granulation signal in the TESS passband}).

We also showed that the temporal binning of RV time series (i.e., the technique that is classically used to mitigate the amplitude of this stellar signal) is limited as this signal remains correlated over timescales $\tau>10-15$ min (Sec. \ref{sec32}). In particular, to reach the RV amplitude level of Earth-mass planets in the HZ of our two stars, we estimated that observations would need to be binned over $\tau \ge 180$ minutes for the G star (in agreement with the predictions of \citeads{2015A&A...583A.118M}) and over $\tau \gg 250$ minutes for the F star.

Computing the flicker index indicator, defined as the slope of the {periodograms}, we demonstrated that both this indicator and the flicker frequency are  correlated with the stellar parameters (Sec.~\ref{sec4}), which is in agreement with previous studies \citepads{2020A&A...636A..70S} based on Kepler observations.

The second goal of this study was to test the predictions from 3D hydrodynamic models of convection  (Sec.~\ref{sec5}). We observed photometric amplitudes with CHEOPS that are consistent with the expectation from 3D models: $48$ ppm for the solar-like star and $66$ ppm for the F star.  The flicker index and flicker frequency derived on synthetic granulation time series generated by these 3D model predictions are also in agreement with both Kepler and CHEOPS observations.

The last goal of this study was to link the spectroscopic and photometric signatures of convection for main-sequence stars with high-precision observations (Sec.~\ref{sec6}). We observed a very good match between the {periodograms} of CHEOPS and ESPRESSO observations for the F star in the frequency regime dominated by stellar granulation.
This was quantitatively confirmed by similar flicker indexes inferred on both datasets and this indicates that the spectroscopic signal of stellar granulation can be predicted by photometric observations.
Unfortunately, the RV precision of the ESPRESSO observations of the solar-like star HD 67458 was not good enough to precisely characterize this stellar signal (the errorbars were between $0.4$ and $2.7$ m/s against a signal with amplitude $<1$ m/s). Additional RV observations with precision $<1$ m/s are then still needed to confirm the strong link that we already observed in the Sun with VIRGO and GOLF observations.

Comparing the RV RMS predictions from photometric measurements lead us to invalidate the empirical relations derived in \citetads{2014AJ....147...29B}, while confirming the relations derived in \citetads{2014ApJ...780..104C}. We however demonstrated the high sensitivity of the data binning -- used to extract the photometric granulation signal -- on these RV RMS predictions. We argued that the amplitude of this stellar signal needs to be adapted on the stellar dependent granulation timescales and high-frequency noise level that affects its characterization. A perspective of this study will be to use a more robust technique to do so, as the one developed in \citetads{2018A&A...620A..38B}.
Indeed, in the context of the upcoming ESA PLATO mission and the extreme precision radial velocity (EPRV) surveys, the interplay between the photometric and spectroscopic observables will be key for mitigating this stellar noise source that limits the detection and characterization of both stellar oscillations and exoplanet signatures.  In line with \citetads{2020A&A...636A..70S}, we expect a high sensitivity to this stellar phenomenon with the future high-precision photometric PLATO observations (at least in the $24$-camera mode).

Finally, we also explored whether it was possible to detect the impact of granulation-induced changes on the CCF shape, corresponding induced RV shifts and brightness changes (Sec.~\ref{sec7}). This analysis was complicated by the presence of p-modes excited by the granulation and occurring on similar timescales. The RV shifts induced by the p-modes dominate over the granulation and efforts to mitigate and/or disentangle the p-modes likely also squash the granulation signal. Moreover, the long timescales required to bin or filter out the p-modes for the hotter F star (HD~88595) mean that even though the granulation signal is also expected to be larger, it is still equally tricky to disentangle. Nonetheless, there are potential hints that the CCF bisector curvature may provide information on the granulation signal, at least for the F star. If the p-modes are sufficiently filtered (i.e., with a cutoff frequency corresponding to $\sim$100-150~minutes or more), there also appears to be a correlation observed between flux and RV for the F star that is consistent across nights. Unfortunately, the G star (HD~67458) observations seem to limited by signal-to-noise as the apparent correlations between shape and RV are signal-to-noise dependent. Future observations are needed to further determine the impact of granulation on the overall brightness, CCF shape and net RV shifts. Additionally, a more refined treatment of the p-modes may also be key to further unveiling the granulation behavior; this is currently under analysis and will be the subject of a forthcoming paper.


\begin{acknowledgements}

The authors thank the anonymous referee for her/his helpful comments that improved the quality of the manuscript. Overall, the paper has been significantly improved in terms of both science and writing.

CHEOPS is an ESA mission in partnership with Switzerland with important contributions to the payload and the ground segment from Austria, Belgium, France, Germany, Hungary, Italy, Portugal, Spain, Sweden, and the United Kingdom. The CHEOPS Consortium would like to gratefully acknowledge the support received by all the agencies, offices, universities, and industries involved. Their flexibility and willingness to explore new approaches were essential to the success of this mission. 
The authors acknowledge the ESPRESSO project team for its effort and dedication in building the ESPRESSO instrument. This paper used data taken during the program IDs: 106.2186.001, 106.2186.002, 106.2186.003, 106.2186.004, 106.2186.005, 106.2186.006. 
This paper includes data collected by the TESS and Kepler missions, which are publicly available from the Mikulski Archive for Space Telescopes (MAST). Funding for these missions is provided by NASA's Science Mission directorate.
The VIRGO and GOLF instruments onboard SoHO are a cooperative effort of scientists, engineers, and technicians, to whom we are indebted. SoHO is a project of international collaboration between ESA and NASA. 
We acknowledge X. Dumusque, Astronomy Department Geneva Univ., Switzerland, for the public HARPS-N data that have been made available.\\
This research has made use of the VizieR catalog access tool, CDS, Strasbourg, France (DOI : 10.26093/cds/vizier). The original description of the VizieR service was published in 2000, A\&AS 143, 23\\
We acknowledges M. Oshagh for sharing the dataset from Oshagh et al. (2017).
SS acknowledges support from CNES, {the Programme National de Planétologie (PNP), and the Programme National de Physique Stellaire (PNPS) of CNRS-INSU.} 
M.L. and S.S acknowledge support of the Austrian Research Promotion Agency (FFG), under project number 859724 (GRAPPA).
M.L. acknowledges support of the Swiss National Science Foundation under grant number PCEFP2\_194576. The contributions of M.L. and A.K. have been carried out within the framework of the NCCR PlanetS supported by the Swiss National Science Foundation.
H.M.C. acknowledges support from the UKRI under a Future Leader Fellowship (grant number MR/S035214/1).
Funding for the Stellar Astrophysics Centre is provided by The Danish National Research Foundation (Grant agreement no.: DNRF106). L. F. R. D. acknowledges support from the Independent Research Fund Denmark (Research grant 7027-00096B).
The numerical results presented in this work were obtained at the Centre for Scientific Computing, Aarhus \url{http://phys.au.dk/forskning}. Part of the calculations have also been performed using the OCA/SIGAMM mesocentre. This research was undertaken with the assistance of resources provided at the NCI National Facility systems at the Australian National University through the National Computational Merit Allocation Scheme supported by the Australian Government.
V.V.G. is an F.R.S-FNRS Research Associate. 
ACC acknowledges support from STFC consolidated grant numbers ST/R000824/1 and ST/V000861/1, and UKSA grant number ST/R003203/1. 
PM acknowledges support from STFC research grant number ST/M001040/1. 
LBo, GBr, VNa, IPa, GPi, RRa, GSc, VSi, and TZi acknowledge support from CHEOPS ASI-INAF agreement n. 2019-29-HH.0. 
MF and CMP gratefully acknowledge the support of the Swedish National Space Agency (DNR 65/19, 174/18). 
S.G.S. acknowledge support from FCT through FCT contract nr. CEECIND/00826/2018 and POPH/FSE (EC). 
ACC and TGW acknowledge support from STFC consolidated grant numbers ST/R000824/1 and ST/V000861/1, and UKSA grant number ST/R003203/1. 
SH gratefully acknowledges CNES funding through the grant 837319. 
This project has received funding from the European Research Council (ERC) under the European Union’s Horizon 2020 research and innovation programme (project {\sc Four Aces}. 
YA and MJH acknowledge the support of the Swiss National Fund under grant 200020\_172746. 
We acknowledge support from the Spanish Ministry of Science and Innovation and the European Regional Development Fund through grants ESP2016-80435-C2-1-R, ESP2016-80435-C2-2-R, PGC2018-098153-B-C33, PGC2018-098153-B-C31, ESP2017-87676-C5-1-R, MDM-2017-0737 Unidad de Excelencia Maria de Maeztu-Centro de Astrobiologí­a (INTA-CSIC), as well as the support of the Generalitat de Catalunya/CERCA programme. The MOC activities have been supported by the ESA contract No. 4000124370. 
S.C.C.B. acknowledges support from FCT through FCT contracts nr. IF/01312/2014/CP1215/CT0004. 
XB, SC, DG, MF and JL acknowledge their role as ESA-appointed CHEOPS science team members. 
ABr was supported by the SNSA. 
This project was supported by the CNES. 
This project has received funding from the European Research Council (ERC) under the European Union's Horizon 2020 research and innovation programme (project {\sc Four Aces}, grant agreement No. 724427).  It has also been carried out in the frame of the National Centre for Competence in Research ``PlanetS'' supported by the Swiss National Science Foundation (SNSF). A.De. acknowledges the financial support of the SNSF. 
The Belgian participation to CHEOPS has been supported by the Belgian Federal Science Policy Office (BELSPO) in the framework of the PRODEX Program, and by the University of Liège through an ARC grant for Concerted Research Actions financed by the Wallonia-Brussels Federation. 
L.D. is an F.R.S.-FNRS Postdoctoral Researcher. 
This work was supported by FCT - Fundação para a Ciência e a Tecnologia through national funds and by FEDER through COMPETE2020 - Programa Operacional Competitividade e Internacionalizacão by these grants: UID/FIS/04434/2019, UIDB/04434/2020, UIDP/04434/2020, PTDC/FIS-AST/32113/2017 \& POCI-01-0145-FEDER- 032113, PTDC/FIS-AST/28953/2017 \& POCI-01-0145-FEDER-028953, PTDC/FIS-AST/28987/2017 \& POCI-01-0145-FEDER-028987, O.D.S.D. is supported in the form of work contract (DL 57/2016/CP1364/CT0004) funded by national funds through FCT. 
B.-O.D. acknowledges support from the Swiss National Science Foundation (PP00P2-190080). 
DG gratefully acknowledges financial support from the CRT foundation under Grant No. 2018.2323 ``Gaseousor rocky? Unveiling the nature of small worlds''. 
M.G. is an F.R.S.-FNRS Senior Research Associate. 
KGI is the ESA CHEOPS Project Scientist and is responsible for the ESA CHEOPS Guest Observers Programme. She does not participate in, or contribute to, the definition of the Guaranteed Time Programme of the CHEOPS mission through which observations described in this paper have been taken, nor to any aspect of target selection for the programme. 
This work was granted access to the HPC resources of MesoPSL financed by the Region Ile de France and the project Equip@Meso (reference ANR-10-EQPX-29-01) of the programme Investissements d'Avenir supervised by the Agence Nationale pour la Recherche. 
This work was also partially supported by a grant from the Simons Foundation (PI Queloz, grant number 327127). 
IRI acknowledges support from the Spanish Ministry of Science and Innovation and the European Regional Development Fund through grant PGC2018-098153-B- C33, as well as the support of the Generalitat de Catalunya/CERCA programme. 
GyMSz acknowledges the support of the Hungarian National Research, Development and Innovation Office (NKFIH) grant K-125015, a a PRODEX Experiment Agreement No. 4000137122, the Lend\"ulet LP2018-7/2021 grant of the Hungarian Academy of Science and the support of the city of Szombathely. 
NAW acknowledges UKSA grant ST/R004838/1.

\end{acknowledgements}

\newpage

\bibliographystyle{aa} 
\bibliography{Bibfile} 

\appendix

\section{Rotation period estimates based on TESS observations}
\label{App_Prot}

To measure the stellar rotation period of HD~88595 and HD~67458, we use the TESS light curves introduced in Sect.~\ref{sec24}. We analyze the light curves using the Generalized Lomb-Scargle periodogram \citep{2009A&A...496..577Z}, after having masked out the most obvious outliers and after applying a linear detrending in time to each sector.

\subsection{HD~88595}

This star has been observed in sectors 9 and 35. For the latter we detect a clear peak around $\sim3$~days, while for the former we get a broader comb of aliases between 1 and 10 days. The joint analysis of the two sectors returns a period of $3.1151\pm0.0003$~days and a photometric amplitude of $83\pm4$~ppm. The detected period would be consistent with rotation and with $v\sin{i} =7.2$~km~s$^{-1}$ assuming that the inclination of the rotation axis is $\sim16^\circ$, meaning the star is seen nearly pole on. In such a scenario, the active regions on the stellar surface are always visible (unless they have nearly equatorial latitudes), and this would explain why the star has a low photometric signal despite being a fast rotator.

\subsection{HD~67458}

This star has been observed in sectors 7, 8, and 34.

The light curve obtained in sector 8 is the most problematic. The data release note\footnote{\url{https://archive.stsci.edu/missions/tess/doc/tess_drn/tess_sector_34_drn50_v02.pdf}} reports noisy quaternions, and the periodogram is difficult to interpret. For all these reasons we exclude this sector from our analysis.

The periodogram of sector 7 suggests a periodicity of $5.80\pm0.03$~days, while for sector 34 we get a period of $10.57\pm0.06$~days. {In both cases the photometric amplitude is $\sim100$ ppm.}
{The joint analysis of the two sectors show significant peaks at these two periods in the periodogram. Since the two periods are roughly in a 1:2 proportion, we thus argue that the period detected in sector 7 is the first harmonic of the true rotation period.}
The detected period of $10.57\pm0.06$~days would be consistent with rotation and with $v\sin{i} =2.179$~km~s$^{-1}$ assuming an inclination of {$\sim26^\circ$}, meaning this star is also seen nearly pole-on.

\section{Exoplanets in the habitable zone}
\label{App_pl}

As a reference in the paper, we evaluate the planetary timescales (transit and in-egress durations, orbital period) and amplitudes (transit depth, RV semi-amplitude) of an Earth-like planet ($1$ $M_\oplus$, $1$ $R_\oplus$) that would orbit in the HZ of HD 67458 and HD 88595. For that purpose, we assume an impact parameter $b=0$, an eccentricity $e=0$, and the inclination to be $i=90^\circ$.
We use the stellar parameters given in Table~\ref{tab_params} and the classical equations given in \citetads{2018exha.book.....P}.

\noindent We compute the transit depth as
$$
\delta = \Big(\frac{R_p}{R_\star}\Big)^2,
$$
and obtain $\delta = 80.6$ ppm for HD 67458, and $\delta = 32.18$ ppm for HD 88596.

\noindent We compute the orbital separation of the planet in the middle of the HZ as 
$$
a = \sqrt{\frac{L_\star}{L_\odot}},
$$
with $L_\star = 4\pi \sigma R_\star^2 T_{\mathrm{eff}}^4$ the stellar luminosity in SI from the Stefan-Boltzmann law, $\sigma = 5.67\times  10^{-8}$  kg s$^{-3}$ K$^{-4}$ the Stefan-Boltzmann constant, and $L_\odot=3.828 \times 10^{26}$ W the solar luminosity according to IAU. We obtain $a= 1.04$ AU for HD 67458 and $a = 1.86$ AU for HD 88595. Using Kepler's 3rd law, we derive the corresponding orbital periods and find $P=402.3$ days for HD 67458 and $P=802.2$ days for HD 88595.

\noindent We express the total transit duration as
$$
T_{T} \approx \frac{P}{\pi}~\frac{(R_\star + R_p)}{a},
$$
and obtain $T_{T} = 14.12$ hours for HD 67458, and $T_{T} = 24.8$ hours for HD 88595.

\noindent The duration of full transit is computed as
$$
T_{F} \approx \frac{P}{\pi}~\frac{(R_\star - R_p)}{a},
$$
which gives a duration of the transit ingress expressed as
$$
T_{in} = \frac{T_{T}-T_{F}}{2}.
$$
We obtain $T_{in} = 7.5$ min for HD 67458, and $T_{in} = 8.4$ min for HD 88595.

\noindent Finally, the RV semi-amplitude of an Earth-mass planet is expressed in m/s as \citepads{2010exop.book...27L}
$$
K = 28.4329 ~\Big(\frac{M_p}{M_J}\Big)  \Big(\frac{M_\star + M_p}{M_\odot}\Big)^{-0.5}  \Big(\frac{a}{1~{\rm AU}}\Big)^{-0.5},
$$
with $M_J$ the Jupiter mass. We obtain $K=9$ cm/s for HD 67458 and $K=5.6$ cm/s for HD 88595.

\section{Solar granulation as seen from different instruments}
\label{App_solar_obs}

Stellar activity is wavelength-dependent. To evaluate this dependence with solar observations, we compare in this section the {periodograms} of VIRGO/SoHO solar irradiance (see Sec.~\ref{sec25}) with HMI/SDO photometric observations (see description in \citeads{2020A&A...636A..70S}). We also compare the GOLF/SoHO spectrophotometric data (see Sec.~\ref{sec25}) with HARPS-N spectroscopic observations\footnote{\href{http://cdsarc.u-strasbg.fr/viz-bin/cat/J/A+A/648/A103}{http://cdsarc.u-strasbg.fr/viz-bin/cat/J/A+A/648/A103}} (\citeads{2019MNRAS.487.1082C}, \citeads{2021A&A...648A.103D}). 

For each dataset, we compute the averaged periodogram based on $1$-day regularly sampled time series of VIRGO (year:1996), HMI (year:2008), and GOLF observations (year:1996). 
Since HARPS-N observations (years:2015-2018) are taken from the ground, they are irregularly sampled and their durations are around $6$-$8$-hours. For this reason, the Lomb-Scargle periodogram \citepads{1982ApJ...263..835S} is used to compute $P_L$  in Eq.~\eqref{eq_PL} for HARPS-N data, while the classical periodogram \citepads{schuster1898} is used for the other datasets. 

\begin{figure*}
\centering
\resizebox{\hsize}{!}{\includegraphics{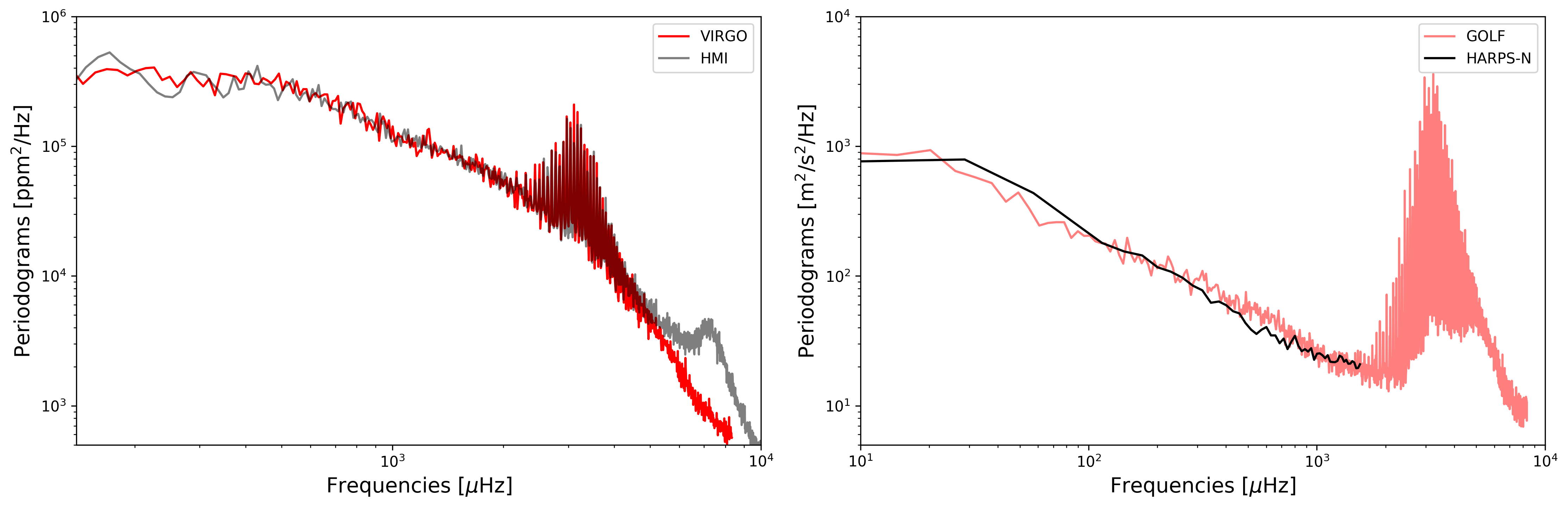}}
\caption{Comparison of averaged periodograms from VIRGO and HMI observations (left), and from GOLF and HARPS-N observations (right). Note the instrumental artifact affecting the high frequencies ($\nu \sim 7000~\mu$Hz) in HMI data.  
}
\label{fig_annSun}  
\end{figure*}

The resulting averaged periodograms are shown in Fig.~\ref{fig_annSun}. We observe a very good match between the slopes of the different periodograms in the frequency region dominated by the stellar granulation signal. While the amplitudes of this signal are wavelength-dependent (see Fig~\ref{fig_f8} for example), the flicker index, which is based on the periodogram's slope, is not {(at least, in first approximation)}. This makes this indicator a useful  diagnostic tool for granulation studies.

\section{F8 scaling relation from 3D stellar atmosphere models}
\label{appD_F8}

\begin{figure}[h!]
\centering
\resizebox{\hsize}{!}{\includegraphics{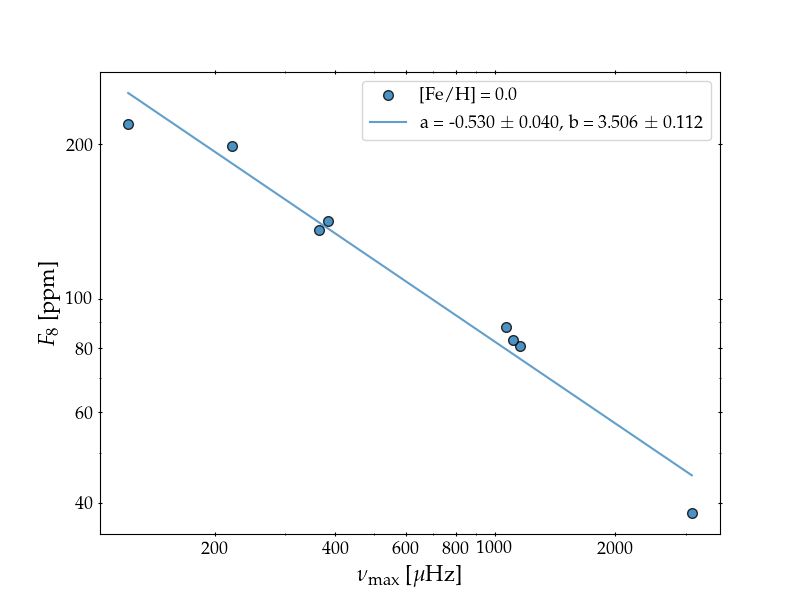}}
\caption{F8 values obtained for 3D stellar atmosphere models at solar metallicity compared to $\nu\rm_{max}$, and the derived scaling relation.
}
\label{F8-3Dmodels}  
\end{figure}

{Since one of the scaling relations provided by \citetads{Rodriguez2022} was for the standard deviation ($\sigma_{HD}$), we calculated a new scaling relation for F8, so we can compare directly with our two CHEOPS targets.

Table~\ref{tab:F8-3Dmodels} summarizes the F8 values for the models used and Fig~\ref{F8-3Dmodels} shows the F8 values for the 3D stellar atmosphere models at solar metallicity and with ${\rm log}g \geq 3.0$, together with a fit for the scaling relation, which is of the form:
\begin{equation*}
    \log_{10}{\rm F8}_{HD} = a\log_{10}{\nu_{\rm max}} + b.
\end{equation*}
We find $a = -0.530 \pm 0.040$ and $b=3.506 \pm 0.112$ leading to the relation F8$_{HD}\propto \nu_{\rm max}^{-0.530}$ that we used, instead on the RMS based relation $\sigma_{HD} \sim \nu_{\rm max}^{-0.567}$ initially published in \citetads{Rodriguez2022}.

}

\begin{table*}[t]\centering 
\caption{Stellar and physical parameters of the 3D stellar models at solar metallicity used to determine the scaling relation of F8. From left to right: model name, target effective temperature (target $T_{\rm eff}$, in K) corresponding to the $T_{\rm eff}$ that we aimed to achieve for each model,  mean $T_{\rm eff}$ (K) from the time series, surface gravity in logarithm scale (${\rm log}g$) (cm/s$^2$), metallicity [Fe/H], $\nu_{\rm max}$ ($\mu$Hz), standard deviation of the brightness fluctuations ($\sigma_{HD}$) (ppm), and F8$_{HD}$ (ppm).}
\label{tab:F8-3Dmodels}
\begin{tabular}{|c|c|c|c|c|c|c|c|} 
\hline
Model & Target $T_{\rm eff}$ & $T_{\rm eff}$ & ${\rm log}g$ & [Fe/H] & $\nu_{\rm max}$ & $\sigma_{HD}$ & F8$_{HD}$ \\
\hline
t50g30m00 & 5000 & 4960 & 3.00 & 0.0 & 121.516 & 219 & 286 \\
t47g32m00 & 4750 & 4727 & 3.25 & 0.0 & 221.351 & 199 & 220 \\
t55g35m00 & 5500 & 5516 & 3.50 & 0.0 & 364.387 & 136 & 145 \\
t50g35m00 & 5000 & 4958 & 3.50 & 0.0 & 384.345 & 142 & 153 \\
t65g40m00 & 6500 & 6413 & 4.00 & 0.0 & 1068.672 & 88 & 83 \\
t60g40m00 & 6000 & 5962 & 4.00 & 0.0 & 1108.355 & 83 & 88 \\
t55g40m00 & 5500 & 5462 & 4.00 & 0.0 & 1157.975 & 81 & 85 \\
t5777g44m00 & 5777 & 5759 & 4.44 & 0.0 & 3106 & 38 & 46 \\ 
\hline
\end{tabular}
\end{table*}

\section{Influence of observational parameters on the flicker index}
\label{App_params}

The flicker index is linked to several observational settings that affect its estimation. In this section, we investigate how the length of the subseries, the temporal sampling, and the level of high-frequency noise impact the inferred flicker indexes.

The flicker index is estimated from the averaged periodogram \eqref{eq_PL}, which is computed based on short-duration subseries. To evaluate the influence of the different parameters, we use VIRGO solar observations (see Sec.~\ref{sec25}).  

\begin{figure*}
\centering
\resizebox{0.75\hsize}{!}{\includegraphics{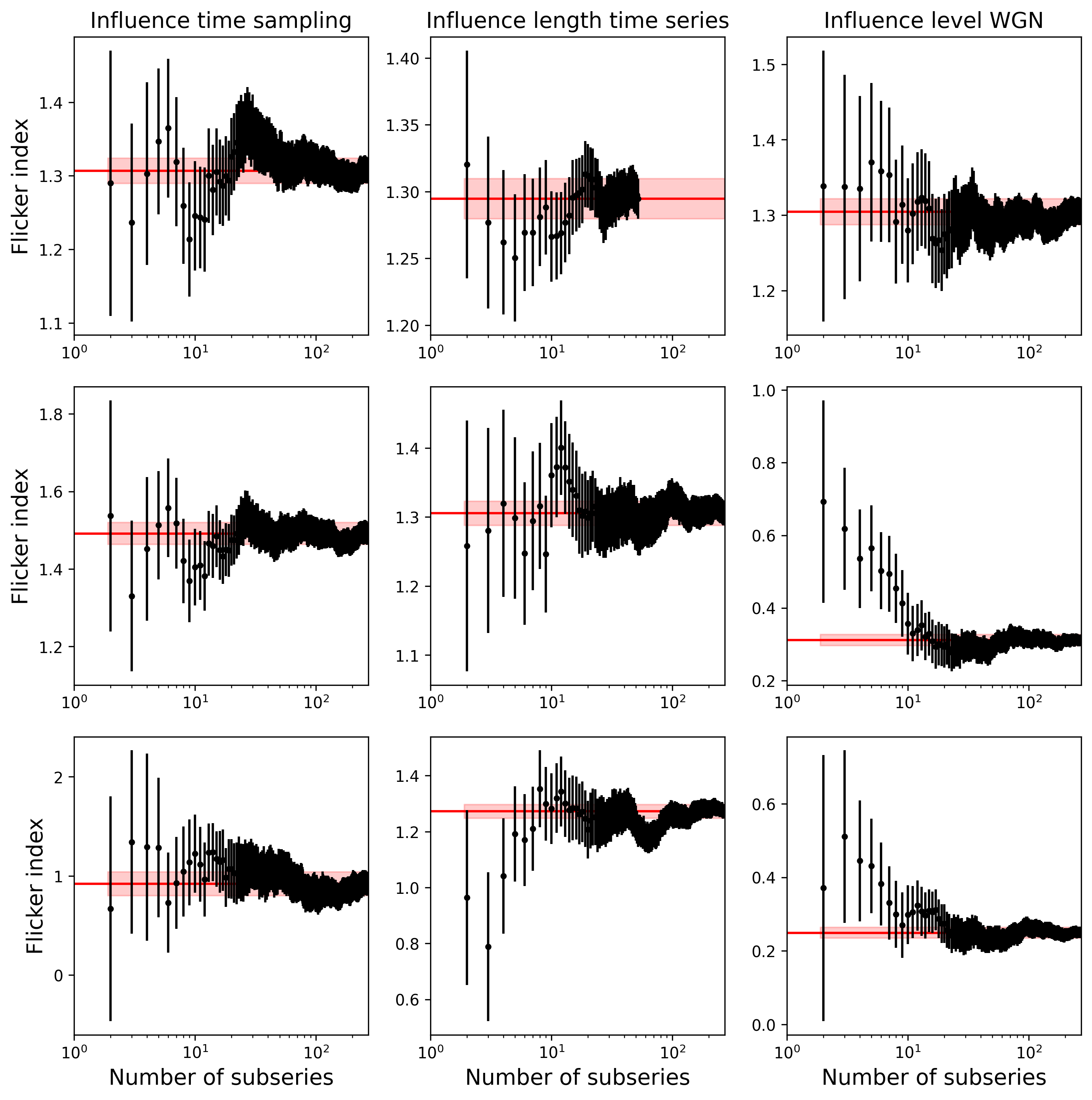}}
\caption{Inferred flicker indexes as a function of the number of subseries ($L$) used to compute the averaged periodogram in Eq.~\eqref{eq_PL} based on VIRGO solar data. \textit{Left column}: influence of the time sampling, $\Delta t=1, 5, 10$ min (from top to bottom). \textit{Middle column}: influence of the duration of the subseries, $T=5$ day, $1$ day, $8$-hours (from top to bottom). \textit{Right column}: influence of the high-frequency noise level, $\sigma_W < 30$ ppm (VIRGO data, top), $\sigma_W = 82$ ppm (level measured on CHEOPS HD 67458 periodogram), $\sigma_W = 92$ ppm (approximately the level measured on TESS HD 67458 periodogram). The red horizontal line indicates the  flicker index value for large $L$ ($\alpha_{g,\infty}$).  
}
\label{fig_uncert}  
\end{figure*}

Fig.~\ref{fig_uncert} shows the flicker index as a function of the number of $1$-day subseries ($L$) for different parameters. In all panels, the reference setting if $\Delta t =1$ min, $T=1$ day and $\sigma_w < 30$ ppm (VIRGO high-frequency noise level). The inferred flicker indexes converge toward a fixed estimate with the increase of $L$ (see red horizontal lines). We note this ``asymptotic'' estimate $\alpha_{g,\infty}$ in the following.

The first column shows the influence of the time sampling on the flicker index. From top to bottom, the subseries are sampled at $\Delta t= 1$ min, $5$ min, and $10$ min. 
For a solar-like star, we see that a too long temporal sampling, as  $\Delta t\ge 5$ min, affects the flicker index value since the high-frequency cut-off is cropped. 
The convergence of the inferred index toward $\alpha_{g,\infty}$ is fast ($L\le 10$), for all $\Delta t$, and no specific bias is observed at low $L$ (the asymptotic index $\alpha_{g,\infty}$ is within the $1\sigma$ errorbars). 
We note that, while $\Delta t$ is not expected to affect the flicker indexes obtained from CHEOPS and the future PLATO space data  (since they both have $\Delta t<5$ min), it can be critical for ground-based RV observations where the exposure time is, in general, $\ge 5$ min.

The second column shows the influence of the subseries duration on the inferred flicker index. From top to bottom, the subseries have duration of $5$ days, $1$ day, and $8$ hours. Again, we observe a fast convergence of the flicker index toward $\alpha_{g,\infty}$ for all $T$. It indicates that the relatively short length of our CHEOPS observations ($8$ hours) are enough to infer accurate flicker index. However, the low number of visits ($3$ to $4$) may slightly bias our inferred values. We note that long observations (e.g., $T=5$ days) does not improve the flicker index convergence toward $\alpha_{g,\infty}$.

The last column shows the influence of the high-frequency noise level on the inferred flicker index. From top to bottom, we added to VIRGO observation a WGN of standard deviation $\sigma_w=0$ ppm (our reference), $85$ ppm ($\approx$ CHEOPS high-frequency noise level), and $96$ ppm ($\approx$ TESS high-frequency noise level). As discussed in Sec.~\ref{sec42}, the index values logically decrease with the level of this high-frequency noise (that was largely dominated by photon noise for our two stars observed by CHEOPS).

\section{Flicker indexes of Kepler and CHEOPS bright targets}
\label{App_flicker}

In Table~\ref{tab:table_flicker}, we show the results from the flicker analyses on some targets described in the present study.   Kepler targets comes from \citetads{2020A&A...636A..70S}. The table shows the cut-off frequencies ($f_c$ and $f_g$) and  flicker indexes ($a_g$) inferred from the MCMC analyses for a time series affected by a high-frequency noise level $\sigma_W$ (see Sec.~\ref{sec41}). The errorbars indicated in the table are the more conservative ones. The flicker index interpolated at the high-frequency noise level of $30$ ppm is indicated by $\widehat{\alpha_g}(\sigma_{30})$. Stellar parameters for Kepler targets ($M_\star$, $R_\star$, $T_{\mathrm{eff}}$, ${\rm log}g$, mag) are from the \textit{Kepler\_stellar17.csv.gz} catalog available at \url{https://archive.
stsci.edu/kepler/catalogs.html}.

\renewcommand{\tabcolsep}{5pt}
\begin{table*}[t]\centering 
\caption{Results from the flicker analyses on some targets described in the present study. This table is available in its entirety in machine-readable form ($315$ targets). We note that values $\widehat{\alpha_g}(\sigma_{30})<0.2$ have been replaced by ``-1'' in the online table.
}
\label{tab:table_flicker}
\small
\begin{tabular}{|c|c|c|c|c|c|c|c|c|c|c|c|} 
\hline
Target & Mission & $f_c$  & $f_g$&  $\sigma_w$ & $a_g$ & $\widehat{\alpha_g}(\sigma_{30})$ & $M_\star$  & $R_\star$  & $T_{eff}$  & $\log{g}$ & mag \\
& &  [$\mu$Hz] &  [$\mu$Hz] & [ppm] &   &   & [$M_\odot$] & [$R_\odot$] &  [K] &  [cgs]  & \\
\hline
HD 67458 & CHEOPS & $3638$ & $480$ & $81$ & $0.526 \pm 0.120$ &  $0.888$ & $0.94 \pm 0.04$ & $1.02 \pm 0.02$ & $5833 \pm 62$ & $4.37\pm 0.10$ & $6.8$ \\
HD 88595 & CHEOPS & $1036$ & $351$ & $77$ & $0.608 \pm 0.493$ & $2.079$ & $1.35 \pm 0.06$ & $1.62 \pm 0.02$ & $6205 \pm 35$ & $3.99\pm 0.06$ & $6.5$ \\
KIC 1430163 & Kepler & $824$ & $378$ & $115$ & $0.61 \pm 0.05$ & $1.35$ & $1.29 \pm 0.08$ & $1.46 \pm 0.06$ & $6586 \pm 85$ & $4.22\pm 0.01$ & $9.6$ \\
KIC 2837475 & Kepler & $837$ & $350$ & $83$ & $0.99 \pm 0.24$ & $1.60$ & $1.34 \pm 0.07$ & $1.58 \pm 0.05$ & $6642 \pm 92$ & $4.16\pm 0.01$ & $8.5$ \\
KIC 3424541 & Kepler & $448$ & $251$ & $142$ & $1.36 \pm 0.14$ & $2.36$ & $1.37 \pm 0.21$ & $2.45 \pm 0.16$ & $6100 \pm 96$ & $3.80\pm 0.02$ & $9.7$ \\
KIC 3427720 & Kepler & $2029$ & $607$ & $95$ & $0.34 \pm 0.06$ & $0.78$ & $1.03 \pm 0.08$ & $1.09 \pm 0.04$ & $6045 \pm 81$ & $4.38\pm 0.01$ & $9.1$ \\
KIC 3456181 & Kepler & $642$ & $392$ & $124$ & $0.60 \pm 0.47$ & $1.15$ & $1.27 \pm 0.18$ & $2.03 \pm 0.15$ & $6372 \pm 77$ & $3.93\pm 0.01$ & $9.7$ \\
KIC 3656476 & Kepler & $1415$ & $385$ & $116$ & $0.65 \pm 0.03$ & $1.37$ & $1.03 \pm 0.10$ & $1.30 \pm 0.06$ & $5666 \pm 76$ & $4.22\pm 0.01$ & $9.5$ \\
KIC 3733735 & Kepler & $1320$ & $360$ & $84$ & $0.86 \pm 0.04$ & $1.52$ & $1.30 \pm 0.07$ & $1.38 \pm 0.05$ & $6676 \pm 80$ & $4.27\pm 0.01$ & $8.4$ \\
KIC 3735871 & Kepler & $2176$ & $357$ & $110$ & $0.27 \pm 0.01$ & $0.65$ & $1.05 \pm 0.08$ & $1.09 \pm 0.04$ & $6108 \pm 92$ & $4.39\pm 0.02$ & $9.7$ \\
\hline
\end{tabular}
\end{table*}

\section{Bisector inverse span and curvature optimization}
\label{App_bis_curve_opt}

We explore the impact of optimizing the BIS and curvature region definitions by searching for the regions of the bisector most sensitive to changes in net RV. This analysis is performed only on the binned data in Sec.~\ref{sec71} as the CCF shape could potentially be impacted by the p-modes in the individual data. For the optimal BIS and curvature range exploration, we employ two approaches: in the first instance we employ an MCMC (implemented with \texttt{EMCEE}; \citealt{emcee}) to explore the top and bottom (middle) definitions, and in the second case we search a fixed grid. 

For the MCMC approach, we start 100 walkers in a tight Gaussian ball near the standard definitions and let the walkers explore the parameters space for 1000 steps (excluding a burn-in of 100 steps). We set uniform priors to avoid the continuum and CCF core (excluding the top 10\% and bottom 95\% of the CCF) and to ensure the various regions are separate. We aim to minimize the residuals between the measured RV and that predicted by a linear fit with the given indicator. We note that this assumes a linear relationship between RV and indicators or flux, which may well not be the case; it also does not account for any potential time lags between RV and shape indicator. As such, this approach may not identify the regions of the bisector most sensitive to granulation, but it is a good first parameter space to search. 

In the second approach, we loop over various top and bottom (middle) region definitions in steps of 5\% of the CCF depth, requiring the range of a given region to be at least as large as one step size and that the regions do not overlap. Similar to the MCMC approach, we limit the search to 10-95\% of the CCF depth. The ``optimal'' region definitions were then those that maximized the correlation (as determined by the Spearman's Rank test) between the BIS or curvature and RV. This approach has the advantage that we maximally search all possible region definitions (and do not need to worry about convergence), but the disadvantage of having a fixed step size. If the relationship between BIS or curvature and RV is linear, then this is equivalent to the previous approach.

We consider instances where both approaches agree to be the most reliable. We find there are often multiple top, bottom, and middle definitions that could be considered optimal and the sign of the correlation can change depending on either the region definition or the night of observations. We note that a change in correlation sign can happen simply from moving to different regions within the bisector and should not be cause for alarm. Nonetheless the inconsistencies across nights indicate that we likely do not identify the regions of the CCF bisector most sensitive to granulation (and may be sensitive primarily to instrumental or telluric noise and/or low-number statistics). Nonetheless, we outline the individual results below.

\subsection{BIS results}
\label{App_sub_bis}
\begin{center}
\begin{figure}[t!]
\centering
\includegraphics[trim=4cm 0cm 4cm 0cm, clip, scale=0.5]{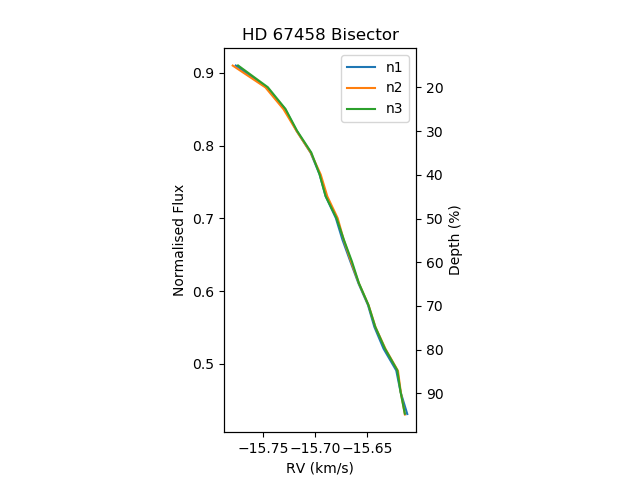}
\includegraphics[trim=4.cm 0cm 4.cm 0cm, clip, scale=0.5]{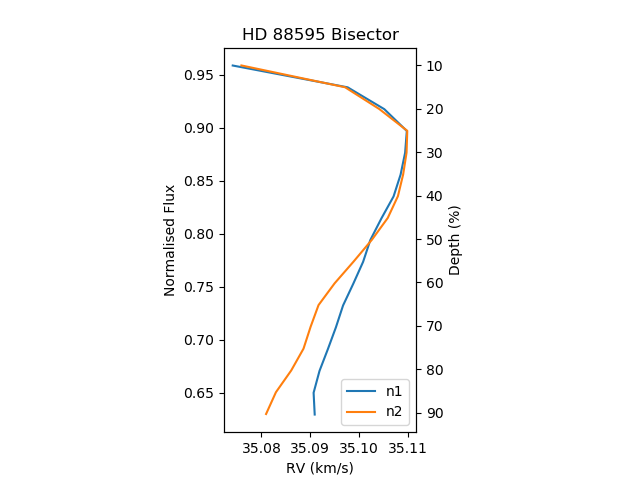}
\caption{Bisectors of the binned CCFs for HD~67458 (left) and HD~88595 (right) for all nights (night 1 indicated by `n1' etc.). See Section~\ref{sec71} for binning details.}
\label{fig:ccf_bisectors}  
\end{figure}
\end{center}

The CCF bisector for HD~67458 is fairly straight with a slight blue-ward bend near the top, and does not change significantly between nights (see Fig.~\ref{fig:ccf_bisectors}). Both nights 1 and 2 are optimized by probing a `top' region very high up the bisector ($\sim$15-20\% the CCF depth) and a `bottom' region defined just below the blue-ward bend in the CCF ($\sim$20-40\%). For night 3, this changes slightly to probe the bottom of this bend ($\sim$25-35\%) and near the middle of the bisector ($\sim$35-55\%). Oddly, for nights 1 and 3 we find a similar correlation, with a Spearman's rank correlation coefficient of $\sim$0.9, but for night 2 the correlation is opposite in sign and weaker, $\sim$ -0.6 to -0.8 depending on the exact range used. The reason for these differences is unclear, but given the correlations seen with S/N in Sec.~\ref{sec7} we may need future observations under more ideal observing conditions to reach any firm conclusions. Alternatively, an optimized CCF mask or a line-by-line approach may be needed as the current CCF template smears out too much of the bisector curvature; this is the subject of a forthcoming analysis for a follow-up paper. 

As noted in Sec.~\ref{sec7}, the bisectors for HD~88595 change significantly in shape between the two nights observed (see Fig.~\ref{fig:ccf_bisectors}). The bisector on night 1 is more vertical, with less blue-shift near the CCF core compared to night 2; this is in agreement with the fact that the FWHM is larger and contrast shallower for night 1 (see Fig.~\ref{fig:shape_rv_hd88}). The broader CCF in night 2 might be linked to the lower RV precision we find for this night. The backwards bend in the bisector is far more prominent for this star, indicative of the more vigorous convection with larger contrasts between granules and inter-granular lanes; this bend occurs at a similar depth in both nights. Since binning to help mitigate the effects of the p-modes only leaves 6 points per night, efforts to find the bisector ranges most sensitive to granulation via maximizing the correlation with RV should be taken with extreme caution. For night 1, the MCMC approach points toward a `top' region of $\sim$10-30\% (above the bend in the bisector) and a `bottom' region of $\sim$70-75\% and a Spearman's rank correlation coefficient near -0.9, while the nested loop approach points toward $\sim$20-40\% (top) and $\sim$30-60\% (bottom) with a correlation coefficient closer to -1. Interestingly, ranges of 55-60\% and 60-65\% can lead to a positive correlation coefficient near 1. For night {2}, both approaches find optimal regions near 35-45\% and 70-80\% for the top and bottom, respectively, and a positive correlation coefficient near 1. With the nested loop approach we find a similar correlation sign and strength with a narrow region defined very near the CCF core (85-90\% and 90-95\%) that might simply be tracing centroid shift rather than any true asymmetry change. The differences between lines is probably being driven by whatever is driving the larger shape changes between the nights, potentially linked to observing conditions since a broader CCF also leaders to poorer RV precision. 

\subsection{Bisector curvature results}
\label{App_sub_bicurve}
For HD~67458, on night 1 we find strong correlations between RV and bisector curvature regions (with a Spearman's rank coefficient near 1) if the top, middle, and bottom regions are approximately 15-20\%, 35-50\%, and 55-60\% of the CCF line depth. For night 2, the strongest Spearman's rank correlation coefficient is near -0.8, with regions of 10-15\%, 35-40\%, and 40-60\% (found with the nested loop approach -- the maximum correlation found with the MCMC approach was near 0.7 for regions near 30-35\%, 60-70\%, and 75-80\%). Similar to the BIS results, night 3 behaved somewhat similar to night 1 with a positive correlation coefficient close to 0.9, for regions near 20-30\%, 35-75\%, and 80-95\%. Interestingly, the optimal regions for nights 1 and 2 probe the top two thirds of the CCF, while for night 3 nearly the whole bisector is probed; this is somewhat similar to the BIS results where nights 1 and 2 favor higher up regions in the bisector as compared to night 3.  

For HD~88595, there are numerous combinations of top, middle, and bottom region definitions that yield similar results, likely owing to the low number of data points for each night of the binned data. For night 1, the optimal combinations found with the MCMC approach were narrowly defined top regions at the blue-ward bend of the bisector ($\sim$25-30\% of the CCF depth), a large middle region spanning most of the bisector ($\sim$40-80\%), and a narrow region near the bottom ($\sim$80-90\%); this combination yielded Spearman's rank correlation coefficients near -0.9. The nested loop approach yielded 272 different combinations that were equally strongly correlated, both positively and negatively, between RV and curvature; combinations that probed the top two thirds were negatively correlated, while combinations probing the bottom two thirds were positively correlated. For night 2, the MCMC approach led to optimal regions defined with a narrow regions above the blue-ward bisector bend ($\sim$10-20\%), a region just below the bend ($\sim$35-55\%), and a region near the bottom of the bisector ($\sim$70-95\%), with a correlation coefficient near -0.9. The nested loop approach yielded 26 combinations of similar strength; 25 of which led to anticorrelations near -1 (the exception, with a positive correlation, was regions 40-55\%, 65-75\%, and 75-80\%). 

On one hand, since we can optimize the BIS and bisector curvature regions to reveal strong correlations with RV, it is possible that we might be probing the impact of granulation. However, since the bisectors do not seem to behave in a coherent way across the nights, it is difficult to confirm the driving force behind the changes we see. Forthcoming work will include a refined treatment of the p-modes and an optimization to mitigate the effects of the current CCF template masks smearing out the individual line shapes and curvatures. This work would also benefit from further observations, {potentially at higher resolution}. 

\vspace{-25pt}

\end{document}